# Interference Management in 5G and Beyond Networks

Nessrine Trabelsi[a], Lamia Chaari Fourati[b] and Chung Shue Chen[c]

[a]*Laboratory of Signals Systems Artificial Intelligence Networks (SM@RTS/CRNS), University of Sfax, Sfax, 3021, Tunisia*
[b]*Laboratory of Signals Systems Artificial Intelligence Networks (SM@RTS/CRNS), University of Sfax, Sfax, 3021, Tunisia*
[c]*Nokia Bell Labs, 12 Rue Jean Bart, Paris-Saclay, France*

## ABSTRACT

During the last decade, wireless data services have had an incredible impact on people's lives in ways we could never have imagined. The number of mobile devices has increased exponentially and data traffic has almost doubled every year. Undoubtedly, the rate of growth will continue to be rapid with the explosive increase in demands for data rates, latency, massive connectivity, network reliability, and energy efficiency. In order to manage this level of growth and meet these requirements, the fifth-generation (5G) mobile communications network is envisioned as a revolutionary advancement combining various improvements to previous mobile generation networks and new technologies, including the use of millimeter wavebands (mm-wave), massive multiple-input multiple-output (mMIMO) multi-beam antennas, network densification, dynamic Time Division Duplex (TDD) transmission, and new waveforms with mixed numerologies. New revolutionary features including terahertz (THz) communications and the integration of Non-Terrestrial Networks (NTN) can further improve the performance and signal quality for future 6G networks. However, despite the inevitable benefits of all these key technologies, the heterogeneous and ultra-flexible structure of the 5G and beyond network brings non-orthogonality into the system and generates significant interference that needs to be handled carefully. Therefore, it is essential to design effective interference management schemes to mitigate severe and sometimes unpredictable interference in mobile networks. In this paper, we provide a comprehensive review of interference management in 5G and Beyond networks and discuss its future evolution. We start with a unified classification and a detailed explanation of the different types of interference and continue by presenting our taxonomy of existing interference management approaches. Then, after explaining interference measurement reports and signaling, we provide for each type of interference identified, an in-depth literature review and technical discussion of appropriate management schemes. We finish by discussing the main interference challenges that will be encountered in future 6G networks and by presenting insights on the suggested new interference management approaches, including useful guidelines for an AI-based solution. This review will provide a first-hand guide to the industry in determining the most relevant technology for interference management, and will also allow for consideration of future challenges and research directions.

## 1. Introduction

Over the last decade, mobile technology has greatly transformed and shaped our daily lives. According to the Ericsson mobility report published in November 2021 [1], the deployment of 4G networks has generated more than 5.5 billion new smartphone subscriptions worldwide since 2011 and a nearly 300 times growth in mobile data traffic. This explosive growth will probably continue and be even faster, with the deployment of 5G networks, than for any previous mobile generation. 5G, which is on track to become the fastest deployed mobile network ever, has already accounted for 660 million subscriptions at the end of 2021 and is forecasted to cover about 75 percent of the world's population in 2027 with 4.4 billion subscriptions globally.

Compared to previous mobile generations, 5G and beyond networks are aimed to be a revolutionary leap forward to meet the demands of different users for higher data rates, lower latency, and increased reliability [2]. To fulfill the ambitious requirements, 5G New Radio (NR), which development started in 3GPP Release 15, relies on several key technologies including the use of higher frequency bands (mm-wave bands) [3] offering larger bandwidths, the massive use of multiple beams antennas [4], new waveforms [5] with multiple numerologies, network densification [6, 7] and dynamic time division duplex [8], etc. Unlike the one-size-fits-all 4G network, the 5G and beyond network is expected to be ultra-flexible to dynamically adapt to the various emerging use cases [2]. However, this flexibility comes at a cost, as the increasing deployment of heterogeneous network infrastructure and the design of disparate technology models generate unprecedented levels of interference that can severely limit network performance. Therefore, the design of effective interference management schemes is undoubtedly a hot research topic.

Prior to this work, there were several survey papers which presented the interference management for 4G or/and 5G networks. Some recent surveys include new approaches for 6G networks. To begin with, we use Table 1 to list and provide a quick comparison. We will introduce each of them in the following.

In [9, 10], inter-cell interference management schemes in Orthogonal Frequency Division Multiple Access (OFDMA)-based cellular networks were considered with a focus, especially in the latter study on Inter-Cell Interference Coordination (ICIC) avoidance approaches in the Downlink (DL). In [9], different interference management techniques in OFDMA femto cell networks to cope with co-tier and cross-tier interference were presented and qualitatively compared. In [10], various ICIC avoidance schemes were classified into static and dynamic schemes and extensively reviewed. Y. Zhou et

# Nomenclature

| | | | | | |
|---|---|---|---|---|---|
| 3GPP | 3rd Generation Partnership Project | IC-IBI | Inter-Cell Inter-Beam Interference | PHY | PHYsical layer |
| ABS | Almost Blank Subframe | ICI | Inter-Cell Interference | QoS | Quality of Service |
| ACI | Adjacent Channel Interference | ICIC | Inter-Cell Interference Coordination | RAN | Radio Access Networks |
| AI | Artificial Intelligence | INI | Inter-Numerology Interference | RB | Resource Block |
| AP | Access Point | IoT | Internet of Things | RF | Radio Frequency |
| BER | Bit Error Rate | IRS | Intelligent Reflecting Surface | RI | Remote Interference |
| BLER | BLock Error Rate | | | RIM | Remote Interference Management |
| BS | Base Station | ISAC | Integrated Sensing And Communication | RRM | Radio Resource Management |
| BWP | BandWidth Part | ITU | International Telecommunication Union | RS | Reference Signal |
| CA | Carrier Aggregation | | | RSMA | Rate Division Multiple Access |
| CC | Component Carrier | IW | Interference Whitening | RSRP | Reference Signal Received Power |
| CCI | Co-Channel Interference | LTE | Long Term Evolution | | |
| CF mMIMO | Cell-Free massive MIMO | LTE-A | LTE-Advanced | SA | Stand-Alone |
| CLI | Cross-Link Interference | MA | Multiple Access | SC | Small Cell |
| CoMP | Coordinated Multi Point | MARL | Multi-Agent Reinforcement Learning | SC-FDM | Single Carrier Frequency Division Multiplexing |
| CP | Cyclic Prefix | MC | Macro Cell | SC-FDMA | Single Carrier Frequency Division Multiple Access |
| CRE | Cell Range Extension | MDP | Markov Decision Process | SCMA | Sparse Code Multiple Access |
| CS | Coordinated Scheduling | MIMO | Multiple Input Multiple Output | | |
| CSI | Channel State Information | mm-wave | millimeter wave | SCS | SubCarrier Spacing |
| D-TDD | Dynamic TDD | mMIMO | massive MIMO | SDN | Software-Defined Network |
| D2D | Device-to-Device | MMSE | Minimum Mean Square Error | SFR | Soft Frequency Reuse |
| DL | DownLink | mMTC | massive Machine Type Communications | SI | Self Interference |
| DRL | Deep Reinforcement Learning | | | SIC | Successive Interference Cancellation |
| eICIC | enhanced ICIC | ms | millisecond | SINR | Signal-to-Interference-plus-Noise Ratio |
| eMMB | enhanced Mobile Broadband | MUI | Multi-User Interference | SRS | Sounding Reference Signal |
| eNB | eNodeB | NGMA | Next Generation Multiple Access | STIN | Space-Terrestrial Integrated Network |
| FD | Full-Duplex | NOMA | Non-Orthogonal Multiple Access | TD-LTE | Time Division LTE |
| FDD | Frequency Division Duplex | NR | New Radio | TDD | Time Division Duplex |
| FeICIC | Further enhanced ICIC | NSA | Non-Stand-Alone | THz | TeraHertz |
| FFR | Fractional Frequency Reuse | NTN | Non-Terrestrial Networks | UAV | Unmanned Aerial Vehicle |
| FR | Frequency Reuse | OFDM | Orthogonal Frequency Division Multiplexing | UDN | Ultra Dense Network |
| FR1 | Frequency Range 1 | | | UE | User Equipment |
| FR2 | Frequency Range 2 | OFDMA | Orthogonal Frequency Division Multiple Access | UL | Uplink |
| GB | Guard Band | | | URLLC | Ultra Reliable Low Latency Communications |
| gNB | gNodeB | OMA | Orthogonal Multiple Access | | |
| HetNets | Heterogeneous Networks | PC | Power Control | ZF | Zero Forcing |
| IA | Interference Alignment | | | | |
| IBFD | In-Band-Full-Duplex | | | | |
| IBI | Inter-Beam Interference | | | | |
| IC | Interference Cancellation | | | | |

al., in [12], overviewed the principal inter-cell interference management techniques used in multiple generations from 2G to 4G and introduced some advanced schemes for 5G systems. In [13, 14], ICIC techniques applied in LTE/LTE-A networks were surveyed. In the former study, the focus was on traditional frequency reuse schemes. Simulations were performed to compare the performance of various methods with frequency reuse 1. The second survey investigated the evolution of inter-cell interference management schemes from ICIC to enhanced ICIC (eICIC) and Further eICIC (FeICIC) to more advanced techniques such as Coordinated Multi-Point (CoMP) in LTE-A Heterogeneous Networks (HetNets). Resource allocation based on Carrier Aggregation (CA) was also highlighted as an efficient inter-cell interference management scheme. While in [14], the author overviewed several standardized ICIC techniques, the study in [15] focused on time domain eICIC schemes and their joint optimization with other RRM methods such as user association and power control (PC) in LTE HetNets.

| [Ref] + Year | Network Environment | Scope |
|---|---|---|
| [9] 2012 | OFDMA femto cell networks | Interference scenarios and resource management |
| [10] 2013 | OFDMA-based cellular networks | ICIC avoidance schemes in the downlink |
| [11] 2014 | Multi-tier networks | Cell association and power control schemes |
| [12] 2014 | From 2G to 4G networks + Insights on 5G networks | ICI management schemes in 2G-4G networks and promising schemes for 5G networks |
| [13] 2015 | LTE networks | A comprehensive survey on ICIC techniques and their performance study compared to frequency reuse-1 system |
| [14] 2015 | LTE/LTE-A HetNets | An extensive survey on ICIC management techniques |
| [15] 2019 | LTE HetNets | eICIC standardization and time-domain eICIC schemes |
| [16] 2019 | 5G HetNets and D2D communications | Interference management techniques (ICIC, CoMP, CS) |
| [17] 2019 | SDN-5G/6G networks | Partly about interference management schemes and a general software-defined based framework for interference control |
| [18] 2020 | D-TDD networks | Cross-link interference mitigation |
| [19] 2020 | 5G HetNets | Radio resource management schemes |
| [20] 2020 | LTE-A and 5G | A comprehensive survey on self interference |
| [21] 2020 | 4G and 5G networks | Radio resource management schemes |
| [22] 2021 | FDD LTE-A and 5G | Self interference cancellation |
| [23] 2021 | 5G HetNets | Resource allocation (models, algorithms) |
| [24] 2021 | 5G and Beyond (HetNets, Relay Nodes, D2D, IoT | Interference management |
| [25] 2023 | SDN-5G and Beyond networks | Routing-based interference mitigation with different SDN-based architectures |
| [26] 2023 | 5G networks and IoT | Study on interference impact and possible optimization techniques |
| [This survey paper] 2023 | 5G and Beyond (with a focus on HetNets) + Insights on 6G networks | A comprehensive survey on interference classification and interference management schemes for each interference type |

Table 1: Most Recent Survey Papers Related to Interference Management in Mobile Networks

While in [11], interference management was discussed from the perspective of user cell association and power control mechanisms in multi-tier networks, the authors in [16] reviewed interference mitigation techniques such ICIC, CoMP, and Coordinated Scheduling (CS) in the context of Device-to-Device (D2D) communications and 5G HetNets. The authors in [19] and [23] also focused on 5G Heterogeneous Networks. In [23], resource allocation models and algorithms were investigated thoroughly, whereas, in [19], an overview of radio resource management (RRM) issues including spectrum allocation and interference mitigation was presented. T. Akhtar et al. also provided in [21] an extensive review of the key challenges of RRM and the different proposed approaches in LTE and 5G networks.

The surveys in [18, 20, 22] focused on a single type of interference. On one hand, in [18], cross-link interference mitigation schemes in dynamic TDD (D-TDD) networks were classified and intensively discussed through academic work and 3GPP specifications. On the other hand, in [20] and [22], the authors provided insight into Self Interference (SI) perceived in LTE and 5G transceivers, and on SI mitigation architectures and mechanisms. In one of the most recent surveys related to interference management in 5G networks, the authors presented in [24] various types of interference, their effects, and management schemes, with a focus on HetNets, Relay Nodes, D2D, and IoT.

While all mentioned surveys focused on interference management in 4G or/and 5G networks, [17] considered 5G and 6G software-defined network (SDN) frontier technology, including system architecture, mobility management, and current interference management techniques in SDN-5G/6G-based wireless networks [27]. The study classified interference as a negative or positive factor, according to whether we can avoid or exploit it, and presented a general software-based framework for the management of interference. SDN-based interference mitigation was also studied in [25], where the authors analyzed the potential of SDN by evaluating current solutions for routing-based interference mitigation in B5G networks. In [26], the impact of interference on different IoT services was studied, as well as the most important factors to take into account in assessing the performance of these services in the face of interference.

Most of the mentioned studies dealt with inter-cell interference without studying the intra-cell interference [9, 10]. Many surveys focused on one interference type [18, 20, 22], such as self interference or cross-link interference. No survey from the above-listed studied remote interference or inter-numerology interference. Some survey papers only considered one or a few interference management categories [13, 15]. Although many research works considered interference management in recent years, there is no available survey that summarizes all interference types with corresponding management schemes.

In this review paper, we seek to provide a unified classification of the main types of interference that can degrade the performance of 4G/5G and beyond networks, as well as a taxonomy and the most recent research work addressing advanced interference management schemes. To the best of our knowledge, no previous survey has so broadly and comprehensively investigated and reviewed interference management schemes for 5G and beyond networks. The main contributions of this paper are listed below:

- Section 2 provides an overview of the major 5G technologies and their potential impact on interference generation. The principal differences with 4G LTE are also highlighted.

- A unified classification and nomenclature of the prominent interference types in 4G/5G networks, along with a detailed description and explanation of each interference, i.e., when it occurs and its effects on the network, are given in Section 3. In the same section, we also provide a comprehensive taxonomy to describe and categorize various interference management approaches.

- Section 4 explains the interference measurements reports and signaling in 5G NR according to the newest 3GPP specifications.

- From Section 5 to Section 11, we present for each previously identified interference type, a survey of recent research work on appropriate management schemes.

- Section 12 highlights the open issues and future research directions in 5G networks and beyond. Envisioned key features of 6G networks and related interference challenges are thoroughly discussed, along with the main novel paradigms for advanced interference management.

The structure of this paper is presented in Figure 1.

## 2. 5G NR: An Overview

5G New Radio is the new radio access technology specified by 3GPP to be the global standard for a high-performance unified radio interface of 5G mobile networks to meet the International Telecommunication Union (ITU) requirements set in International Mobile Telecommunications-2020 (IMT-2020). 5G NR technical work was launched in 3GPP in early 2016 based on a kick-off meeting in the autumn of 2015. The first NR specifications were published in 2018 as part of Release 15, followed by Release 16 in 2020 and Release 17 which was frozen in mid-2022. The 3GPP RAN plenary recently agreed on a work package for Release 18, which is considered to be the first release of 5G Advanced. The extensive 5G research and studies carried out over the past years and the continuous specifications development efforts reveal several pivotal technologies that have enabled the ambitious network and service requirements to be met. In the following, we overview the major 5G enabling technologies and their potential impact on interference generation. The main differences with 4G LTE are also emphasized.

### 2.1. Network Architecture & Targeted Use Case Classes

To fulfill the stringent requirements of the mobile networks, the 5G NR is envisioned to support different use cases in three main categories [28, 29, 2]:

- Enhanced Mobile Broadband (eMBB): eMBB aims for services with high data rates. It is an evolution of 4G LTE to have higher spectral efficiency and improved throughput performance. The usage covers a great variety of application scenarios, including wireless wide area networks and hotspots.

- Massive Machine Type Communications (mMTC): mMTC aims for machine-type devices and supports wireless devices interconnected as a part of the Internet of Things (IoT). Devices are usually of low cost and low power consumption for long battery life.

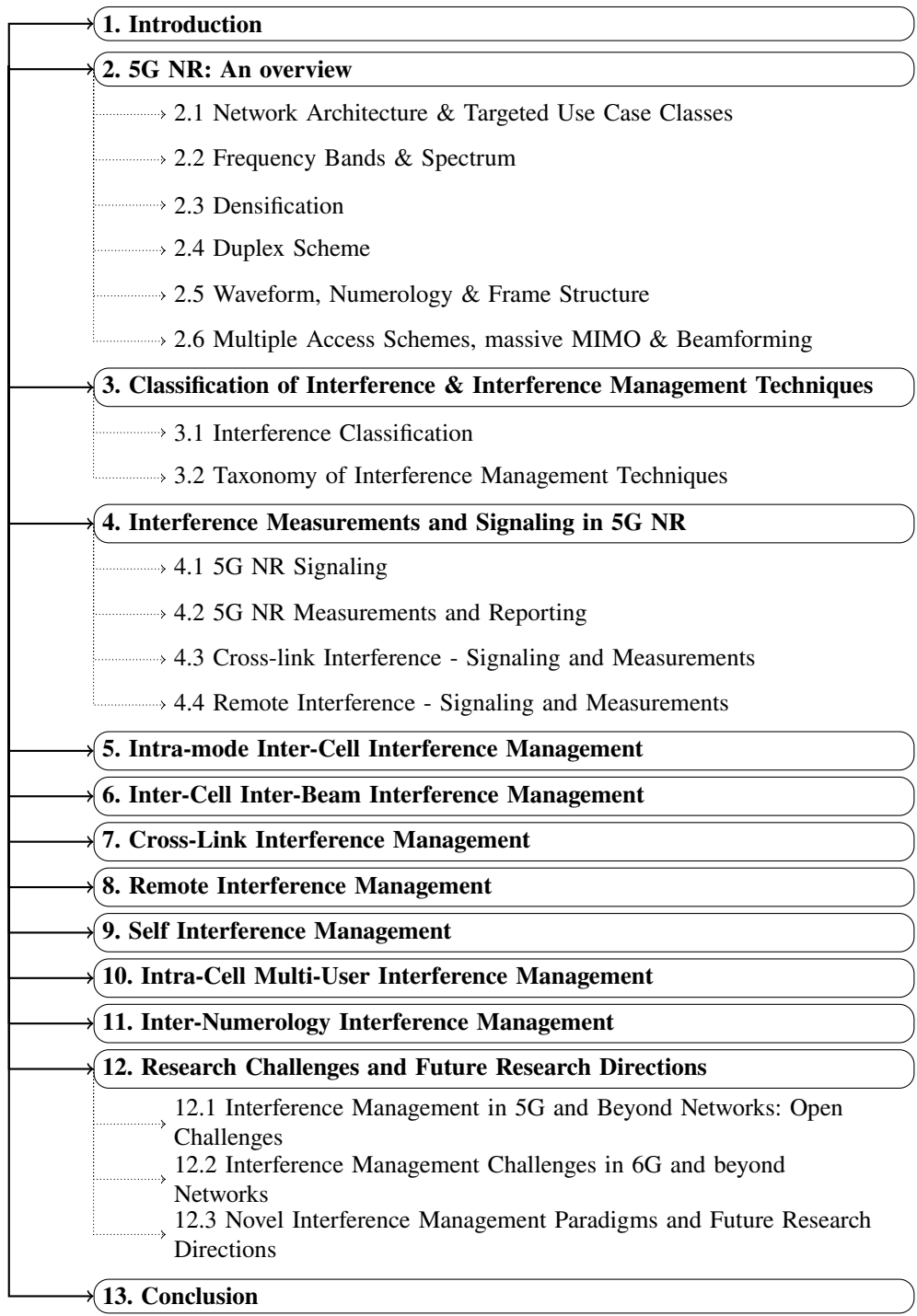

Figure 1: Paper Structure.

- Ultra-Reliable Low Latency Communications (URLLC): URLLC aims for low latency communications between devices under high network availability and reliability (very low packet error rate). Examples of use cases include autonomous driving, remote surgery, and public safety.

5G networks are first deployed in a Non-Stand-Alone mode (NSA)[1][30] using 4G LTE mobile core network for the control plane, while 5G NR is used for the user plane. It is expected that 4G LTE and 5G NR will co-exist for a long time before the maturation of the Stand-Alone (SA) mode [31] of 5G core network (See Figure 2). The SA mode is the end-to-end full 5G network with its own dedicated 5G Packet Core. The full 5G is envisaged to have lower cost, and higher efficiency and to cater to a wide range of new usage scenarios.

---

[1]They are not 5G standalone due to the use of 4G LTE control plane and are considered as the transition towards 5G NR.

Interference Management in 5G and Beyond Networks

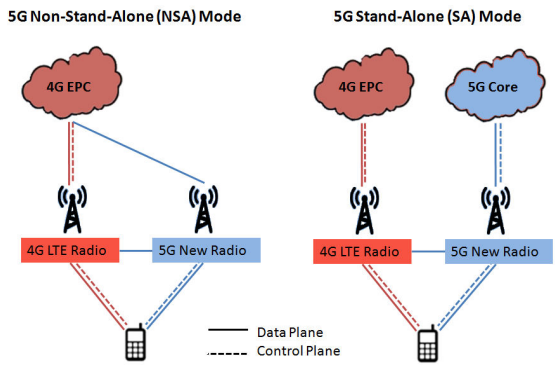

Figure 2: 5G Deployment NSA and SA Modes.

## 2.2. Frequency Bands & Spectrum

Millimeter wave is considered a key 5G technology and also an enabler for eMBB user access. The reasons can be summarized as follows:

- Deployed mobile networks are using microwave bands (below 6 GHz) in which the spectrum resources are almost exhausted. In contrast, the frequency bands above 6 GHz are often under-utilized or even abundant [2].

- From the perspective of link capacity, a large carrier bandwidth of around 1 GHz may carry a data rate of 10 Gbps or higher, compared to that of a carrier bandwidth of 100 MHz, which may carry a data rate of 1 Gbps in the low-frequency spectrum [28].

- The size requirement for the antenna is lower which allows for smaller antennas inside the fuselage and greatly increases the number of antennas.

However, there are also some issues with the high frequency bands which offer limited coverage due to the higher radio propagation loss. Therefore, in 5G, lower frequency bands can be used for providing wide-area nationwide coverage, whereas the millimeter wave or above 6 GHz bands can provide more targeted coverage in smaller areas with very high throughput.

According to 3GPP specifications, frequency bands for 5G NR are split into two different frequency ranges: Frequency Range 1 (FR1) and Frequency Range 2 (FR2). The range of FR1 is 450 MHz-6 GHz, which is typically called a sub-6 GHz frequency band. The FR2 is concentrated in 24.25-52.6 GHz and is generally called the millimeter wave. Within each frequency range, 3GPP has defined different operating bands associated with different bandwidths ranging from several MHz to a few GHz. To suit various spectrum configurations while keeping implementation complexity to a minimum, 5G features a variety of channel bandwidths, i.e., the bandwidth of the NR carrier, from 5 to 400 MHz. One of the core features introduced in 3GPP Release 15 to reduce the power consumption of 5G NR devices is the BandWidth Part (BWP), which is a set of contiguous resource blocks configured within a channel bandwidth. With BWP, the carrier can be subdivided and support multiple services, such as eMBB and mMTC [32].

To further improve network capacity and expand coverage in the medium and high-frequency bands, 5G NR supports Carrier Aggregation, which increases bandwidth by combining multiple carriers. While carrier aggregation is not specific to 5G and has already been used in 4G LTE, it is envisioned as a key 5G technology that creates opportunities far beyond those of 4G networks. Specifically, 5G NR supports carrier aggregation of a maximum of 16 contiguous and non-contiguous component carriers on multiple spectrum bands in FR1 and FR2.

## 2.3. Densification

One of the best-known approaches to improve the capacity of wireless networks is the deployment of smaller cells. This enables higher frequency reuse to improve network capacity, also called cell splitting gain. A traditional macro cells (MCs) system may also be made denser by increasing the number of sectors per macro site or by implementing more base stations, but in practice, it is often much more complicated and costly to acquire macro cell BS sites. Based on cell densification through small cells (SCs), heterogeneous networks consisting of MCS and SCs have been proposed in 3GPP LTE/LTE-A to fit various use scenarios and to deal with spectrum scarcity. In general, in homogeneous cellular networks, the deployment of macro base stations is carefully planned to minimize the overlap of neighboring BSs while ensuring seamless coverage for all terminals in the system. Heterogeneous networks inherently modify this concept through the overlay of the existing homogeneous macro cell networks, usually referred to as the macro layer, with extra low power, low-complexity BSs, known as small cells, while keeping infrastructure costs low. It is expected that mm-wave will operate in small cell architecture and be especially favorable for hotspots that require targeted coverage and demand for very high data rates. Network densification involves the pre-planned deployment of macro BSs that usually operate at high power levels, overlaid with multiple types of small BSs (e.g., pico and femto cells), which operate at significantly lower power levels [33, 34]. Since macro base stations already provide coverage, SCs are typically implemented in densely crowded regions to increase the capacity of the wireless network or to cover certain coverage gaps. Due to their weak transmitter power and small physical size, smaller BSs can provide great flexibility in site acquisition and significantly reduce network CAPEX/OPEX. Small cell technology is very important to millimeter wave communications for 5G networks [35].

As illustrated in Figure 3, among the several types of SCs, we can have :

- Femto cell (also called home BS): It is initially intended for domestic use, but can also be used in an office or enterprise. Its radio coverage reach is often inferior to 50 m with a transmission power inferior to 23 dBm.

- Pico cell: It is intended for indoor businesses and public places, but can also be used outdoors. It typically serves a few dozen users in a 300 m radio coverage area with a transmission power typically between 23 dBm and 30 dBm.

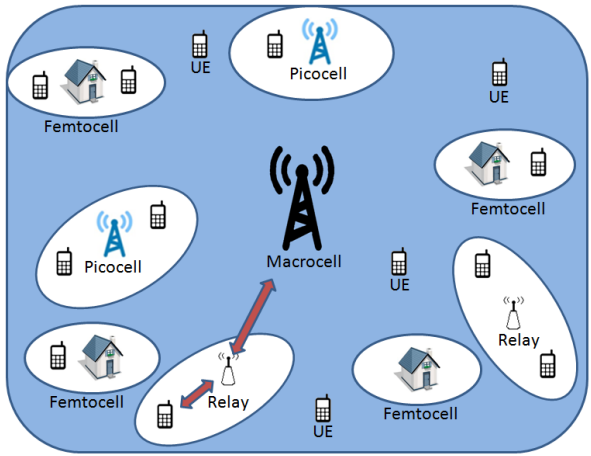

Figure 3: HetNets Small Cell Types.

- Relay: It can be used to route data from macro cells or can be used to extend the radio coverage range of a mobile network. Its transmission power is between 23 dBm and 33 dBm for outdoor use, and 20 dBm or less for indoor use.

Since the low-power BSs are deployed on top of the macro cell layer, the spatial reuse of time-frequency resources is greatly enhanced, resulting in increased network capacity. We refer to this as area spectral efficiency. Network densification reduces the average distance between UEs and the nearest base station, which decreases the transmitted signal loss and improves link gain and channel capacity. An important advantage of SC deployments is that they allow some UEs to be offloaded from MCs to SCs, thus balancing the traffic load and improving throughput and overall network efficiency. However, increasing cell density may result in increased intercell (cross-tier and intra-tier) interference that needs to be effectively managed.

In recent years, ultra densification has further been adopted in mobile networks, where the BS density potentially reaches or even exceeds the user density to the point where the Ultra Dense Network (UDN) capacity grows sub-linearly, due to the inevitably growing impact of severe inter-cell interference. Due to the massive number of adjacent cells and edge users in UDN, traditional interference distributed coordination approaches can not effectively mitigate ICI, as they may cause significant additional signal exchange and rely on simple and obsolete interference information. Therefore, accurate interference modeling in UDN is essential, along with adapted advanced interference management approaches.

### 2.4. Duplex Scheme

The two key duplex schemes used by mobile operators are called Frequency Division Duplex (FDD) and Time Division Duplex (TDD). TDD architecture uses one frequency band both for uplink (UL) and downlink transmissions and assigns alternative time slots for transmitting and receiving, making it more suitable when a paired spectrum is not available. FDD uses two different frequency bands, one for downlink and one for uplink, with a sufficient frequency gap allocated between them, which is called a duplex distance. Both FDD and TDD have benefits depending on the application. In general, FDD is considered better for coverage, while TDD is better for capacity.

As 4G LTE networks employ both FDD and TDD with FDD as a primary duplex scheme, 5G networks will also support both of them, using a slightly different approach. In 5G NR, the millimeter frequency band can use Time Division Duplex while the lower frequency bands (low and mid) may still use FDD. This is due to different reasons:

- Since FDD has been a winner in the implementation of legacy mobile networks, the FDD spectrum has become costly and scarce.

- As FDD requires a larger duplex distance between downlink and uplink bands, this can be very difficult to achieve when carrier frequency increases.

- Since most futuristic use cases of 5G NR operate at millimeter waves, TDD can provide the required flexibility to adjust downlink/uplink network resources depending on the customer's needs.

Contrary to LTE which uses two different frames for FDD and TDD, 5G NR networks will operate in both models using the same frame structure for both duplex schemes, as it will be more explained in the next sub-section. 5G TDD can be static or dynamic. In Static TDD, the UL/DL ratio is decided only once. In dynamic TDD, resource allocation between DL and UL can be dynamically adjusted which can provide significant performance improvements. However, dynamic TDD also brings out additional interference between UL and DL that may degrade system performance.

It is also important to note that Full-duplex (FD) (or In-band-Full-Duplex IBFD) technology is a futuristic option for 5G and beyond networks as it enables the terminal to transmit and receive signals in the same frequency band at the same time. FD can effectively double the capacity and the efficiency of the spectrum. However, the main issue of IBFD technology is how to effectively cancel the strong self interference [36, 37].

### 2.5. Waveform, Numerology & Frame Structure

The 5G and beyond mobile networks are envisioned to cater to multiple use cases with various requirements in terms of latency and data rates. Defining the adequate waveform to be used in 5G RAN is one of the essential elements of the PHY layer design.

The waveform defines the physical shape of the signal carrying the modulated information [38]. Orthogonal frequency-division multiplexing (OFDM), the most popular multicarrier modulation scheme used in the downlink 4G LTE and the IEEE 802.11 networks, has many advantages, but also some drawbacks. On one hand, it utilizes the spectrum efficiently as the channel is divided into a set of independently modulated and orthogonal subcarriers. OFDM can also be easily implemented and partly diminishes the inter-carrier and inter-symbol interferences. On the other hand, OFDM systems are more sensitive to synchronization errors than single-carrier systems. If the orthogonality is lost, the leakage from other subcarriers causes interference between carriers. Similarly, the timing offset causes interference between symbols or carriers.

Although many waveforms were considered as strong contenders to replace OFDM in 5G, 3GPP eventually decided to stick to OFDM, for the simple but powerful reason of backward compatibility to 4G LTE system. More precisely, as specified in 3GPP Release 15, the standardization organization decided to use the Orthogonal Frequency Division Multiplexing with cyclic prefix (CP-OFDM) in the downlink and CP-OFDM and Discrete Fourier Transform-spread-OFDM with CP, better known as Single Carrier Frequency Division Multiplexing (SC-FDM), in the uplink [28].

As already mentioned, 5G systems are expected to support a wide range of deployment scenarios: from macro cell deployments with low-band frequency to smaller cell sizes with higher carrier frequencies. A subcarrier spacing (SCS) of 15 kHz, as in an LTE system, may be sufficient to handle the delay spread for the low range of carrier frequencies. However, for implementation limitations in the mm-wave band, like the phase noise, and to mitigate inter-carrier interference, a higher SCS, and a shorter cyclic prefix are suitable. As having a single numerology, i.e., subcarrier spacing in the frequency domain, is not efficient or possible, 5G NR allows the use of multiple numerologies with different SCS ($\Delta f$) to provide different services while satisfying related requirements [28, 39]. The scalable OFDM numerology specifies SCS, CP, and the number (or duration) of slots. As shown in Table 2, the 5G numerology is based on SCS between 15 kHz and 240 kHz ($\Delta f = 2^\mu \times 15$ kHz, where $\mu = \{0,1,2,3,4\}$) with a proportional change in CP duration.

| $\mu$ | $\Delta f$ (kHz) | N slots / sub-frame | CP Type | Support for Data |
|---|---|---|---|---|
| 0 | 15 | 1 | Normal | Yes |
| 1 | 30 | 2 | Normal | Yes |
| 2 | 60 | 4 | Normal - Extended | Yes |
| 3 | 120 | 8 | Normal | Yes |
| 4 | 240 | 16 | Normal | No |

Table 2: Numerology Structures in 5G

The multiplexing of different numerologies to meet various requirements and frequency bands can be done in two different ways: non-overlapping numerologies, or overlapping with spectrum sharing [40]. In the non-overlapping mixed numerology system, the multiplexing of numerologies can be done in the time or frequency domain. In time-domain multiplexing, thanks to the waveform design, orthogonality is maintained between consecutive blocks. In non-overlapping frequency domain numerology multiplexing, the entire system bandwidth is subdivided into BWPs, where each part of the bandwidth has its numerology and signal characteristic, enabling a more efficient spectral and power utilization. Therefore, frequency multiplexing has better forward compatibility and support for multi-service with different requirements coexistence in comparison to the time domain counterpart [32, 41]. However, due to the difference in SCS, combining multiple numerologies in one frequency band may destroy inter-numerology orthogonality and result in out-of-band-emission [40, 42]. In the overlapping mixed numerology system, also called spectrum sharing multi-numerology system, different numerologies are combined in the same time and frequency resource, resulting in the loss of orthogonality between overlapping numerologies [42]. Generally, the flexibility provided by mixed numerologies may destroy the inter-carrier orthogonality of multiple numerologies. Achieving symbol alignment in the time domain can also be an issue.

Different from LTE, 5G NR defines a unique frame structure that supports both TDD and FDD. It consists of a 10ms radio frame formed by 10 subframes. A 1ms subframe is divided into $2^\mu$ adjacent slots, depending on the numerology. Each slot of a duration $1/2^\mu$ ms comprises 14 symbols (or 12 symbols in the case of extended CP). For the numerology 0, a 5G NR slot is similar to an LTE sub-frame, for backward compatibility. Figure 4 illustrates the 5G NR frame structure for numerologies that support data.

Each OFDM symbol in a slot can be downlink, uplink, or flexible. If it is DL or UL, it means that no transmission in the other direction is allowed. Flexible symbols can be adapted for DL or UL transmission. The allowed symbol combinations by 3GPP are similar to LTE TDD, with an important difference: the division of UL and DL was based on subframe but in 5G it is based on OFDM symbols.

Along with flexible numerology, 5G NR supports an efficient approach to allow immediate transmission of low latency communication, referred to as "mini-slot" transmission using the shortest scheduling units of 2, 4, or 7 OFDM symbols.

### 2.6. Multiple Access Schemes, massive MIMO & Beamforming

From the first generation to the fourth generation of wireless networks, orthogonal multiple access (OMA) schemes, such as frequency division multiple access (FDMA), time division multiple access (TDMA), code division multiple access (CDMA), and orthogonal frequency division multiple access (OFDMA) have been employed to allocate orthogonal resources in frequency, time, or code domains, to multiple users, and hence avoiding, in theory, intra-cell multi-user interference using relatively cost-efficient receivers [43]. In particular, in the same way, as 4G LTE, OFDMA and SC-FDMA, where the radio resources are orthogonally partitioned in the time-frequency grid, were also adopted in the first releases of 5G as DL and UL multiple access schemes respectively. However, although such schemes simplify the transceiver design, they lead to low spectral efficiency in general and it is not a trivial task to realize massive access in 5G and beyond wireless networks [44]. The exploding demand for wireless capacity and the spectrum resource scarcity have led to the deployment of multiple-input multiple-output (MIMO) antennas at the base stations and to the application of a new non-orthogonal multiple access scheme, referred to as space division multiple access (SDMA), or multi-user MIMO (MU-MIMO). Exploiting the spatial dimension created by the multi-antennas transmitter, SDMA allows to project narrow beams and to serve many users in the same time-frequency resource thus improving the network transmission capacity. Each user decodes its intended data stream by treating other users' interference as noise [45]. To further improve the spectral efficiency, other multiple access schemes have been studied in the literature, specifically power-domain Non-Orthogonal Multiple Access (PD-NOMA) and code-domain NOMA (CD-NOMA), allowing to superpose multiple users in the same time-frequency resource by using complex algorithms for multi-user detection and interference cancellation. While these schemes have also been considered in 5G standardization, they have not yet been adopted. Research is still ongoing for optimal multiple access scheme design. In the following, we will focus on SDMA and the use of beamforming in mm-wave massive MIMO networks.

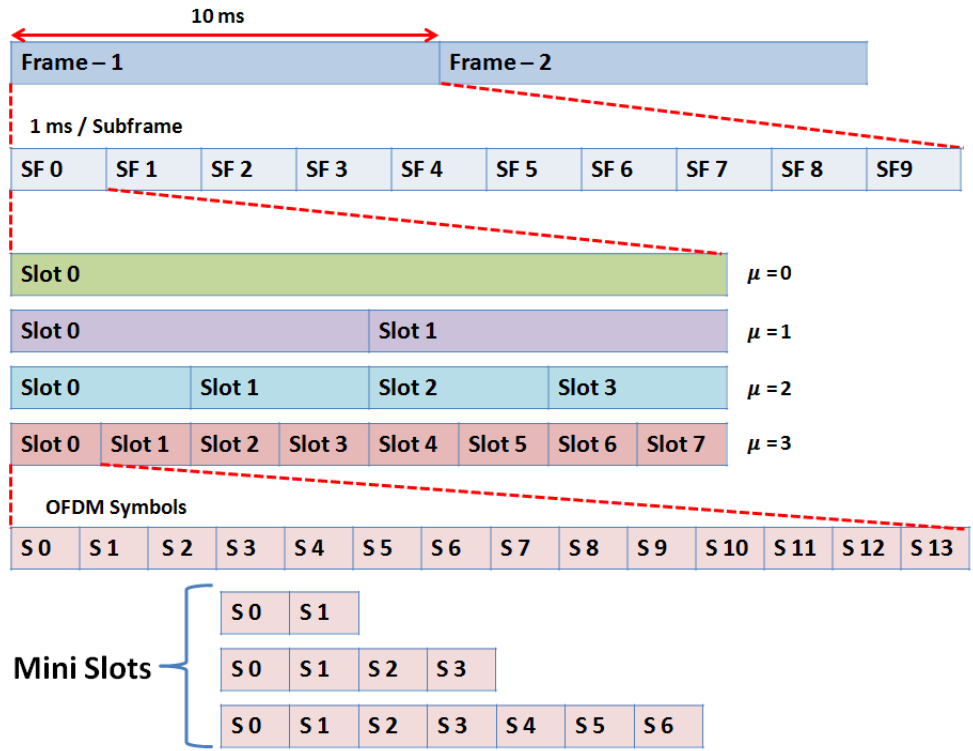

Figure 4: 5G Frame Structure.

Based on the use of multiple antennas for transmission and reception, massive MIMO is considered to be one of the key technologies of 5G [4]. The most commonly seen definition is that mMIMO is a system that deploys a very large number of antennas on the BS. Some of the factors that drive 5G's massive MIMO adoption can be summarized as follows [2, 28, 46]:

- It is the only wireless technology that can multiply system capacity without additional spectrum, as it enables serving many devices with the same time-frequency resource. Multiple users, separated in the spatial domain, receive simultaneous transmissions using very directive signals.

- The use of higher frequencies and smaller wavelengths enables the deployment of large-scale antenna elements that can be grouped into sub-arrays at the base station. These sub-arrays provide array gain to overcome the effects of noise and fading, which results in enhancing the network data rate and link reliability.

- Massive MIMO networks will utilize beamforming and combining technology for transmission and reception respectively, which can enhance spectrum and energy efficiency through various domains. Typically, beamforming is achieved by proper phase shifting and amplitude scaling of the signal transmitted to create multiple spatial beams.

In essence, in mm-wave mMIMO networks, based on channel quality measurements sent from the terminals, the BS applies distinct precoding[2], usually selected from a codebook to reduce the complexity, for the data stream of each UE where the location of all UEs are taken into account. This aims to optimize the signal for the target UE and to minimize inter-user interference.

Generally, beamforming can be divided into three categories: analog beamforming, digital beamforming, and hybrid beamforming. Considered as the simplest, most straightforward, and cost-effective implementation, with analog beamforming, using a single radio frequency (RF) chain, the same analog signal is first split into several paths and fed to each antenna. Then, using analog phase-shifters, each signal is steered, i.e., phase-adjusted and amplified to create one directional beam per set of antenna elements. While analog beamforming enhances the cell coverage, it is still limited in performance for two main reasons: (i) a single data stream is fed to each antenna, and (ii) the maximum number of beams that can be formed is hardware-dependent. Offering more flexibility than analog beamforming, digital beamforming enables the adjustment of the phase and amplitude of the signal in baseband processing before analog transmission. Different signals with different powers and phases are then fed to the multiple antennas, allowing the formation of multiple beams, one per user, from each set of antenna elements. In digital beamforming, one different TRX chain is required in the BS for each co-served user, which in turn limits the maximum number of simultaneous beams and co-scheduled users. While digital beamforming further enhances flexibility, performance, and cell capacity, it induces more complexity and baseband processor's cost and power requirements. Combining digital and analog processing in the baseband and RF domains respectively, hybrid beamforming is a promising solution for mm-wave mMIMO networks, which is under development using different approaches in the literature.

---
[2]The terms of precoding and beamforming are used interchangeably in this article.

**Discussion:** In this section, we provided an overview of the key pivotal 5G technologies and the main differences with 4G LTE networks. Mainly, the distinctive feature of the 5G and Beyond network is its intrinsic flexibility as (i) it supports several use cases with different QoS requirements (eMBB, URLLC, and MMTC), (ii) it uses low, medium and high-frequency bands, licensed and unlicensed spectrum, and supports carrier aggregation, (iii) it uses multiple numerologies and new flexible waveforms are currently studied, (iv) is envisioned as a multi-tiered HetNets network supporting multiple mobile devices in each layer, which can be deployed randomly, and reusing the same frequency resources, (v) it supports both TDD and FDD and Full duplex is considered as a promising and attractive solution for the near future, (vi) it supports carrier aggregation and massive MIMO, etc. However, this flexibility may result in unprecedented levels of interference that can severely limit network performance.

## 3. Classification of Interference & Interference Management Techniques

The term "interference" is commonly used to describe the addition of unintended signals to a desirable signal. It can be caused for example by the reuse of the time–frequency resources of different base stations or user devices, or due to the unintentional leakage of a transmit signal. Interference is one of the most dominant factors, which could lead, if not properly mitigated, to considerable performance deterioration.

Despite the inevitable advantages of all 5G key technologies previously introduced in Section 2, the variability and flexibility in 5G could introduce a non-orthogonality into the network and generate consequent interference. To alleviate the severe and sometimes unpredictable interference in 5G, it is critical to design efficient interference management techniques. Therefore, in this section, we will first discuss and propose a unified classification and nomenclature of the prominent interferences in 5G and beyond cellular networks. Then, we develop a comprehensive taxonomy for describing the different interference management schemes.

### 3.1. Interference Classification

As shown in Figure 5, there are two major categories of interference for cellular mobile networks: intra-cell interference and inter-cell interference. Next, we will highlight the various types of interferences for these two categories, starting with inter-cell interference.

#### 3.1.1. Inter-Cell Interference (ICI)

Inter-cell interference, generally caused by using the same resources in neighboring cells, is a classical problem for mobile cellular networks, which could limit the overall network performance and degrade the signal quality of edge users. With the massive deployment of small power low-complexity BSs within the macro coverage and with the reuse of scarce spectrum resources, ICI became more and more serious [12].

A) *Co-tier and Cross-tier Interference*
While multi-tier networks improve significantly network capacity, they face important challenges, especially in managing interference between neighboring cells [9]. Generally, two types of inter-cell interferences may degrade the performance of heterogeneous networks: either co-tier interference between network entities in the same tier or cross-tier Interference (also called inter-tier interference) between network entities from different tiers of the network, for example, interference between SCs and MCs.

B) *Adjacent Channel Interference (ACI)*
In general, inter-cell interference is either due to operating on the same channel which is referred to as Co-Channel Interference (CCI), or due to the overlapping between adjacent channels which is called Adjacent Channel Interference. ACI mainly occurs due to imperfect receiver filters, allowing adjacent frequencies to enter the pass-band, and to amplifiers' non-linearity. One critical issue of OFDM networks is the high out-of-band emissions that result in important interference [46]. As the problem of adjacent channel coexistence is relatively common in mobile networks, regulators have already established stringent rules concerning adjacent channel leakage and undesired emissions in the operating band to limit ACI [47]. For example, the effects of ACI can be reduced by (i) carefully choosing the bandpass filter at the receiver, (ii) using advanced modulation schemes and signal processing techniques, (iii) allocating a guard band, and (iv) allocating adjacent channels to the cells in different geographical regions.

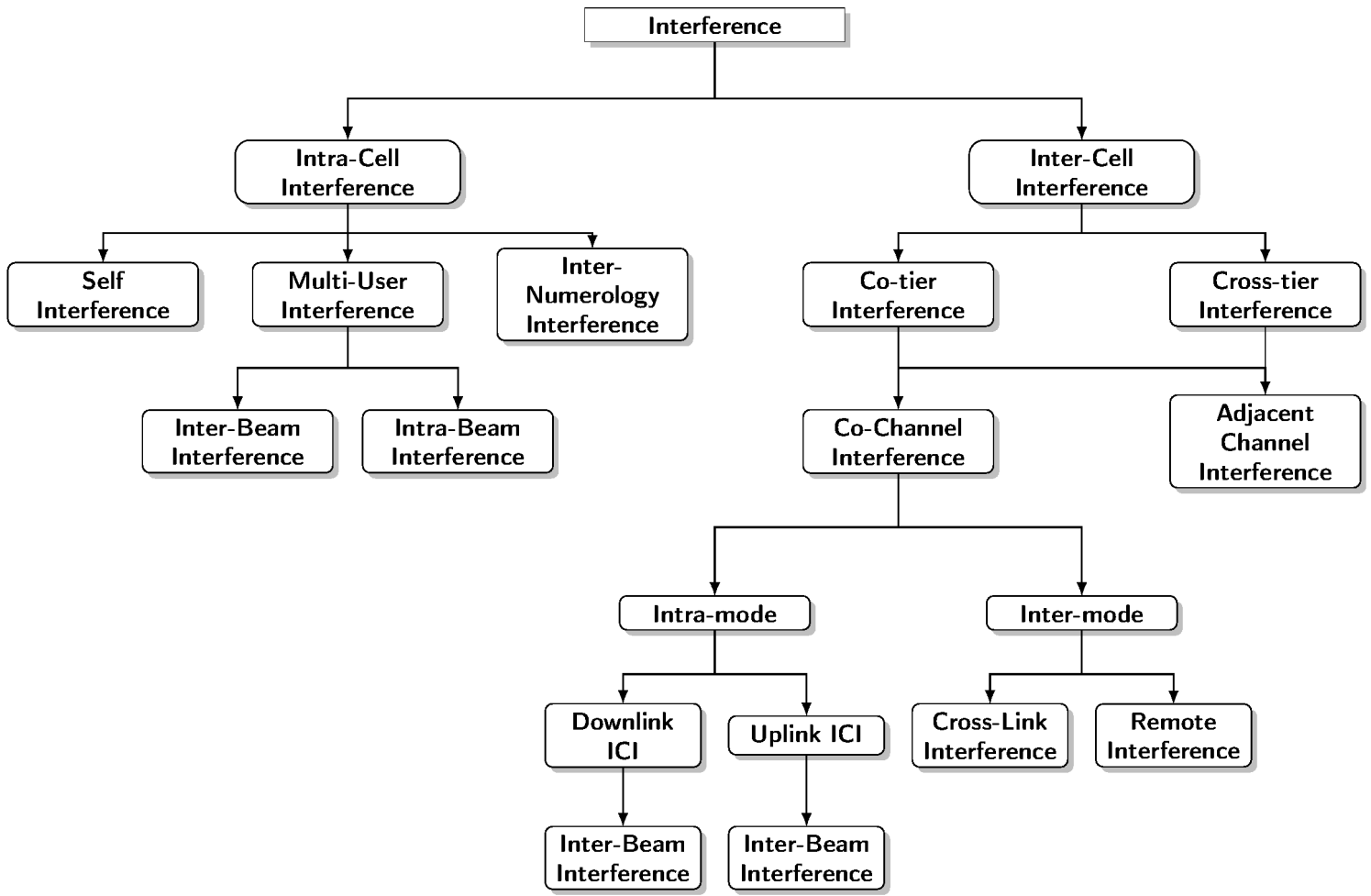

Figure 5: Interference Classification in 5G and beyond Networks.

C) *Co-Channel Interference (CCI)*
Co-Channel Inter-Cell Interference is the most studied interference type in the literature. In an FDD or static and perfectly synchronized TDD network, scientific research concentrates on intra-mode interference. On the other hand, inter-mode interference (or cross-mode interference) generally occurs in dynamic TDD or Full Duplex networks.

(a) *Intra-mode Interference (Downlink and Uplink ICI)*
In Intra-mode Co-Channel Inter-Cell Interference (referred to as Intra-mode), there are two types of interference depending on the transmission direction:

- Uplink Inter-Cell Interference: The UE is the aggressor and causes uplink interference to a neighboring base station, which is the victim.
- Downlink Inter-Cell Interference: The base station is the source of downlink interference to a UE from a neighboring cell.

Figure 6 further explains the intra-mode interference for both cross-tier and co-tier cases.

A sub-category of Intra-mode interference for both downlink and uplink ICI is known as **inter-beam interference**. In a multi-beam transmitting and receiving antenna system, millimeter-wave base stations and users steer the phases of their antenna array to transmit and receive signals in certain directions, in contrast to conventional networks where inter-cell interference is caused by the omnidirectional transmission of base stations [48]. Despite improved spatial diversity thanks to beamforming, beams from different base stations can interfere with each other and cause Inter-Cell Inter-Beam Interference (IC-IBI), i.e., interference between beams formed by neighboring BSs [49]. Even though the vast majority of scientific research mentions inter-cell interference rather than inter-beam interference, it is important to identify and highlight this type of interference [50].

With the increasing bandwidth demands, mm-wave frequency bands are considered the optimal solution in 5G NR. As the sub-band above 6 GHz is subject to significant propagation losses limiting the coverage of BSs, multi-beam antenna systems based on applying beamforming/combining along with massive-MIMO technology are used to compensate for the attenuation losses and to provide sufficient channel gains [51]. To minimize the high cost and energy consumption of mixed circuits, either a fully analog or a hybrid beamforming architecture is being pursued in these systems to exploit this potential. However, adopting these architectures also brings a number of challenges for subsequent signal processing, including channel estimation. Consequently, pre-defined codebooks are generally employed for initial access and data transmission. However, since they are predefined, these beams are typically designed with a focus on enhancing beam-forming/combining gain from specific directions only, regardless of interference. This poses problems when interfering UEs in the surrounding area are transmitting on the same time and frequency resources. These interference-unaware beams can lead to a high level of packet delivery failure and cause severe inter-beam interference that needs to be efficiently handled [52]. It is worth noting that even when the downlink interfering beam is not directly pointing at the UE, which is referred to as the flashlight effect, IC-IBI is still considerable because of the dense and ultra-dense deployment of base stations [53].

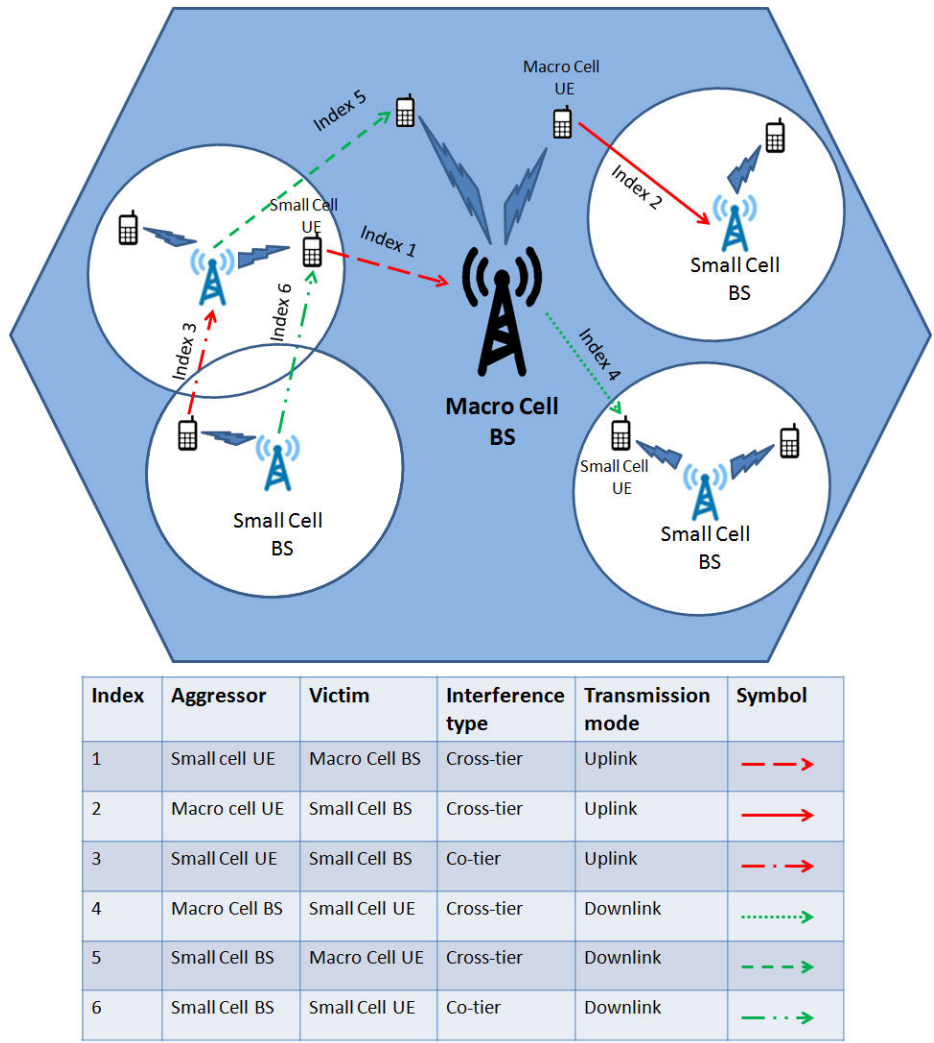

Figure 6: Inter-Cell Interference Intra Mode.

(b) *Inter-mode Interference (Cross-Link and Remote Interference)*
Generally, Time Division LTE (TD-LTE) networks adopt synchronous transmission in the whole network, with all cells with overlapping coverage applying the same slots configuration for uplink and downlink transmission. Moreover, a guard period is specified in the special subframe [54]. This time alignment and guard period permit us to avoid the inter-mode inter-cell interference, i.e., downlink-to-uplink interference (base station-to-base station), and uplink-to-downlink interference (terminal-to-terminal) [55]. However, synchronous multi-cell operation with a shared UL/DL configuration is not convenient in 5G networks, where the amount of traffic may be very different between cells for downlink and uplink transmissions. As seen in the previous section, 5G NR supports both FDD and TDD for paired and unpaired spectrum respectively. For traffic adaptation, it is necessary to adopt Dynamic-TDD to allow every cell to have customized subframe configurations according to its traffic situation with no need to be synchronized [18]. With the adoption of D-TDD, spectrum utilization can be enhanced and latency reduced. However, the inter-mode interference represents a serious challenge to deal with. Obviously, these interference scenarios are also noticed in In-band-Full-duplex networks.

- Cross-link interference (CLI) also called cross-slot interference is an inter-mode inter-cell interference, which happens generally between adjacent (i) BSs or (ii) UEs using different transmission directions as displayed in Figure 7. Both interference scenarios need to be managed in macro and small-cell deployments.

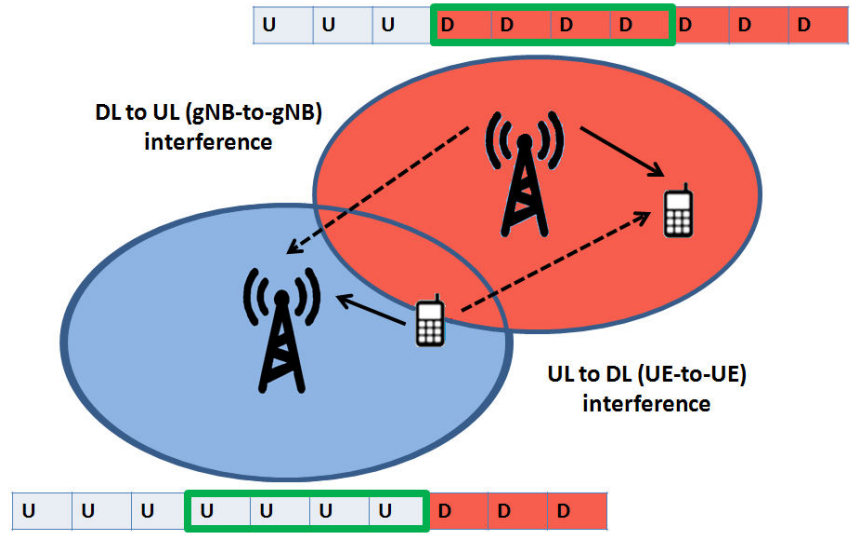

Figure 7: Cross-Link Interference (CLI).

The issue of DL-to-UL (base station-to-base station) interference is particularly noticeable in macro-deployments for two reasons. First, the transmission power in macro BS is much higher than in mobile terminals. Second, the macro BS antennas are usually deployed over rooftops which can imply line-of-sight propagation. Therefore, the downlink transmission of a macro base station would highly impact the uplink signal in neighboring BSs [56]. On the other hand, in small-cell deployments, UEs and small BSs have similar transmission power and their antennas are approximately at the same level, thus the small-BS-to-small-BS interference is not that critical and is more similar to intra-mode uplink interference from UE to the base station. In such small-cell deployments, and since the user density of 5G devices will be very high, two closely located UEs and served by different gNBs, may experience critical cross-link UL-to-DL interference. This means there is a possibility that at any given instant, a UE may be transmitting, and a neighboring UE will be receiving at the same time.

- Remote Interference (RI) is a downlink-to-uplink interference that may occur between very distant BSs, even in static synchronized TDD networks due to very large propagation delays. In certain weather conditions, the refraction index gradient decreases rapidly in the low troposphere, forming anomalous atmospheric ducts [57]. Such ducts impact electromagnetic waves and send the trapped transmitted signals very far with much lower attenuation than in a natural atmosphere (see Figure 8). In these rare scenarios, the propagation delay goes beyond the guard period which is set to manage adjacent BSs interference. Consequently, the strong copy of the DL transmission of the aggressor BS may critically interfere with the UL reception at a victim distant base station. This remote interference may affect the UL data, as well as the reception of uplink reference signals, such as sounding reference signals (SRS) used by the victim BS to estimate the uplink channel's quality. In 5G TDD systems, relying on channel symmetry and UL/DL channel reciprocity characteristics, the SRS also permits the estimation of the DL channel's quality. When the SRS signal is affected by remote interference, this will directly lead to the deterioration of KPI indicators such as call drop rate and handover success rate, resulting in poor user experience. According to statistics, many countries such as China and the United States have suffered from RI for a long time [58]. Although ducting is a rather rare event, whenever it does occur, it can affect thousands of base stations, and the network performance can be seriously impacted. The damaging effects of atmospheric ducting and remote interference have drawn the interest of industry and academia alike. ITU recommended to consider atmospheric ducts in channel modeling and 3GPP initiated an RI project in the standardization process of 5G. Moreover, many telecom companies started to monitor and analyze RI effects.

To manage inter-mode interference (cross-link and remote interference), 3GPP Release 16 introduced two enhancements, namely Cross-link interference handling and Remote Interference Management (RIM) [59], to mainly enhance the detection of such interference using specified signaling and measurements. While CLI handling is mainly aimed at small-cell deployments based on dynamic TDD, RIM targets large macro-cell deployments where interference scenarios can be managed in an automated way. These two enhancements will be further explained in Section 4.

### 3.1.2. Intra-Cell Interference

The use of OFDMA (and SC-FDMA) in LTE networks made the intra-cell interference almost negligible, due to the orthogonality among subcarriers. In 5G and beyond, the multiplexing of several numerologies, the use of the same time/frequency resource to serve two or multiple users in

Page 13 of 63

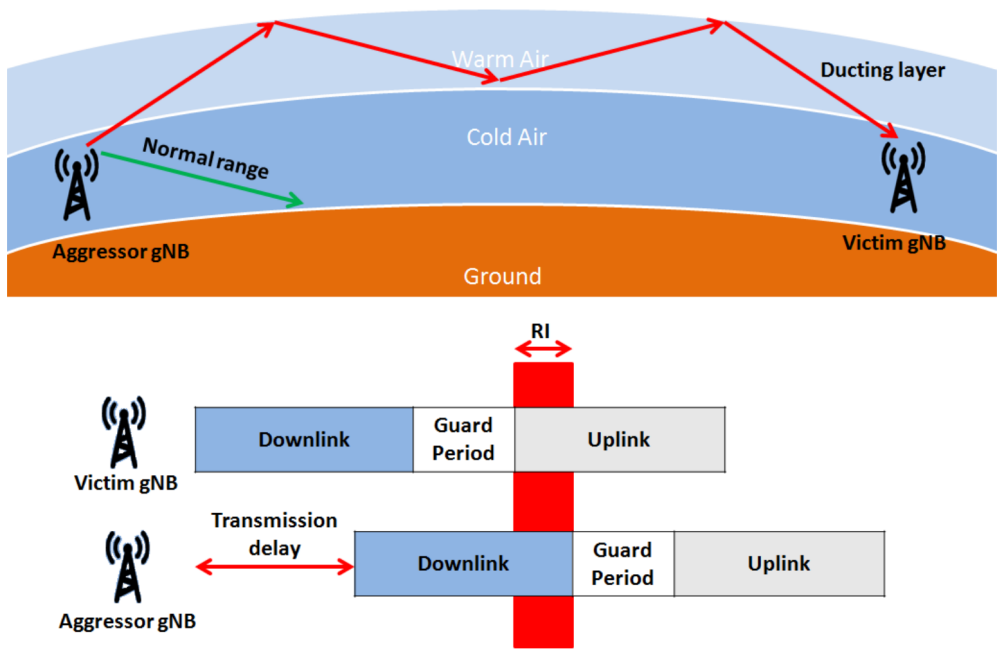

Figure 8: Remote Interference (RI).

the same cell via NOMA, SDMA, and other multiple access schemes, and the deployment of In-Band-Full-Duplex transceivers, may cause strong intra-cell interference, besides the very well known inter-cell interference.

A) *Self Interference (SI) - also called intra-system interference*
While LTE/LTE-A and 5G are mainly built upon FDD or TDD, IBFD has become an attractive field of research to enhance network performance and double the spectral efficiency or capacity. However, the lack of frequency isolation for Uplink and Downlink transmission leads to severe self interference [22]. In an FD transceiver, the highly powerful transmitted signal is propagated into the local receiver through a direct path between the transmitting and receiving antennas, and through multiple reflection paths caused by the scattering of the transmitted signal by surrounding objects. These transmitted signals combine and modulate at the local receiver's RF front-end, becoming self interference. Because it is emitted locally, this undesirable energy is billions of folds (100 dB+) higher than the intended received signal and can disrupt or even saturate the reception chain [60, 61]. Self interference can be then defined as interference directly generated by the transmitter to the receiver antenna in the same device and may result in the most severe performance degradation of the receiver, compared with other interfering signals [20]. Thus, self interference should be minimized, at most under the noise power level of the receiver.

Even though self interference mostly affects full-duplex devices, transceivers that operate in FDD may also be affected by signal attenuation due to SI effects [22]. Although Tx and Rx signals are spaced apart by the duplex spacing determined by the band specifications, and a relative isolation of 50 dB to 55 dB between these signals is provided by front-end filters, Tx leakage is still likely to have a significant power level in Rx chains, given the huge Tx-Rx power difference of up to 120 dB [31]. The use of carrier aggregation in FDD can further complicate the problem, as several Rx and Tx chains operate at the same time.

Considerable progress has been achieved in both academic and industry circles in the field of SI cancellation and some of the proposed schemes combining analog and digital countermeasures will be explained later on.

B) *Multi-User Interference (MUI)*
Such intra-cell interference may occur between two different beams, namely inter-beam interference, or between different users covered by the same beam, i.e., intra-beam interference. As already addressed, beamforming is a key technology in 5G cellular networks, especially in mm-wave communications. Based on Massive MIMO and using spatially narrow beam directive signals to maximize the sum of data rates, it permits the scheduling of a great number of mobile devices in the same time-frequency resource, constituting spatial division multiple access [4, 62]. When scheduled in the same time slots, the adjacent beams of the same cell may be overlapped to prevent coverage gaps in the angular domain and thus cause intra-cell inter-beam interference [63, 64]. Furthermore, in mMIMO mm-wave systems, with large antenna array size, an additional propagation delay at each antenna element is introduced leading to a frequency-dependent variation in the beam direction, referred to as beam squinting. This important issue may further result in additional intra-cell IBI [65]. Due to beamforming considerations, beam squint, and the spatial distribution of the achievable beam-pair association on the network, the multi-user allocation problem formulation for millimeter-wave is inherently distinct from that for microwave networks and therefore necessitates novel interference management schemes [66].

When NOMA is applied, two or several UEs can be served by the same beam which can result in intra-beam interference between the users. Such interference is usually reduced via superposition coding in the transmitter and interference cancellation techniques in the receiver to decode the signal and subtract the perceived multi-user interference [67].

C) *Inter-Numerology Interference (INI)*

In a single-numerology system such as LTE, intra-cell interference is materialized as Inter-Carrier Interference and Inter-Symbol Interference. The former defines the interference generated among subcarriers in the frequency domain, and the latter defines the interference among symbols in the time domain. However, Inter-Carrier and Inter-Symbol interferences are not sufficient to capture all the impairments incurred in a multi-numerologies system such as 5G. While the use of multiple numerologies enables high flexibility for the different services, it destroys the orthogonality principle defined for single-numerology systems, which in turn results in interference among the multiplexed numerologies, i.e., inter-numerology interference [68, 69, 70].

Non-orthogonality can result either from partially or completely overlapped numerologies, or nonoverlapping numerologies for synchronous communications [71]. In non overlapping mixed numerologies system, the principal source of non-orthogonality is out-of-band transmission, as the subcarriers of one numerology can receive energy leaked from the subcarriers of other numerologies. Apart from destroying the orthogonality in the frequency domain, multiplexing different numerologies may also result in symbol misalignment in the time domain. However, the scalable numerology design of 5G permits the multi-numerology symbols to be aligned over the least common multiplier symbol duration as discussed in [72].

In an overlapping mixed numerologies spectrum sharing system, achieving reliable co-existence among different numerologies at the same time-frequency resources is more challenging due to the non-orthogonality originating from overlapping subcarriers. Interference management is, therefore, more complicated in mixed numerology overlapping systems than in single numerology or non-overlapping systems [40].

To take full advantage of the multi-numerology concept, analyzing the pattern of emitted interference and identifying the various factors contributing to INI is inevitable. Recent studies [69, 73, 74] show that the inter-numerology interference depends on subcarrier spacing, number of subcarriers and bandwidth within each numerology, power offset between numerologies, doppler spread, etc. Managing the inter-numerology interference in 5G networks is definitely one of the most challenging issues. Advanced Interference Cancellation (IC) techniques and signal processing, along with adaptive resource and power allocation may reduce the INI impact. Different approaches will be presented and discussed in Section 11.

### 3.2. Taxonomy of Interference Management Techniques

Designing an appropriate interference management has always been a key factor in mobile cellular networks. This continues to be the case in 5G and beyond cellular networks where flexibility comes with a cost. Legacy interference management schemes are a good point to start from and make improvements to suit such an ultra-dense and dynamic network. In this subsection, we will develop a comprehensive taxonomy for describing the different interference management schemes, giving some examples to illustrate. From Figure 9, the taxonomy of interference management can be divided into five categories as follows.

#### 3.2.1. Basic Objective

Interference management schemes can be sorted based on their basic objective into three classes:

1. Interference avoidance schemes: the main idea is to prevent and avoid interference occurrence. The interference avoidance techniques are applied before signal transmission and usually require some coordination between the transmitter and receiver to divide the available resources and avoid collisions on simultaneous transmissions. There are multiple examples, such as frequency reuse-based schemes and inter-cell interference coordination techniques.

2. Interference suppression/reduction schemes: the applied approaches aim to eliminate the interference when it occurs, either during the transmission or at the receiver side that decodes the interference or treats it as noise [45]. For example:

- Conventional precoding techniques [75] which take advantage of the Channel State Information (CSI) available at the transmitter and direct the signal spatially with varying amplitude and phase, and thus offer maximum beamforming gain in the required direction, and reduce interference in other directions. Popular precoding schemes include Zero Forcing (ZF), Maximum Ratio Transmission, Minimum Mean Square Error (MMSE) and Dirty-Paper Coding [76]. Interference Alignment (IA) [77] is also a well-known linear cooperative precoding scheme that exploits the multiple signaling dimensions to align and constrain interfering signals into an interference sub-space so that they can be easily decoded and filtered out at the receiver.

- Advanced receivers applying filtering mechanism or interference cancellation schemes such as Successive Interference Cancellation (SIC) [78, 79], Parallel Interference Cancellation [80] and hybrid mechanisms that combine both approaches [81], to decode the received signal and subtract the interference.

3. Interference exploitation schemes [82]: While interference is traditionally viewed as a performance limiting factor in wireless networks, which is to be avoided or eliminated, a third line of work has argued that interfering signals can add up constructively at the receiver side and further improve the system performance. Constructive interference precoding schemes allow interference between spatial flows to be constructively correlated, rather than completely decorrelated as in traditional approaches. A few examples are listed below.

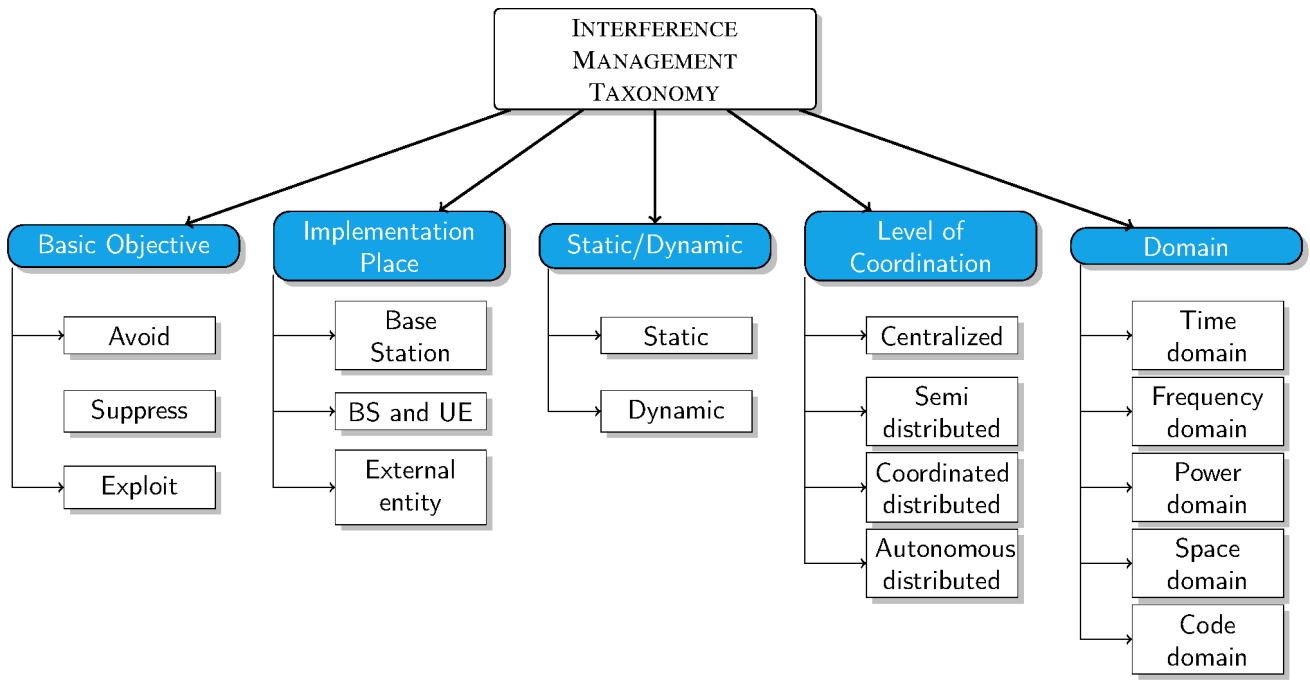

Figure 9: Proposed Interference Management Taxonomy.

- Adaptive MIMO beamforming and combining [83] where the signals are spatially multiplexed and integrated in a way that enhances the transmitted or received signal strength to or from a certain direction.

- Symbol-Level Precoding (SLP) [84] is a type of nonlinear precoding scheme for multi-antenna transmission. Unlike traditional linear precoding schemes where the precoding matrix is only based on channel state information and applied to a block of symbols, SLP makes use of not only the transmitter's CSI but also the symbol constellation structures. The SLP is able to control both the strength and steering of interfering signals at the symbol level so that the interference can be used as an extra signal strength source.

### 3.2.2. Implementation Place

Interference management schemes can be implemented either:

1. In the base station which organizes the transmitting signals in a way that avoids interference (for example subframe muting) or facilitates interference detection at the receiver (for example: precoding schemes). Interference randomization [85] is an additional example where the transmitter randomizes the interference scenario and achieves frequency diversity by spreading the users' transmission over the spectrum for instance.

2. In both the terminal and the BS such as applying power control, filtering, decoding, or advanced receiver cancellation schemes.

3. In an external entity that coordinates the BS transmissions and applies some restrictions to avoid inter-cell interference.

### 3.2.3. Static or Dynamic

Interference management schemes can be implemented either in a static or dynamic fashion.

Static approaches are based on a fixed pre-configuration in the different concerned base stations. For example Fractional Frequency Reuse (FFR) scheme [86] is based on cell partitioning into spatial regions that are assigned different frequency bands with or without different power levels. Although static interference management techniques can be easily implemented, they are unable to handle changing data traffic and are generally not convenient for 5G and beyond networks.

Based on periodical measurements and interference information from neighboring BSs, dynamic interference management schemes can periodically adapt base station configurations to deal with dynamically changing traffic loads in the network. In some cases, static schemes can be adapted to dynamically manage the interference. For instance, adaptive FFR as explained in [87] permits the dynamic allocation of frequency bands and power levels taking into account the channel conditions changes. Coordinated Multi Point [88] transmission/reception (including Coordinated Scheduling/Coordinated Beamforming, Dynamic Point Selection, Dynamic Point Blanking [89] and Joint Transmission [90]) is the most pertinent example of dynamic interference mitigation schemes as it is based on dynamically coordinating the transmission and reception of multiple cells.

### 3.2.4. Level of Coordination

Based on the level of coordination, interference management schemes can be categorized into four categories:

1. Centralized [17]: A central entity (which can be an external controller or a macro base station) is in charge of collecting different measurements and required information from concerned cells (channel gains, users' QoS requirements, power and frequency patterns, etc.) and managing and allocating the available resources according to a utility function (maximize capacity, sum data rates, energy efficiency, etc.). Generally, these schemes deliver optimal global resource allocation, however, due to frequent exchanges between multiple cells and the centralized coordinator, they result in high backhaul signaling and computational complexity, which renders this architecture infeasible in large-scale networks.

2. Semi-distributed: Instead of centralizing all the required processing, a functional split [91] is performed such that interference management is performed at both a centralized external entity and at the BS. Based on partial network information sent from the base stations, the coordinator allocates a subset of the resources (time, frequency, power levels..) to each BS which is responsible for performing users' scheduling on a finer level. Compared to centralized schemes, the overhead and computational complexity of these two-level schemes are reduced. An example of a semi-distributed scheme can be found in [92].

3. Coordinated-distributed: In this category, all the processing is performed at the BS using local information received from its associated users. However, to perform global inter-cell interference management, the neighboring base stations exchange minimal amounts of information [93]. Although coordinated-distributed interference management schemes permit the minimization of the backhaul signaling and computational complexity, their implementation is limited due to constraints on communication between BSs.

4. Autonomous-distributed: No coordination is required between the different base stations. Based on local reported measurements, each BS applies an individual strategy to manage the inter-cell interference, for instance, to reduce power level transmission of selected resources with low SINR [94, 95]. The advantage of such schemes is that there is no signaling overhead and each BS can adapt its scheduling strategy very quickly to changing channel conditions. However, designing completely distributed algorithms that converge rapidly to optimal solutions is rather complicated [96].

### 3.2.5. Domain

From a domain perspective, interference management schemes can be categorized into four categories:

1. Time domain: These schemes enable time sharing of spectrum between neighboring cells in a way that minimizes inter-cell interference. Time domain eICIC technique is a well-known example where a BS (generally a macro BS) mutes its transmission during certain subframes referred to as Almost Blank Subframes (ABS) so that neighboring BSs (generally small cells) schedule their edge users in the protected subframes [97]. Various interference management schemes using ABS have been suggested in the literature [98].

2. Frequency domain: In the same way as time domain schemes, the frequency resources are divided between interfering BSs to reduce resource collision probability.

3. Power domain: These schemes enable better use of time-frequency resources, as the base stations are allowed to use the available spectrum during all subframes. However, to minimize the inter-cell interference level, the BSs apply some power control to reduce the generated interference. Further eICIC (FeICIC) is based on the same concept as eICIC, with the difference that during ABS, the BS is allowed to transmit data using relatively low power. Besides DL power control, it is also possible to perform UL-PC to optimize the power of associated UL channels and limit the effects of interference especially for cell edge users. This UL-PC can be performed via an open-loop PC, where the user determines its adequate uplink transmission power based on some DL measurements, or via a closed-loop PC where explicit power-control commands are transmitted by the BS based on previous measurements of the received UL power. To note that 5G NR has the capability for beam-based power control as an extension of LTE.

4. Space domain: Interference alignment, interference whitening (IW) and adaptive beamforming are powerful interference management schemes that can serve as examples for this category. For instance, beamforming exploits the spatial domain created by several antennas to focus signal energy on the receiver side, which increases the signal-to-interference ratio [99]. Treating interference as colored Gaussian noise, spatial-domain IW is a low-complexity scheme using the estimated covariance matrix of interference plus noise to whiten the received signal and estimated MIMO channel matrix [100]. Blind interference alignment is applied to serve multiple users in each cell and mitigate interference at the cell-edge users [101].

5. Code domain: Through the use of well-designed spreading codes, these schemes can manage interference between the users with similar channels. Sparse code multiple access (SCMA) [102] is an example of code-domain non-orthogonal multiple access schemes, which explores the codebook sparsity to eliminate the multi-user interference.

Several research works address the interference problem from multiple domain perspectives [103].

**Discussion:** In this section, we first classified the prominent interference in 5G and beyond cellular networks into two main categories, namely intra-cell interference and inter-cell interference. Then, we further organized the various types of interference for these two categories and provided a detailed description and explanation of each interference, i.e., when it occurs and its effects on the network. We also developed a comprehensive taxonomy to describe and categorize various interference management approaches according to their basic objective, implementation place, static or dynamic nature, level of coordination, and domain.

## 4. Interference Measurements and Signaling in 5G NR

Before providing a survey of the different interference management approaches for each interference type, it is necessary to have a complete understanding of the signaling for interference measurements and the different information exchanges between all the nodes in 5G systems. While academia is generally more focused on the study of interference management schemes than on signaling and measurement reports, standardization has been more active on this topic. Thus, this section mainly deals with a survey of 3GPP specifications of interference measurements and signaling in 5G NR, with specific subsections dealing with cross-link and remote interference-related enhancements.

### 4.1. 5G NR Signaling

For DL/UL channel sounding, beam management, synchronization, channel demodulation, and so on, Reference Signals (RS) are exchanged between the gNBs and UEs. In previous cellular technologies, always-on cell-specific synchronization and reference signals were spread over the entire spectrum for precise channel estimation. 5G NR uses a completely new approach, as it only broadcasts a minimum amount of cell-specific signals and the only always-on signal is the synchronization signal block (SSB). The main DL and UL synchronization and reference signals are introduced below.

- Channel State Information Reference Signal (CSI-RS) is sent by the base station towards the UE and used for DL CSI acquisition. Compared to Cell-specific Reference Signal (CRS) which was introduced in LTE Release 8, CSI-RS is configured per UE or beam for any configurable frequency range, has less overhead as it is not necessarily transmitted continuously, and supports a larger number of antenna ports. The resource elements carrying the CSI-RS can be configured to be either Zero Power CSI-RS (ZP-CSI-RS) or Non Zero Power CSI-RS (NZP-CSI-RS). CSI-RS also supports interference estimation.

- Sounding RS (SRS) is transmitted by each user periodically or aperiodically to the base station and utilized for UL channel quality estimation over full bandwidth or for a certain segment of the frequency band depending on its configurations specified by the gNB. Its role is similar to CSI-RS used for DL but in the opposite direction. Assuming channel reciprocity in TDD networks, SRS can be helpful for both UL and DL scheduling based on UL channel estimation.

- DeModulation RS (DMRS) is used in virtually each UL or DL physical channel instead of always-on CRS, for decoding the specific channel and providing information about a limited frequency range.

- Synchronization signals and physical broadcast channel including its DMRS, called SS/PBCH block (SSB) and transmitted over a limited bandwidth with a periodic but lower duty cycle than CRS, may be used for DL channel quality estimation to establish an initial connection or perform beam management [104]. To cover the whole cell, every gNB transmits one SSB per beam, which can be static or semi-static. While using SSB to acquire CSI reports does not lead to additional overhead, it is rather limited in terms of frequency and time configuration flexibility compared to CSI-RS.

Typically, to measure the interference from other neighboring BSs, the user may directly measure the reference signal power from the interfering signal directly, or use the ZP-CSI-RS which are special empty resource elements that can be configured for channel state information-interference management (CSI-IM) estimation [105]. With the latter approach, the network configures the CSI-IM to puncture the PDSCH from the serving cell creating a transmission gap and to collide with data transmission of the neighboring cell. In this way, the UE can perform interference measurements and provide feedback.

For a reasonably accurate estimate of the channel quality, ideally, orthogonal pilot signals should be assigned to different beams or users. However, considering the scarcity of orthogonal sequences, such a transmission approach becomes infeasible [106]. Consequently, non-orthogonal RS are used to satisfy the requirements, resulting in a significant problem, which is inter-pilot interference, referred to as pilot contamination [107]. This issue arises both in the same cell or between different cells. In multi-cell systems, the non-orthogonal pilot sequences in adjacent cells, especially those of cell edge users, may overlap with each other, leading to a poor UL channel estimation [108]. Furthermore, due to the sporadic massive access in 5G mMIMO networks, several users in the same cell may probably be assigned the same pilot for UL training resulting in severe intra-cell pilot contamination [106]. Besides its detrimental impact on UL coherent demodulation, in TDD networks, pilot contamination may also lead to inaccurate DL channel estimation and a degradation in DL throughput. The issue of pilot contamination, its possible sources, and mitigation techniques has been investigated by a wide body of existing literature. We refer the interested reader to [109] for more detailed information.

### 4.2. 5G NR Measurements and Reporting

When the users and gNBs receive reference signaling, they perform some measurements of the quality of the channel, to better manage the available resources and handle the interference. Typically, a UE measures the quality of downlink signals, such as CSI-RS and SSB, while the base station measures the quality of uplink reference signals such as SRS. However, it's also possible for a UE to measure uplink SRS sent by other UEs. The main UE measurement quantities include: (i) SS reference signal received power (SS-RSRP) and CSI-RSRP that represent the linear average power received from resource elements corresponding to the respective reference signal, (ii) SS signal-to-noise and interference ratio (SS-SINR) and CSI-SINR that represent the ratio of the linear average wanted signal power to the linear average interference plus noise power received from resource elements corresponding to the respective reference signal and (iii) NR Received Signal Strength Indicator (NR-RSSI) and CSI-RSSI that represent the linear average power over certain OFDM symbols from all sources including co-channel serving and neighboring cells, noise and ACI. If capable, a 5G user may also measure and report CLI-RSSI and SRS-RSRP representing DL-to-UL interference from a neighboring gNB and UL-to-DL interference from neighboring UEs respectively. For the gNB, it essentially measures UL SRS-RSRP.

Using the previously mentioned measurements, some lookup and mapping tables to discretize the measured entities and map them to a set of predefined values, the user generates its reporting which consists of the feedback concerning the channel quality and measured interference, and sends it back to the gNB in a periodic, aperiodic or semi-persistent manner depending on the configuration. While all measurements are done in the physical layer, the reporting can be sent via layer 1 in the CSI reporting or via layer 3 RRC in the measurement report message. In NR, the CSI report generally includes the following indicators: (i) Channel Quality Indicator (CQI) providing a measure of the channel strength, (ii) Precoding Matrix Indicator (PMI) to reconstruct or calculate the DL precoders, (iii) CSI-RS Resource Indicator (CRI) precising the index for strongest beamformed RS for beam training, (iv) Strongest layer Indicator (SLI), (v) Rank Indicator (RI) for multi-stream transmissions, and (vi) L1-RSRP [110].

Generally, the CSI reporting contains the feedback of codewords indexes, which refer to the optimum scheduling parameters, including the DL precoders, for the concerned user depending on its channel measurements. A set of different pre-defined codewords represents a codebook. Relying on the codebook design, the CSI feedback enables the BS to calculate the precoding matrix for adaptive beamforming and user scheduling [111]. Acquiring accurate CSI at the transmitter is of the utmost importance, especially for codebook-based beamforming, traditional resource scheduling schemes, and conventional interference management techniques. However, in practice channel estimation errors, along with feedback delay often occur in dynamic mobile networks, leading to imperfect and outdated CSI, which further complicates the design of adequate radio resource management approaches. In TDD networks, assuming channel reciprocity, the BS may use the UL SRS-RSRP channel estimation, to deduce the DL propagation channel, compute the DL beam weighting factors, and allocate the DL resources. While this approach permits the design of flexible dynamic beams following the users in a very accurate way in contrast to CSI-dependent codebased beamforming, it requires significant baseband processing on the gNB side.

Considered as one of the most important interference management schemes, designing adaptive beamforming approaches that are robust to channel aging, delay feedback and inaccurate estimation, etc., in dynamic mobile environments is still ongoing research. Some of the proposed schemes in the literature will be reviewed in the next sections.

### 4.3. Cross-link Interference - Signaling and Measurements

To mitigate cross-link interference in dynamic TDD-LTE systems, 3GPP Release 12 proposed different approaches for the enhanced interference mitigation and traffic adaptation (eIMTA) [54], including power control, cell clustering, UL/DL dynamic configuration, interference cancellation, and resource allocation and scheduling schemes.

To allow more flexible resource allocation in dynamic TDD networks, Release 16 introduced enhancements to better manage CLI [56, 59] that consist of:

- special measurements and reports on the terminal side: this is done by extending the RSRP measurement/reporting done by the device such that it is possible to measure uplink SRS of other nearby terminals. In this way, the network can gain knowledge about potential interference between devices in different neighboring cells.

- inter-gNB signaling to enable inter-cell coordination and interference mitigation: Resources can be either fixed resources, i.e., assumed to be used in either DL or UL, or flexible which implies that the gNB can use them in both transmission directions. The enhanced inter-gNB signaling is used by a gNB to announce its fixed and flexible resources to its neighboring gNBs. Using this information, a gNB may update its scheduling strategy, for instance, to schedule critical DL transmission in fixed resources where the interference levels from neighboring gNBs are known, while less critical transmission can be scheduled in flexible resources.

### 4.4. Remote Interference - Signaling and Measurements

There has been no standardized solution for managing the remote interference in TD-LTE [112]. However, in commercial LTE TDD networks, degradation of network coverage and connection success rate due to remote interference has been observed, thus RIM has been a mandatory feature by the operators to ensure reliable network performance. In 5G NR, with large-scale deployment of TDD networks, it is expected that the impact of remote interference will be much more severe not only in one country but also across borders. furthermore, as current RIM schemes implemented in LTE networks rely generally on manual control via operation and management system, it was important to have a standardized solution that minimizes manual intervention and improves the network robustness. Therefore, based on the study in [57], NR Release 16 included tools to support automatic RIM [56]. Specifically, two new RIM-reference signals type 1 and type 2, having the same basic design with different objectives, were specified.

- RIM-RS type 1 is used by a victim gNB and sent to aggressor gNBs to indicate that a ducting phenomenon

occurs and that RI is detected. This reference signal can convey information about (i) the id of the gNB (or group of gNBs) causing interference, (ii) the number of symbols affected by the RI, and (iii) whether or not the interference has already been mitigated so that a progressive remote interference management scheme can be applied to the aggressor.

- RIM-RS type 2 is transmitted by the aggressor gNB after applying the RIM procedure, to inform the victim that a ducting phenomenon continues to occur.

Three different frameworks were proposed: centralized and two distributed ones, which differ in terms of where in the network the decision of RIM is applied and how the gNBs communicate with each other. It is important to note that the actual schemes to mitigate remote interference are not specified by 3GPP.

**Discussion:** This section dealt with a survey of 3GPP specifications of interference measurements and signaling in 5G NR. The main DL and UL synchronization and reference signals were first introduced, including CSI-RS, SRS, DMRS, and SSB. Then, we introduced the problem of inter-pilot interference, referred to as pilot contamination, which is due to the use of non-orthogonal RS, and briefly discussed its detrimental impact on both UL and DL. A detailed explanation of 5G NR measurements and reporting performed by both the users and gNBs, was then provided, along with some insights on channel estimation and CSI feedback challenges. Finally, we discussed some NR Release 16 enhancements specifically dealing with cross-link and remote interference signaling and measurements.

## 5. Intra-Mode Inter-Cell Interference Management

In this section, we will present various research works that deal with intra-mode co-channel inter-cell interference management for downlink and uplink transmissions in LTE 4G and 5G networks. A more focused review of inter-cell inter-beam interference management will be provided in the next section. To alleviate the severe inter-cell interference, several interference management schemes have been proposed in the literature, including power-domain schemes [113, 114, 115, 116], time-domain approaches [117, 118], frequency-domain schemes [119, 120, 121], spatial-domain schemes [122, 123, 100], code-domain schemes [124] and multi-domain schemes [125, 126, 127]. In the following, we review and compare some of the proposed techniques.

A) *Power-domain schemes*
Assuming that macro co-tier interference is avoided through some form of frequency repartition, the work in [113] only focused on mitigating cross-tier DL and UL interference in an OFDMA-based HetNet with multiple pico cells overlaid on a macro cell. To expand small cell coverage, the authors first introduced closed-form expressions to calculate the proper cell offset values for two different range expansion strategies. To avoid both downlink and uplink interference and to improve pico cell performance, a cooperative scheduling based on a power transmission scheme was then proposed. The main idea of this method is that pico BS would inform the macro BS, via the X2 interface, about the set of resources allocated to all the users in the expanded region. Hence, the macro would lower its transmit power in these RBs to ensure the desired SINR for pico edge UEs. Simulation results provided insights on appropriate cell bias values and demonstrated that the proposed cooperative scheduling power transmission scheme enhances the downlink and uplink average network rates by 36% and 7%, respectively compared to ICIC approaches.

A similar approach to [113] was presented in [114], where a distributed DL resource allocation algorithm based on transmit power minimization, requiring some coordination between the base stations, was proposed. However, in the former studies, only co-tier DL interference in OFDMA femto cell networks was considered. Firstly, every station dynamically and independently selects a modulation and coding scheme, a resource block, and a transmission power for its UEs, to minimize its total DL transmission power, and still meet the UEs' QoS requirements. Then, it informs its neighboring cells of the RBs assigned to its cell edge users, so that they minimize their transmit power on those specific RBs. This inter-cell coordination was proved to provide efficient spatial reuse and interference management. Extensive simulation results demonstrated that downlink power control on a per-RB basis provides a 20% improvement in network throughput.

While in the last studies, the intra-mode inter-cell interference management was performed via a distributed coordinated optimization on a per-RB basis, in [115], the authors developed a Deep Reinforcement Learning (DRL)-based centralized solution for DL-PC in a multi-cell 5G network optimizing power transmission per cell. The problem was first formulated as a Markov Decision Process (MDP), then the well-known deep Q-networks (DQN) algorithm and some of its recent extensions were employed to learn optimal downlink cells' power based on global information such as certain user statistics of SINR and the harmonic-mean of cell throughputs, and aiming to maximize user rates while ensuring fair resource allocation. With the help of a 5G-compliant system-level simulator, several DQN agent variants with different extensions and hyperparameters were trained and validated. The best model attaining the highest gain in throughput of the cell-edge UEs in the validation stage was then selected for the test stage, along with vanilla DQN. The performance of the two models was tested and compared against the fixed power allocation approach in a set of scenarios different from the training and validation sets. The results showed that: (i) the proposed scheme permits the mitigation of the ICI experienced by cell edge UEs, and thus significantly improves their throughput, (ii) the vanilla DQN agent is slightly more efficient than the DQN agent with extension but transmits 20% more power on average and (iii) both agents exhibit good generalization capabilities due the use of stochastic policies to generate training data and random initialization of cell powers.

Interference Management in 5G and Beyond Networks

Power control schemes have also been frequently proposed to mitigate uplink inter-cell interference, such as in [116], where Deb et al. proposed a measurement data-driven machine learning approach, called LeAP, for power control in OFDMA-based heterogeneous networks. The proposed semi-distributed framework permits the interference pattern to be modeled while taking into account variations in traffic load as well as propagation geometry and topology. However, the effects of fast fading were not adequately handled. Based on the measurement of modeled interference patterns, a stochastic learning-enabled gradient algorithm was proposed to identify the best power control settings and maximize uplink network performance. The authors carried out in-depth evaluations based on radio network plans from a real LTE network. The results demonstrated that, in comparison with conventional methods, LeAP offers an x4.9 gain in the 20th percentile of user data rate and an x3.25 gain in the median data rate.

B) *Time-domain schemes*

Time domain eICIC scheme was studied in [117] and [118] where a two-tier HetNets, containing macro and pico cells, was considered. In the former study, users were sorted as center and edge by comparing their RSRP measurements with a fixed threshold. The authors first presented a stochastic geometric analysis of spectral efficiency, then they used it to derive a constrained optimization problem formulation of the Almost Blank Subframe configuration for both macro and pico cells. The goal was to minimize the DL cross-tier interference impacting edge users' performance and to maximize the derived energy efficiency objective. Using the Lagrangian function and Kuhn–Karush–Tucker (KKT) conditions, the authors provided a partial closed-form solution. It was assumed that macro and pico BSs exchange their ABS patterns. The cell edge users are then scheduled during the ABS corresponding to the other tiers. The numerical results provided some insight into the system's behavior. For instance, guidelines on whether to use ABS for macro, pico or both tiers were presented. It was also shown that macro ABS implementation provides better energy efficiency. Moreover, the impact of cell range extension bias on the choice of ABS patterns was given and an intuition of the value of the cell bias was provided.

In [118], Ayala-Romero et al. developed an online learning algorithm to jointly control energy efficiency and interference coordination using eICIC in a two-tier HetNets. The two-level framework based on a constrained contextual bandit problem, aimed to maximize the energy efficiency, while meeting users' requested QoS. At the higher level, a centralized coordinator, in charge of an MCs cluster and operating with reduced state and action spaces, selects global control eICIC parameters, i.e., ABS ratio and CRE bias, for the group of MCs. At the lower level, multiple controllers, one per macro cell, translate each global configuration into local configurations. Then, each macro aggregates the feedback concerning energy consumption and attained QoS received from its corresponding pico cells, and sends it back to the coordinator. Using these performance observations, the global coordinator updates its knowledge allowing it to learn stage by stage how to select better policy maximizing the energy efficiency with a minimum required QoS threshold. According to the numerical results, the proposed algorithm achieves an energy saving of 25% and a QoS fulfillment ratio of 100%.

C) *Frequency-domain schemes*

The authors in [119], presented a novel FFR scheme, called FFR-3SL, to avoid both intra and inter-tier DL interference in HetNets, through a frequency partitioning mechanism. The MC coverage zone is divided into three sectors, themselves subdivided into three layers (central, middle, and outer), and the total bandwidth is split into seven sub-bands. One sub-band is reserved for macro UEs of the central layer. The other six sub-bands are assigned to macro users in the middle and outer layers of the three sectors, and can also be reused by femto cells. When a new femto cell is turned on, the system recognizes its position and tries to allocate the best available sub-band. A Monte Carlo simulation is carried out to evaluate the performance of the FFR-3SL algorithm. The simulation results showed that the proposed FFR-3SL approach outperformed other FFR approaches in terms of throughput and efficiency. However, one of the limitations of this algorithm is the pre-assignment of a dedicated sub-band to the macro central layer, which can affect cell edge users.

Adopting a centralized heuristic approach with partially static frequency subband assignment to mitigate time-varying intercell interference is generally not effective because of its lack of adaptability and generalizability for the dynamic environment. Therefore, in [121], an intelligent adaptive distributed ICI approach for autonomous SBS was proposed. Specifically, with reduced signaling interactions, the small BSs sense the environment and exploit partially observable network information to allocate sub-channels to every UE at each TTI in such a way that mitigates ICI and maximizes the long-term throughput. The sequential decision-making processes of the decentralized SBSs were first modeled as a distributed inverse reinforcement learning problem following the non-cooperative Partially Observable MDP games. Then a non-prior knowledge-based self-imitating learning algorithm was proposed to imitate the expert strategies, generate adversarial training samples, and perform the decentralized resource scheduling. Simulation results confirmed the effectiveness of the proposed method to mitigate ICI and improve the overall throughput compared to other known benchmark algorithms.

While the last studies dealt with DL interference, in [120], UL multi-tier interference in HetNets was mitigated through a heuristic physical-layer technique. Specifically, the authors proposed to assign to multi-priority users a novel Frequency Hopping (FH) pattern with multi-level Hamming correlations. The numerical and simulation findings showed that the prioritized radio-access scheme based on the FH sequence set might minimize UL cross-tier interference in HetNets while ensuring high transmission quality and spectral efficiency for multi-tier users, even at the cell boundaries of HetNets.

Particularly under the imperfect power control over the cross-tiers, the proposed FH method outperforms the traditional FH technique. The proposed approach only considered two levels of priorities among users, but a general case of FH applied to flexible multi-level priorities was not investigated.

D) *Spatial-domain schemes*

Considering both cross-tier and co-tier interference, an interference management approach based on a hierarchical precoding scheme was proposed in [122] to mitigate interference from the macro and FD-enabled SCs with perfect self interference cancellation capabilities. The authors considered a dense deployment of multi-antennas macro and small BSs using co-channel TDD protocol, where SCs serve both macro and small cells' users. The problem of joint load balancing and interference mitigation was formulated as a network utility maximization problem taking into account dynamic backhaul constraints, traffic load, and imperfect CSI. Random matrix theory was applied to derive closed-form for the objective and constraints and Lyapunov optimization was then used to decouple the original problem into several solvable sub-problems. Using the successive convex approximation method, the problem was converted to a convex problem, and an optimal solution for user association, beamforming design, and power allocation was derived. Numerical results revealed that the proposed framework can achieve a 5.6 times gain in cell-edge users' performance compared to un-optimized ultra-dense networks with 350 SCs per $km^2$.

While a precoding scheme was proposed to manage ICI in the last study, [123] and [100] investigated spatial-domain interference whitening (IW) using Machine Learning. IW permits the estimation of the colored interference-plus-noise covariance matrix using demodulation reference signals. However, when the interference is rather small in comparison with the noise, interference whitening may degrade the performance gain due to inaccurate covariance matrix calculation based on a few DMRS samples. To address this issue, in [123], the authors proposed a distributed RL-based interference suppression scheme for adaptive IW control. More specifically, using the Q-Learning algorithm, each agent, i.e., the UE, selects the best IW mode, i.e., ON or OFF, based on the estimation of its average channel condition including interference and noise. To ensure that each UE learns autonomously and that only local information is exploited, the reward value is defined as the difference between the channel's capacity and a specified target threshold. The experimental results demonstrated that RL-IW improved the performance gains in terms of block-error rate (BLER) and DL throughput compared to conventional interference whitening.

While in the last paper, the authors chose to activate/deactivate IW according to the performance gain, in [100], the problem of computing the interference-plus-noise covariance matrix from a sparsely located DMRS was addressed using a supervised learning (SL) based algorithm aiming to minimize both the BLER and the whitening complexity. Under various interference scenarios characterized by different channel models, interference occupancy, Signal-to-Interference Ratio (SNR), etc., a single neural network is trained using the quasi-Newton method to select an appropriate IW option to minimize the BLER. Specifically, the proposed SL algorithm decides (i) whether or not to calculate the covariance matrix by averaging over the entire bandwidth and (ii) whether the covariance matrix can be approximated with a diagonal matrix or not. Results showed that the proposed scheme can reduce the BLER, even under unseen interference scenarios while minimizing the complexity of whitening.

E) *Code-domain schemes*

In [124], a non-orthogonal sparse code multiple access-based uplink inter-cell interference cancellation technique was proposed to enhance the performance of the cell-edge UEs in a multi-cell scenario, by jointly decoding the useful signal and the interference signal at the BSs. Through the design of sparse SCMA codewords, the users' data are spread among the shared time-frequency RBs which permits the achievement of frequency and interference diversity and a near-optimal moderate complexity multi-user detection at the base station. To optimize the BER even further, the SCMA scheme was combined with the Alamouti code to provide full space-time-frequency diversity. Simulation results proved the ICI cancellation efficiency of the suggested SCMA-based scheme, especially when combined with the MIMO technique.

F) *Multi-domain schemes*

Several studies were based on multi-domain schemes to effectively mitigate inter-cell interference, such as [125, 126, 127].

M. Kaneko et al. suggested an autonomous distributed resource allocation in [125] for frequency-power domains based inter-cell interference management. Specifically, the authors considered an OFDMA-based macro cell/femto cell overlaid Heterogeneous Network. Each femto cell is assumed to operate in the closed access mode and to share the spectrum frequency with macro cells. Unlike conventional methods of cross-tier interference management that relied on the CSI provided by macro BSs as well as macro users allocation mapping, and causing a significant overhead increase, the work in [125], proposed a distributed resource allocation method based on femto BS' overhearing of local CSI and prediction of the sub-channels likely to be used by the nearby macro UEs. Using the predicted low-interference sub-channels, the femto BS performs the resource allocation to its associated UEs while considering the maximization of the overall system throughput without requiring additional control overhead between BSs. The simulation results demonstrated that the proposed approach efficiently manages to avoid cross-tier downlink interference from femto BS to nearby macro UEs, as it provides a trade-off between macro and femto BSs throughput while reducing required signaling overhead.

In addition to the power domain, the proposed interference mitigation approaches in [126] were also based on the time domain. More specifically, Liu et al. considered a randomly generated HetNet where macro BSs would offer muted or reduced power subframes, while the pico BSs always transmit on all subframes. They assumed that only the pico BSs may change their offset bias value to extend their coverage and assumed that the interference range of a base station is within a few of its nearby cells. Using an exact-potential game framework appropriate for performing eICIC and FeICIC optimizations, the authors presented a scalable distributed coordinated scheme to optimize cell offset and ABS patterns (muted or reduced power) using better response dynamics. The objective utility function could be flexibly chosen. Moreover, a downlink scheduler based on a cake-cutting approach was proposed and compared with well-known schedulers. It was proved via simulations, that the proposed framework achieves a significant gain in energy efficiency using eICIC and further improvement with FeICIC. Higher cell edge throughput and user fairness were also observed.

While the previous work focused on eICIC/FeICIC optimization, the study in [127] proposed a frequency-time-power domains scheme to handle the DL co-tier interference between SCs and improve the overall energy efficiency. Specifically, M. Osama et al., proposed a static Soft Frequency Reuse (SFR) scheme and a dynamic on/off switching of the small cells based on their Interference Contribution Rate (ICR) values, while considering the irregular nature of the 5G HetNets. Cross-tier interference was not considered, as MCs and SCs are assumed to use two different frequency bands. In this paper, a centralized controller, which can be the MC, gathers the information from the different SCs in its coverage, statically allocates the sub-bands, and then controls the SC switching based on their load and their generated interference. Based on SFR, each small cell is divided into two areas: center, and edge and the frequency band is partitioned into seven sub-bands. The controller uses an iterative algorithm to allocate for the edge zone of each small cell, one sub-band that has not been allocated to neighboring SCs edge area. The rest of the sub-bands can be used by the center area with lower transmitted power to reduce the interference between small cells. After assigning the resources to the users by their serving cells, the controller performs the SC on/off switching based on the ICR concept. On the other hand, MCs are maintained active continuously to ensure network coverage. The results showed that the suggested approach could decrease the level of co-tier downlink interference, as well as the power consumption and traffic loss.

**Discussion:** Table 3 provides a comparison between all previously listed papers dealing with intra-mode inter-cell interference management. The majority of the studies dealt with DL interference as it is the most affecting the network performance. Few other papers studied the UL or DL and UL inter-interference. Diverse domains were applied in the different proposed schemes in the literature, with a special focus on power-domain approaches or multi-domain approaches combining time/frequency/power schemes to avoid inter-cell interference. Centralized methods usually incur high overhead signaling while autonomous or coordinated distributed methods require zero or limited information exchange. With the dynamic time-varying channel conditions and users' mobility, static interference coordination schemes can no longer adapt and manage effectively severe and unpredictable interference. Recent emerging machine learning algorithms, such as deep reinforcement learning and multi-agent DRL algorithms, in particular, are excellent tool for tackling these challenges.

# 6. Inter-Cell Inter-Beam Interference Management

This section will be dedicated to the in-depth study of the proposed approaches managing the inter-cell inter-beam interference from the beamforming transmissions of neighboring BSs in 5G and beyond networks.

To avoid inter-cell inter-beam interference, different approaches have been recently studied, including: interference-aware beam selection schemes [62], interference-aware beam power allocation schemes [128], interference-aware beam selection and beam power allocation schemes [129, 49, 130], time-domain beam scheduling schemes [52, 131] and precoding design schemes [132, 133, 134, 135].

In the following, we review and compare some of the proposed techniques.

A) *Interference-aware beam selection schemes*
The study in [62] considered the joint optimization of user association and beam pair link (BPL) allocation in multi-user multi-cell mm-wave networks to maximize the network sum throughput, and mitigate the spatial interference.

| Ref | Interference Type | Co-tier/ Cross-tier | Network Environment | Proposed Scheme | Control Level | Limits |
|---|---|---|---|---|---|---|
| [126] | Downlink ICI | Cross-tier & Co-tier | LTE-A HetNets | eICIC / FeICIC + CRE | Distributed coordinated | eICIC and FeICIC applied only on macro BSs |
| [119] | | | LTE-A HetNets | Fractional Frequency Reuse with Three Sectors and Three Layers | Centralized | pre-assignment of a dedicated sub-band to the macro central layer |
| [122] | | | TDD 5G HetNets | Hierarchical precoding scheme | Centralized | - |
| [117] | | Cross-tier | Two-tier HetNets | eICIC ABS + CRE | Centralized | disjoint ABS and CRE optimization |
| [118] | | | Two-tier HetNets | eICIC ABS | Semi-distributed (2 levels control) | reduced space dimension of possible configurations |
| [125] | | | FDD HetNets | Local CSI overhearing and resource allocation on predicted low interference sub-channels | Autonomous distributed | co-tier interference among secondary users was not considered |
| [127] | | Co-tier | FDD HetNets | SFR + ICR based on/off Switching | Centralized | Cross-tier interference is not considered |
| [114] | | | OFDMA Femto cell Networks | Power control on a per-RB basis | Distributed coordinated | Cross-tier interference is not considered |
| [115] | | | 5G Macro layer | DQN-based Power control | Centralized | The power control is done on a per-cell basis |
| [121] | | | OFDMA SBS Networks | SIL algorithm which consists of WGANs) and Double DQN methods to schedule sub-channels to individual users | Autonomous distributed | - |
| [123] | | | MIMO-OFDM multi-cell | QL based IW | Autonomous distributed | The definition of the reward may induce a decrease in the throughput |
| [100] | | | MIMO-OFDM multi-cell | Neural network based IW | Autonomous distributed | - |
| [113] | Downlink & Uplink ICI | Cross-tier | OFDMA HetNets | Cell offset and cooperative scheduling | Centralized | Overhead Signaling |
| [116] | Uplink ICI | Cross-tier & Co-tier | OFDMA HetNets | Power control based on the measurement of modeled interference patterns | Semi-distributed | the effects of fast-fading were not adequately handled |
| [120] | | Cross-tier | HetNets | A prioritized radio-access system based on a frequency hopping technique | - | Only two-level access priorities applied to HetNets uplinks |
| [124] | | Co-tier | Multi-cell network | An SCMA-based Uplink ICI | Centralized | - |

Table 3: List of Papers dealing with Intra-mode Inter-Cell Interference Management Schemes

While prior works in the literature generally relied on statistical channel models [136] and idealized antenna patterns which may severely underestimate the spatial interference, [62] used site-specific 3D directional mm-wave channel data with realistic building and antenna models for a more accurate networking study that fully characterizes the spatial intra and inter-cell interference in SDMA networks. The study further assumed an analog beamforming model using a predefined beam codebook. Two heuristic interference-aware schemes based on a centralized joint-BS and a distributed Per-BS approaches were proposed. While the former considers all possible BPLs on all base stations for each UE, the latter is executed on each base station, taking into account only a small set of candidate BPLs on the serving base station for every connected user. In the two approaches, the set of candidate BPLs for every UE was first determined. Then, every approach sequentially verifies the UEs and all of their candidate BPLs and computes beamforming weights for each new candidate BPL considering all previously allocated BPLs at the serving BS. After updating the performance metrics, the candidate BPL is either added to the UE's provisional BPL allocation set or discarded depending on the comparison of previously allocated BPLs and a predefined coverage threshold. Finally, the candidate BPL in the provisional set with the best SINR is assigned to the final BPL allocation set. The performance of the proposed strategies was evaluated against the interference-agnostic Max-RSS baseline and load-balancing WCS benchmark. On one hand, the results demonstrated that the centralized approach substantially surpasses all other schemes, although it is more complex and requires inter-BS coordination and frequent up-to-date CSI from several base stations. On the other hand, the distributed Per-BS scheme achieves a comparable performance to the centralized WCS, while being less complex, as it only takes into account the BPLs assigned to the serving BS, and only operates based on CSI-RS updates from the serving base station, as in 5G-NR's default beam management.

B) *Interference-aware beam power allocation schemes*
To mitigate uplink inter-cell inter-beam interference and enhance the total data rate on the UL channel, the authors in [128], proposed a coordinated distributed Multi-Agent Reinforcement Learning (MARL) beam-based uplink power control framework which is compliant with the 5G NR specifications in Release 15. They considered a multi-cell system where both BSs and UEs were equipped with multiple antennas. It was assumed that all the cells reuse the same frequency band and that UEs periodically measure their associated beam pairs. In this study, intra-cell interference was not considered as only a single active UE per time-frequency resource in a cell was allowed. The proposed approach consists of a joint power optimization of multiple antenna ports per UE. Based on uplink measurements of the SRS transmissions of the associated UEs and using Q-Learning which is an off-policy value-based RL method, each agent, i.e., the BS, learns the most appropriate power allocation on the uplink and specifies to its users how to change their transmit power levels via Transmit Power Control (TPC) commands which are limited to the set {-1, 0, +1, +3}. The learning process is achieved based on the exchange of UL measurements inter-BSs and the computation of a reward function as a convex parameterized weighted sum of all the cells' data rates. Such reward describes the direct impact of the BS's decision on its associated UEs and the data rate of the other cells of the network. The determination of the most suitable design of the parameterized weight in the reward function is scenario and operator-objective dependent. Owing to the results in the paper, when it is below 0.6, it results in the most cooperative behavior of the system, which efficiently reduces the interference, despite the small increment of the transmit power. Overall, the simulation results showed that the proposed uplink power control framework provides near-optimum performance in terms of total transmit power, total data rate, and network energy efficiency.

C) *Interference-aware beam selection and beam power allocation schemes*
In [129], the authors considered a multi-access OFDM system, from a multi-antenna BS to a single-antenna UE. The sub-6 GHz band was used for voice and the mm-Wave band for data. Starting from the question if there is a way to jointly optimize beamforming, power control, and ICI which achieves the upper bound on SINR, the authors formulated the joint optimization problem as a non-convex problem maximizing the achievable sum rate of the users. The authors demonstrated that there is no closed-form solution for the formulated problem. Therefore a centralized coordinated scheme for voice and data bearers using deep reinforcement learning was proposed. Based only on the reporting of the user coordinates to the BS, the selection of the beamforming vector of a selected user is first performed, then the centralized coordinator adjusts the beam power of the BS to this particular user. It was demonstrated that the proposed approach enhances the performance measured by SINR and total throughput capacity and that its run time is considerably quicker than that of the upper bound algorithm.

Using a similar interference management approach as the previous study, in [49], an online multi-agent Q-learning algorithm was proposed to jointly optimize the user association and beam power allocation in a NOMA mm-wave network, where multiple users per cell were served on the same time/frequency resources. The intra-cell intra-beam interference was supposed to be canceled via Successive Interference Cancellation at the UEs. Furthermore, as different beams in the same cell were allocated different frequency bands, the intra-cell inter-beam interference was avoided. However, while the intersection of coverage areas of multiple beams from different gNBs offers more options in the association of users, inter-cell inter-beam interference mitigation was essential to increase the network sum rate. In the proposed scheme, each gNB was considered as an agent that chooses user association for only cell edge users and inter-beam power levels. The binary states of each agent were defined according to the average SINR, reflecting the level of interference in the environment, compared to a predefined threshold. After each learning step in the Q-Learning algorithm, each agent receives a +1 or -1 reward depending on the achieved SINR and previously mentioned threshold. The proposed machine learning algorithm was compared to a baseline algorithm that allocates equal power to the beams and results revealed a sum rate improvement of 13% and 30% corresponding to the lowest and highest traffic loads, respectively. However, there were no reported results on cell edge users' performance.

While in [49] a joint optimization of beam selection and power allocation was performed, in [130], the optimization was performed in a disjointed manner and a multi-cell least beam interference (MLBI) algorithm and a multi-beam prioritized power allocation (MPPA) algorithm were presented to manage inter-cell inter-beam interference, guarantee the QoS of edge UEs and enhance the overall rate. The authors considered a multi-cell mm-wave network where each BS has M beams selected from the codebook and orthogonal to each other, therefore intra-cell inter-beam interference was not considered. However, the use of two beams with the same index by two adjacent BSs would lead to IC-IBI for edge users. In this study, for simplicity, only the main lobe interference was considered (side-lobe interference was ignored). As the optimization problem is a mixed-integer nonlinear programming problem, two sub-problems were formulated. For the first problem, each cell is divided into sections, and users are classified into central and edge users for each section. The purpose of MLBI is to select the candidate beam set and corresponding UEs with the least interference, based on the average power allocation. To avoid IC-IBI, it is important to schedule different beam sets for edge users of adjacent sectors. Based on the result of the first sub-problem, the MPPA algorithm including two scenarios - considering fairness or not - was introduced to allocate power to multiple beams to maximize the rate and improve the QoS. This problem was solved using the water-filling algorithm and using a low-complexity sub-optimal algorithm. Simulations demonstrated the efficiency of the proposed solutions to improve the achievable rate, even though it is sub-optimal.

D) *Time-domain beam scheduling schemes*
Such schemes allow the coordination of the beam transmission of different cells on the time domain to avoid inter-cell inter-beam interference. In [52], two transmission coordination approaches aiming to maximize the aggregate throughput in a dense mm-wave wireless network were introduced. The first one is a BS-density coordination approach defining an interference range based on the user and BS locations and SINR threshold, and deriving an optimal transmission probability of BSs. The second one is a pairwise beam collision-measurement-based approach that exploits the downlink beamforming information exchanged inter-BSs to define an inter-base-stations collision matrix and a threshold for the packet delivery failure probability-based transmission policy.

In [131], a graph theory-based time-domain beam scheduling scheme, referred to as least beam collision (LBC), was developed to avoid DL inter-cell inter-beam interference in mm-wave networks. It is assumed that every cell only uses one beam at every time slot. The network was modeled as a graph in which every cell is symbolized by a node. The beams were sorted depending on the closest adjacent cell they were likely to interfere with, and thus formed an edge between the two network elements. Beam collision was defined as the simultaneous use of the beams that are represented by an edge connecting a pair of adjacent cells. The purpose of the proposed LBC algorithm is to minimize the number of beam collisions while scheduling the beams in different time slots to guarantee the users' demands. Monte Carlo simulations verified that the proposed scheme ensures the acquisition of the global minimum amount of beam collisions and mitigates the strong inter-cell interference of the nearest cell. However, taking into account only one beam interferer is not always sufficient in dense networks.

E) *Precoding design schemes*
To mitigate the inter-beam interference caused by adjacent beams in multi-cell multi-user environments, in [132, 133], four different digital precoding techniques with a hybrid beamformer were proposed. The authors further analyzed the pros and cons of each technique in terms of required RF chains in the BS and UE, and network overhead and they compared their performance in terms of the achievable rate and BER. It has been shown that type 2/type 3/type 4 precoders with the estimated effective network channel may substantially decrease inter-beam interference in millimeter-wave cellular networks with stationary users.

Existing codebooks-based beamforming/combining are generally pre-defined and only concentrate on enhancing the target user's beamforming gain, without considering interference, which leads to severe performance loss. Therefore, in [134], the design of an interference-aware analog combiner that maximizes the achievable rate of the target user without knowing the explicit channel knowledge was investigated. The study leveraged the powerful exploration capability of reinforcement learning to propose an actor-critic-based DRL approach that iteratively optimizes the beam pattern to suppress the interference without explicit CSI knowledge nor inter-BSs coordination, and based solely on power measurements. The proposed analog beamforming/combining method also complies with the main hardware constraints, including the quantized phase shifter constraint. Simulation results proved that the proposed approach was able to learn an effective beam pattern suppressing inter-beam interference at the cost of negligible beamforming/combing gain reduction. In [135], the idea of the last study was further extended with a sample-efficient digital twin essentially used to provide the RL agent with a simulated environment to interact with and learn optimal beam pattern design. The adoption of the digital twin enhances the sampling efficiency by taking better advantage of the underlying relationship between signals and by allowing data sharing among different beam learning agents.

**Discussion:** Table 4 provides a comparison between all previously listed papers dealing with inter-cell inter-beam interference management. Generally, to manage inter-cell inter-beam interference in mm-wave networks, there are two principal approaches: (i) to perform interference-aware beam selection/beam power allocation/time beam scheduling and (ii) to design multi-user precoding/combining for IBI cancellation. The first method depends on codebook-based analog beamforming with limited CSI feedback. The second approach however is generally based on hybrid beamforming consisting of analog beamforming using a limited number of radio frequency chains coupled and digital baseband beamforming using full channel estimates, which may lead to extra CSI feedback and delay requirements. As traditional fully digital beamforming schemes require one RF chain per antenna which is not cost and power-efficient to implement, and as analog beamforming only enables one single stream transmission per baseband chain, hybrid beamforming offers a compromise on the performance and cost trade-off. However, designing hybrid precoding for very large arrays and bandwidths is still challenging.

## 7. Cross-Link Interference Management

This section will be dedicated to the in-depth study of the main proposed schemes and research studies related to cross-link interference management in 5G and beyond networks.

Z. Shen et al., in [55], provided an interesting tutorial on dynamic downlink/uplink configuration, adapting to changing traffic needs, and interference management schemes in time-division LTE networks. The opportunities and challenges of D-TDD systems were analyzed and system requirements and performance evaluations were provided. Different approaches have been adopted in the literature to manage CLI, including: power control schemes [137, 138, 139], UL/DL dynamic configuration [140, 141, 142, 143], cell clustering [144, 145], beamforming schemes [146, 147, 148], interference suppression schemes [149, 150, 151] and joint optimization of coordination schemes [8]. In the following, we review and compare some of the proposed techniques.

A) *Power Control* permits the management of CLI by reducing or/and boosting the DL/UL transmission power respectively.

UL power control aims to adjust the power of the uplink UE, either to reduce the UL-to-DL (between UEs) interference or to enhance uplink performance in case of high BS-to-BS interference. For instance, Q. Chen, et al. in [137], proposed a closed-loop uplink power control scheme to reduce DL-to-UL interference among TDD-based small cells. By adequately boosting the UL transmission power according to different interference levels, the authors aimed to improve UL performance. In this proposed scheme, based on geometric location information, the number of interfering base stations is first estimated, then the BSs exchange their UL configuration number. After that, interfering BSs compute the interference level indicator and configure UE power adjustment to enhance their UL performance. From the simulation results, it was shown that the proposed scheme achieves about 5% gains in average DL/UL packet throughput.

Some studies cater to adjusting the DL power control as an interference control mechanism like in [138]. The authors considered local 5G indoor micro operator utilizing dynamic TDD in the 3.6 GHz band and experiencing both inter-operator intra-mode inter-cell interference and inter-operator cross-link interference as cells that belong to separate operators are not coordinated. Assuming that the interfering network is fully loaded, the victim micro-operator BS estimates the total power received from all interfering BS that belong to different micro-operators. The maximum of these measured values is then compared with a predetermined interference threshold to set the transmit power scaling factor for the interfering micro-operator. This simple interference control depends on the appropriate definition of the interference threshold which is extremely scenario-dependent.

Other research works like [139] considered both DL and UL power control. To mitigate cross-link interference and maximize traffic adaptation, a DL/UL transmission power control for D-TDD heterogeneous networks was applied to flexible subframes that can be used either for downlink or uplink. As macro and small BSs were assumed to operate on different frequency bands, only co-tier interference in small cells was considered. To define the best power allocation balance for downlink and uplink transmissions while semi-statically exchanging information among BS, the authors proposed to manage DL-to-UL interference by (i) reducing the BS transmission power within a certain range to maintain an acceptable DL throughput, and (ii) boosting UL transmission power with a constraint on the user device power consumption, in flexible subframes. However, this power adjustment is done using fixed values. The proposed power control scheme achieved a maximum gain of 21% for UL throughput, but slightly degraded DL performance due to the reduction in the base station's power.

B) *UL/DL Dynamic Configuration* is another method to mitigate CLI [140, 141, 142, 143]. It is expected that 5G NR would support both semi-static configuration, to avoid inter-operators interference, and dynamic configuration in SCs.

In [140], the use of a multi-agent Q-learning-based approach to dynamically change the switching point (transition point) from UL to DL transmission in non-cooperative femto cell networks while mitigating CLI and maximizing the total cell average packet throughput, was discussed. In the learning phase, each agent, i.e., each femto BS senses its environment and based on its local quantized information data, including the level of the total received UL and DL interference, and its current switching point, it configures the new UL/DL switching point. As a result of its action, it observes instantaneous performance from UEs and an individual cost taking into account the traffic demands (arrival rates and packets' size) and the total UL and DL service rate.

| REF | PROPOSED SCHEME | CONTROL LEVEL | MAIN IDEA |
|---|---|---|---|
| [62] | Interference-aware beam selection scheme | (i) Centralized and (ii) Distributed | Two heuristic interference-aware strategies using a centralized joint-BS and a distributed Per-BS approach for the joint optimization of user association and beam pair link allocation |
| [128] | Interference-aware beam power allocation scheme | Coordinated distributed | A multi-agent reinforcement learning beam-based uplink power control approach relying on the exchange of UL measurements inter-BSs |
| [129] | Interference-aware beam selection and beam power allocation scheme | Centralized | A deep reinforcement learning scheme for joint beam selection and power allocation for voice and data bearers |
| [49] | | Distributed | An online multi-agent Q-learning algorithm to jointly optimize the user association and beam power allocation |
| [130] | | Distributed | A disjoint optimization of the two sub-problems via a multi-cell least beam interference algorithm and a multi-beam prioritized power allocation algorithm |
| [52] | Time-domain beam scheduling scheme | Coordinated distributed | A BS-density and a pairwise beam collision-measurement based approaches |
| [131] | | Centralized | A graph theory based time-domain beam scheduling scheme to minimize the number of beam collisions, while scheduling the beams in different time slots to guarantee the users demands |
| [132, 133] | Precoding design scheme | Distributed | Four different digital precoding techniques with a hybrid beamformer |
| [134, 135] | | Autonomous distributed | An actor-critic based DRL approach for the design of interference-aware analog-only beamforming/combining + a sample-efficient digital twin for environment simulation |

Table 4: List of Papers dealing with Inter-Cell Inter-Beam Interference Management Schemes

Through a learning phase based on trial and error behavior, each BS learns the optimal configuration to reduce the overall packet transmission delay while considering the asymmetric traffic demands and inter-cell interference. The major concern of the proposed method based on multi-agent Q-learning is the convergence performance, as at each learning step, multiple agents are adapting their switching point and thus making the environment non-stationary from a single femto cell's point of view, which is a well-known problem in multi-agents RL, especially for independent learners who don't share observations or cost function. Simulation results demonstrated that the proposed scheme achieves higher cell average packet throughput compared to fixed TDD and existing queue-aware dynamic TDD approaches, especially in the case of different UL and DL traffic in the different femto cells.

Similarly to the last study using RL algorithms, in [141], Tang et. al proposed a Deep RL-based dynamic UL/DL TDD configuration in a high mobility 5G HetNet, where channel conditions and network traffic demands are dynamically changing. The research first considered three user mobility types: high-speed vehicles, relatively slow-speed pedestrians, and unmanned aerial vehicles with ultra-high speed. Then the UL/DL configuration problem was formulated as a Markov Decision Process. The state of each agent, i.e., each BS, at a considered time slot, considered historical information and was defined as the discrete and normalized values of up/downlink occupancy, buffer occupancy, etc., during the last Nt time slots. Each BS has to choose, according to its local state, one TDD configuration from 6 possible choices that are employed in LTE-release 11 and beyond. After applying the chosen actions (TDD configuration) for all base stations, the network sends back an individual reward value to each agent which is the sum of up and down channel utilities depending on the channel occupancy of the next time slot. In the simulated learning phase, past channel conditions were known at each base station. The proposed DQN-based solution was compared to conventional dynamic TDD resource allocation taking into account only current information, and a tabular Q-learning-based solution. The simulation results proved that the proposed DQN solution improves the network performance in terms of throughput and packet loss rate. Using deep neural networks and historical information as a state's features permits the authors to model the complexity of a HetNet, predict future traffic patterns and thus intelligently change TDD configuration. While this DRL-based algorithm has significant advantages, it is important to note that in this multi-agent setting, each BS learns an individual policy to update its UL/DL configuration and does not explicitly cooperate with other BSs to reduce the cross-link interference.

In another UL/DL configuration approach to mitigate CLI, the authors of [142] proposed a novel centralized and distributed UL/DL reconfiguration method based on an efficient solution of a mixed integer linear program. Considering the multi-cell DL-to-UL and UL-to-DL interference levels and the traffic conditions, the suggested scheme aimed to maximize the overall network data rate. Simulation results indicated that the proposed solution performs better than the conventional static TDD system and Traffic-Only cell reconfiguration schemes. For example, with a system model composed of 4 cells, where each BS serves 10 users, it was shown that the proposed centralized (distributed) scheme improves the users' throughput by 4.65 (2.75) times, as compared to the conventional static TDD.

All of the previous studies aimed to mitigate cross-link interference as it decreases system performance. But what if cross-link interference could be used to enhance network capacity? In [143], it was argued that it is in fact possible. First, the CLI was analyzed and compared to conventional intra-mode interference, i.e., BS-to-UE and UE-to-BS interference. It was shown that UE-to-UE interference is on average smaller than BS-to-UE interference and thus the DL performance is better when the interference is originated from the UE instead of the BS. In contrast, BS-to-BS interference is stronger than UE-to-BS. Fortunately, BS-to-BS interference can be estimated and canceled. Based on these observations, the authors proposed a novel aligned reverse frame structure to avoid BS-to-UE interference and induce more UE-to-UE interference, along with an interference cancellation technique to suppress the inter-BS interference. As it is obviously more efficient to change the nearest BS interference to CLI, neighboring base stations construct a BS pair to share transmission data information and their uplink/downlink configuration. Then, each base station of the BS pair aligns its transmission to the reverse of the other BS, to maximize weak inter-UE interference instead of strong BS-to-UE interference. Using the channel estimation and exchanged data information, an interference cancellation was applied at the BS to cancel the strong DL-to-UL interference. Mathematical analysis and numerical results verified that the proposed approach outperforms conventional dynamic TDD performance in terms of uplink and downlink SINR, using 10dB of interference cancellation (IC). With 20dB of IC, the proposed scheme was as efficient as with perfect IC.

C) **Cell Clustering** consists in regrouping multiple cells in the same cluster based on a specific rule and applying the same uplink/downlink configuration for the group of cells within the cluster to avoid cross-link inter-cell interference [144, 145].

In [144], the authors proposed a centralized cell clustering scheme with an uplink/downlink reconfiguration to mitigate CLI in dynamic TDD HetNets where multiple pico and macro cells are using the same carrier frequency. To perform cell clustering, the mutual coupling loss (MCL) between eNBs was used, as it takes into account antenna gains and propagation loss and thus permits the measurement of the BS-to-BS interference level. If the MCL between two eNBs is below a certain threshold, then the base stations are grouped into a single cluster. Then, the study investigated the best uplink/downlink configuration based on different traffic situations for each cell cluster. In a single cell, to fully utilize the radio resource, an eNB would choose the DL/UL configuration corresponding to the amount of data in UE buffers on every TDD reconfiguration. This metric obviously cannot be used in a cluster with different BSs. In this regard, a new metric of the resource utilization rate of DL and UL was introduced. A center node collects the information concerning the statistical ratio of used RB among uplink and downlink transmissions for all cells belonging to the same group and then selects the optimal configuration that maximizes the total throughput while increasing the radio resource utilization. The new uplink/downlink configuration was then sent to all the cells in the cluster. The simulation results demonstrated that the proposed approach can mitigate BS-to-BS interference. However, excessive enlarging of cluster size results in low downlink performance gain in pico cells.

In [145], to mitigate CLI, the authors proposed two enhancements to existing rules for cell clustering, taking into account interference patterns and traffic load. The proposed metrics were used in a very simple heuristic algorithm and then compared to existing methods. To increase the flexibility of frame configuration, BSs with high traffic should be grouped in different clusters. Therefore, the first proposed rule is based on a differentiating metric with the inclusion of the total traffic of the BSs pair. The second proposed cell clustering metric attempts to track the number of cross-slots. The proposed clustering rules, especially the second one, provided significant improvement in terms of UL SINR, while not improving the DL performance, as expected. The heuristics used for clustering can be improved using meta-heuristic methods, but as they require some time to converge, they should be used for mid to long-term planning.

D) **Beamforming Schemes** consisting of adjusting the beam configuration in the transceiver are another possible approach to mitigate cross-link interference [146, 147, 148].

In [146], Y. Long et al. investigated the impact of an adaptive beamforming scheme with mMIMO in a dynamic TDD network with a realistic network model where transceiver noise at the BS and pilot contamination were taken into account. To derive a tight approximation of SINRs and UL and DL ergodic achievable rates, random matrix theory was used. Based on these approximations, the authors further analyzed the CLI and transceiver noise and proved that mMIMO effectively suppresses both of them. More specifically, according to [146], UL and DL tend to be decoupled under massive antenna arrays, and therefore there is a way to mitigate CLI without having to solve complex optimization problems. A joint transmit power control scheme of UL and DL to maximize overall throughput was also presented and validated by extensive simulation results.

In [147], a distributed pricing-based beamforming approach was proposed to adjust the beam direction and mitigate downlink-to-uplink interference in a dynamic TDD Multiple-Input-Single-Output pico cell network. The main objective was to enhance the uplink SINR by meticulously selecting the BS precoders, without boosting the power of UL users. To this end, the authors formulated an optimization problem that consists of penalizing the base station transmission by the interference on uplink users. To solve this optimization problem, a distributed beamforming algorithm, running in sequential or parallel mode, and associated signaling support, was proposed. Specifically, each BS identifies the appropriate precoder vectors that maximize an objective function that depends on the achieved rate and a periodically updated penalty factor that considers BS-to-BS interference. The beamforming vectors were therefore calculated to find the best equilibrium between transmitted power and induced interference. On one hand, in terms of algorithm implementation, the convergence of the parallel mode scheme was not proven. On the other hand, although the sequential implementation has been shown to converge, it resulted in higher signaling overhead than the parallel implementation. The proposed schemes were tested using a realistic system simulator and they have proven to be an effective way to improve UL performance at the cost of a slight reduction in downlink capacity.

While the previous study dealt with BS-to-BS interference, in [148] the authors investigated the UE-to-UE interference management using coordinated beamforming for MIMO Interfering Broadcast–Multiple Access Channel (IB-MAC) Interference Channel. In particular, they studied zero-forcing transmit beamforming design at initialization with and without water-filling, and the iterative weighted minimum mean-square error (WMMSE) algorithm to maximize the sum rate. They considered a two-cell network with one cell in DL and the second one in UL and supposed that BS-to-BS interference is already managed by exploiting a limited rank BS-to-BS channel and that ZF transmit filters are being used at the base stations to handle the intra-cell interference, thus only cross-link UL-to-DL (UE-to-UE) was being studied in this paper. The proposed solution consisted of constructing and using at the initialization the ZF beamformers at DL and UL UEs so that the UL-to-DL interference is canceled. Then, the WMMSE iterative algorithm was used. Exploiting the water-filling algorithm at initialization was also considered to improve the performance at low SNR.

E) **Interference Suppression Schemes** based on advanced receivers algorithms such as in [149, 150, 151] are also a viable way to mitigate CLI.

In [149], S. Guo et al. considered a multi-user MIMO small cells network operating in D-TDD mode where UL/DL configuration is performed based on the traffic buffer status of each cell and there is no intra-cell inter-user interference. They first provided analytic expressions of minimum mean square error interference rejection and cancellation (MMSE-IRC) receiver and enhanced MMSE-IRC (eMMSE-IRC) receiver with imperfect channel state information for gNB-to-gNB interference suppression. Then, to support the proposed interference suppression schemes, an interference measurement scheme based on the demodulation reference signal information of adjacent cells was also presented. The simulation results confirmed the UL user packet throughput gain of the two proposed interference suppression schemes compared with the baseline MMSE receiver. While the eMMSE-IRC receiver performs better than than MMSE-IRC receiver with imperfect CSI, the simulations showed that their UL SINR performance is near the ideal MMSE-IRC receiver with perfect CSI.

In [150], a high-performance and complexity-efficient DL-to-UL CLI suppression scheme was proposed. With a limited and 3GPP-compliant signaling overhead over the backhaul links, the BSs first exchange the user DL spatial signatures. The base stations then approximate the corresponding orthonormal projector subspace and spatially project the estimated IRC interference covariance onto the projector subspace before decoding. Their improved formulation of the IRC receiver allows it to further direct the user to the effective channel. Following the major assumption for URLLC, system-level simulations were performed and the proposed solution was compared against a few DTDD studies from industry and academia. It was shown that the enhanced IRC receiver offers about 199% gain of the URLLC outage latency performance and improves the ergodic capacity by 156% while requiring a modest inter-BS signaling overhead.

Starting from the observation that conventional channel modeling does not fit cross-link interference cancellation, Tan et.al in [151] first presented a more realistic BS-to-BS MIMO channel model taking into account the hardware-dependent non-linearity characteristics of radio frequency. Then, using this channel model, three types of digital-domain CLI cancellers- one polynomial canceller and two machine learning-based CLI cancellers- were proposed to mitigate BS-to-BS interference. Through analysis and simulations, it was demonstrated that the three cancellers can improve CLI cancellation by 43.4% compared with the traditional CSI-based interference canceller and that the hybrid canceller incorporating the linear and non-linear CLI components achieves the best CLI cancellation performance while keeping the computational complexity very low.

F) *Joint Optimization of Coordination Schemes* as in [8], where M. Ding et al. proposed a unified framework to investigate the technical issues of combining multiple interference mitigation methods in dynamic TDD small cells networks, as well as their possible introduction in HetNets. First, they presented a dynamic resource allocation and scheduling scheme to balance DL/UL traffic. They also investigated the effectiveness of multiple interference mitigation schemes (clustering, interference cancellation, UL power boosting) and their combinations. It was demonstrated that using a clustering scheme permits avoiding most of the inter-cell interference and that there was no important gain to combining a cell clustering scheme with interference cancellation. However, combining UL power control and cell clustering is an efficient strategy to improve the system performance, especially for low-complexity implementations. Furthermore, the authors proposed two new partial interference cancellation schemes, namely the BS-oriented partial IC, where only interference from adjacent BS is canceled, and the UE-oriented partial IC, where only cell edge users are allowed to use IC. Simulation results showed that the BS-oriented partial IC scheme achieves similar performance to the complete IC scheme with lower complexity. Furthermore, the uplink performance of cell center users was also impacted by the DL-to-UL interference, therefore applying partial interference cancellation for cell edge users was not enough, and using BS-oriented partial IC was proven to be more effective. Finally, the joint optimization of cell bias and ABS in dynamic TDD HetNets was studied. Scheduling policies and an algorithm to derive the ABS ratio were proposed.

**Discussion:** Table 5 provides a comparison between all previously listed papers dealing with cross-link interference management. Several approaches have been studied in the literature. Power control for DL or/and UL usually depends on some interference threshold and is done via fixed values, which is obviously sub-optimal and highly scenario-dependent. In some other studies, it was proposed to cluster the cells based on specifically designed rules to apply the same TDD configuration among the BSs in the same group. Another CLI alleviation approach is to adjust the wireless signal transmission strategies via interference cancellation and adaptive beamforming schemes. Many studies also investigated the dynamic configuration of UL/DL to mitigate CLI, using reinforcement learning algorithms. Such approaches permit taking into account the users' mobility and the actual traffic demands per cell to learn how to change dynamically and intelligently the TDD configuration. However, in the existing literature, most UL/DL configurations are cell-centric, i.e., based on the average UL/DL traffic within that cell, without further differentiation of resource demands of the specific UEs. Moreover, when adopting multi-agent settings to distribute the computation, one should pay attention to the coordination and cooperation among the different agents and the convergence of the approach. This can be done via a shared reward between the agents and by centralizing the training of the DRL algorithm.

## 8. Remote Interference Management

To our knowledge, very few scientific papers have dealt specifically with remote interference management. In this section, we overview and discuss the most important research contributions. We classify the studies into two categories : (i) remote interference prediction/detection [152, 153, 154, 155, 58, 156, 157] and (ii) remote interference mitigation [112, 158, 159], and we review and compare some of the proposed techniques in the following.

A) *Remote Interference Prediction/Detection*

At present, there are two principal methods for detecting and estimating the atmospheric duct: (i) theoretical calculations using Ray-optics and Parabolic Equation (PE), and (ii) practical measurements using radar [58]. Because of the abruptness of the atmospheric duct and the spatial distance between interfering BSs, performing time-consuming, multi-regional channel measurements is undoubtedly a tricky issue. Instead of pre-building mathematical models, machine learning and specifically, supervised learning techniques provide an efficient solution to accurately predict remote interference.

| Ref | Proposed Scheme | Control Level | Main Idea | Limits |
|---|---|---|---|---|
| [137] | Power Control | Distributed Coordinated | Closed-loop UL PC based on exchanging UL configuration between closest BSs and computing interference level indicator in each BS | Interference is only estimated based on UL configuration |
| [138] | | Autonomous distributed | Defining a transmission power scaling factor for interfering micro operator after comparing maximum BS-to-BS interference to a defined threshold | The appropriate definition of the interference threshold is highly scenario-dependent |
| [139] | | Distributed coordinated | DL/UL transmission power control applied to flexible subframes with constraints on the UE power consumption and DL throughput | The power adjustment is done by means of fixed values |
| [140] | UL/DL Dynamic Configuration | Autonomous distributed | Multi-agent Q-learning based approach to dynamically change the switching point of each femto cell to minimize transmission time while considering the asymmetric traffic demands and inter-cell interference | The convergence is not guaranteed |
| [141] | | Autonomous distributed | Multi-agent DQN based solution to choose, according to historical local information, one TDD configuration from 6 possible choices | No explicit cooperation between BSs to reduce the CLI |
| [142] | | Centralized / Distributed | An efficient solution of a mixed integer linear program considering DL-to-UL and UL-to-DL interference levels and the traffic conditions | Problem formulation at the expense of additional variables and constraints |
| [143] | | Distributed coordinated | Forming BS pairs using a novel aligned reverse frame structure to avoid BS-to-UE interference and induce more UE-to-UE interference | The need of data exchange and interference cancellation at the BS |
| [144] | Cell Clustering | Centralized | Clustering based on mutual coupling loss between eNBs compared to a threshold + DL/UL configuration based on a new defined metric of cluster resource utilization rate | Accurate threshold design is critical as excessive enlarging of cluster size results in low DL performance gain |
| [145] | | Centralized | Two enhancements to existing rules for cell clustering, taking into account interference patterns and traffic load | DL performance not improved |
| [146] | Beamforming | Distributed | The use of random matrix theory to derive a tight approximation of SINRs and UL and DL ergodic achievable rates | - |
| [147] | | Distributed | A pricing-based beamforming approach, running in sequential or parallel mode, to enhance the uplink SINR | The convergence of the parallel mode not proven + higher signaling overhead in sequential mode |
| [148] | | Distributed coordinated | ZF transmit beamforming design at initialization with and without water-filling + WMMSE algorithm to maximize the sum rate | CLI between BSs not considered |
| [149] | Interference Suppression | Distributed | MMSE-IRC and eMMSE-IRC receivers with imperfect CSI for gNB-to-gNB interference suppression + interference measurement scheme based on the DMRS information of adjacent cells | - |
| [150] | | Distributed coordinated | Enhanced IRC receiver based on the projection of the estimated IRC interference covariance onto the projector sub-space prior to decoding | Some signaling overhead is required |
| [151] | | Distributed | Three types of digital-domain CLI cancellers : one polynomial canceller and two machine learning-based CLI cancellers | - |
| [8] | Joint Optimization of Coordination Schemes | - | A unified framework to investigate the technical issues of combining multiple interference mitigation methods | - |

Table 5: List of Papers dealing with Cross-Link Interference Management Schemes

In [152], a remote interference prediction model based on a feed-forward neural network (FNN) was proposed. The database was generated using Petool, a widely acknowledged simulation platform to acquire channel data. Taking into account signal frequency, top and bottom height of duct, antenna height, down tilt angle, vertical height, and horizontal distance as input, the platform computed the pathloss. By directly learning from the generated data on both the Sub-6 GHz band and 28 GHz band, the FNN-based framework achieved a prediction accuracy of 88.72% and 94.77% on the two bands respectively.

Deep learning models were also applied in [153] to properly model the over-the-horizon propagation of intercity link signals induced by atmospheric duct and tropospheric turbulence effects, taking into account real terrain and land cover. Using 1300 sets of random terrain and landforms, the authors proposed Deep MultiLayer Perceptron (DMLP) and Long-and-Short Term Memory (LSTM) network models to predict the pathloss distribution at the antenna height layer. The two models used fixed radar antenna parameters and atmospheric environment parameters and took as input topography and landform data. The prediction accuracy of the proposed models remained mainly between 0.8 and 1, however making large to moderate errors after 80 Kms using DMLP and LSTM models respectively. While the model training results had certain errors mainly because of the randomness of the generated data, they demonstrated that topography and land cover types highly impact the propagation when a terrestrial duct happens.

While in the previous studies, the data was generated using a simulator, in [154], the authors analyzed the characteristics of remote interference, called atmospheric duct interference, using real data from a TDD LTE network owned by China Mobile. According to [154], when RI occurred in Xuzhou, Jiangsu Province, about 27.6% of cells experienced UL interference higher than -90 dBm, which totally overwhelmed the useful signal and blocked the cell's communication. Based on their analysis and considering the huge amount of input data, the authors proposed an SVM (Support Vector Machine)-classifier, which is a centralized machine-learning algorithm, to forecast if the base stations in the testing set will be affected by the remote interference. They further proposed an implementation of the Alternating Direction Methods of Multipliers (ADMM) framework [160], to perform a distributed SVM prediction scheme that reduces data exchange between different areas. The two prediction-based schemes were compared with the KNN (K-Nearest Neighbor) algorithm where a BS is classified depending on its nearest neighboring BS, whether it is predicted to be affected by RI or not. The simulation results showed that when the number of training samples is 40 000, the precision rate of the two proposed methods can reach 72%, which surpasses the conventional KNN algorithm. With its distributed implementation, ADMM-SVM prediction is thus a more suitable and practical solution for a successful prediction of the remote interference occurrence. An extension of the previous work can be found in [155]. The authors proposed another machine learning algorithm called Random Forest that has a shorter learning time than SVM algorithms, while maintaining similar prediction precision, and is therefore more convenient for operators.

In [58], 5,520,000 TDD network-side data, including the longitude, latitude, time, antenna height, and down tilt angle, were collected by real sensors from 240,000 antennas in Jiangsu Province of China and then used to validate the RI detection accuracy of nine Artificial Intelligence (AI) algorithms. Specifically, the authors proposed a remote interference detection testbed consisting of four modules, including a meteorology and signal module, a data processing module, an AI-based learning module, and a validation module. They identified the most impacting factors of atmospheric duct affecting signal pathloss as follows: temperature, atmospheric pressure, relative humidity, time, longitude, latitude, antenna height, and down tilt angle. After collecting and processing the data, the AI-based learning module is used to train and generate multiple feature models including single model algorithms (kNN, SVM, and Naive Bayes algorithm), and ensemble algorithms (Random Forest, Bootstrap Aggregating, Boosting, and Stacked Generalization), to discriminate the RI of the BS. The selected algorithms were tested and compared against each other on various aspects including accuracy and recall, sensitivity to the data size and interfering ratio, robustness, time complexity, etc. Numerical results demonstrated that the ensemble algorithm can achieve an average accuracy that is 12% higher than that of the single model algorithm.

While previous reviewed studies proposed remote interference prediction/detection schemes based on essentially numerical data, in [156], a reliable remote interference detection model was proposed based on the image recognition ability of a deep convolutional neural network and a high number of PRB interference statistics at the cell level. The RI identification accuracy of the proposed model reached 99%. When the remote interference is detected, an interference optimization algorithm based on daily experience is initiated: users are then associated with the FDD network on the common cell or to a new cell that has not been affected by the atmospheric duct. An automatic optimization algorithm would improve the network performance. Since April 2019, the proposed detection model has been used to identify and optimize remote interference management in Hebei province in China. After five months of adjustment, RI occurrence has dropped from 35% to 7%.

In another approach to efficiently detect RI caused by multiple cell aggressors, two reference signal designs, denoted as 1OS (OFDM Symbol) and 2OS based RIM-RS for RIM were described along with the detection processing framework in the receiver side in [157]. The considered RIM-RS solutions were thoroughly investigated in multiple realistic interference scenarios. Results indicated that the proposed 2OS RIM-RS design achieves modestly better detection performance than the 1OS RIM-RS design in simple configurations involving a low number of interfering BSs. However, when the number of interfering BSs grows, the 1OS design tends to perform better than the 2OS-based design.

B) *Remote Interference Mitigation*

While an accurate prediction/detection of remote interference is of great importance, efficient remote interference management schemes are required to enhance the mobile network suffering from the RI.

In [112], an overview of the atmospheric duct and its effects on radio propagation was presented, along with the 3GPP proposed enhancements to mitigate resulting remote interference. Authors, also classified interference mitigation schemes into four categories:

- Time domain schemes: can be either (i) semi-static where a longer guard period can be configured, or (ii) dynamic by muting the aggressor on DL resources causing remote interference. Currently, time domain solutions are the most used in TD-LTE networks by re-configuring the guard period from 3 to 9 OFDM symbols. This permits extending the protected area from RI from about 64 km to 192 km [161], however, it decreases the DL capacity by 15.7%.
- Frequency domain schemes: both the aggressor and victim BSs can be configured with orthogonal frequency resources in a semi-static way at the expense of the spectral efficiency loss. As a dynamic approach, the victim may stop allocating the frequency resources suffering from high RI.
- Spatial domain schemes: careful network planning, for example by installing antennas at lower altitudes, or adjusting the down-tilt can avoid RI for example. It is also possible to apply beam nulling or beam selection.
- Power domain schemes: by applying power control at the victim or the aggressor.

Besides, according to [112], the polarization domain can also be exploited to attenuate the co-channel RI [162]. To minimize the performance loss due to RI, F. Liu et al. suggested that more effective and intelligent RI mitigation solutions using machine learning should be explored.

The studies in [158] and [159] proposed two spatial domain schemes to manage remote interference. In [158], a complete traceable RIM simulation scheme using meteorology measurement data in terrestrial and maritime systems and communication settings including antenna parameters, was provided. The simulation system combining parabolic equation methods computing pathloss and refractive index estimation could be useful for network planning and remote interference mitigation. With different settings in the testbed, the authors computed path loss in various atmospheric conditions. The findings showed that the adjustment of the antenna tilt and height may minimize the RI. For instance, an increase of 1 degree on antenna tilt can reduce the remote interference by 2 or 3 dB. According to [158], the adjustment of antenna tilt has been used by China Mobile to manage RI, but there is no theoretical analysis to support this technique.

In [159], based on the spatial domain of MIMO networks, the authors applied the existing interference alignment scheme to the RI problem and studied its limitations, as it does not handle the performance degeneration of the aggressor cell. Then, considering the total network performance including the DL performance of the aggressor cell, they proposed a precoding scheme to steer the RI signal toward the null space of the UL channel of the victim cell, instead of aligning the RI signals. The authors modeled the atmospheric duct channel as a wireless channel representing small-scale fading and pathloss and further assumed perfect CSI at the aggressor base stations. It was shown via numerical results that the proposed null space alignment method outperforms the standard RI alignment as it further improves the network performance.

**Discussion:** Table 6 provides a comparison between all previously listed papers dealing with remote interference management. As previously stated, there is a lack of research studies dealing with RI. Some papers investigated the use of supervised learning along with deep learning algorithms to detect atmospheric ducts causing remote interference using synthetic data sets, or real network side data collected by real sensors owned by Chinese mobile telecommunications. Besides numerical data, some studies based their detection model on the image recognition capability of deep convolutional neural networks. Despite the importance of these studies, it is crucial to investigate proactive schemes to predict the occurrence of remote interference and the interfering and victim BSs, to allow for interference avoidance. To mitigate such interference, time-domain solutions, especially guard periods are the most used, despite their negative impact on DL capacity. Furthermore, few papers studied remote interference mitigation through spatial approaches. However, the proposed schemes either lack theoretical analysis, are static or require full CSI knowledge. Therefore, it is important to design more efficient and intelligent schemes, with theoretical support and imperfect CSI. This can be done by the support of machine learning algorithms such as deep reinforcement learning along with digital twin architecture to model the environment and atmospheric duct.

## 9. Self Interference Management

A diverse range of SI management schemes have been proposed to handle self interference by the research community from the simplest isolation to the complex adaptive cancellation. In this section, we will first identify the challenges of SI management scheme design, then we will classify and discuss the main proposed approaches in 5G FDD and IBFD.

First of all, one can ask, if the transceiver knows the transmitted, then why is self interference difficult to subtract? In fact, due to unknown noise and linear and non-linear distortions in the radio TX chain, the transmitted signal looks quite different from the original digital baseband signal after its conversion to analog and up-conversion to the right carrier frequency. Therefore, it is required to account for all the analog distortions and compensate for them at the receiver.

| REF | PROPOSED SCHEME | MAIN IDEA | LIMITS |
| --- | --- | --- | --- |
| [152] | Remote Interference Prediction/Detection | FNN-based model for pathloss prediction taking into account distance and duct and antenna parameters | A simple model using synthetic data |
| [153] | | Deep learning models (DMLP and LSTM) to predict pathloss distribution at the antenna height layer considering real terrain and land cover | the model training results makes large to moderate errors after 80 Kms |
| [154] | | Centralized and distributed SVM model to forecast RI using real network data | Prediction precision is not very high |
| [155] | | An extension of previous paper using Random Forest model | Same prediction accuracy as previous study |
| [58] | | A remote interference detection testbed to validate the RI detection accuracy of nine AI algorithms | - |
| [156] | | A reliable RI detection model based on the image recognition capability of deep convolutional neural network and a large number of PRB interference statistics on the cell level | - |
| [157] | | The design of reference signal along with the detection processing framework in the receiver side | - |
| [112] | Remote Interference Mitigation | An overview of the atmospheric duct and its effects on radio propagation + 3GPP proposed enhancements to mitigate RI + interference mitigation schemes classification | - |
| [158] | | A complete traceable RIM simulation scheme + adjusting the antenna tilt and height | No theoretical analysis to support the applied mitigation scheme |
| [159] | | A precoding scheme to steer the RI signal toward the null space of the uplink channel of the victim cell | - |

Table 6: List of Papers dealing with Remote Interference Management Schemes

Besides the non-linear self interference, there is another effect that must be dealt with in the transceiver design, which is the receiver saturation. This could happen when the input signal is beyond the analog-to-digital conversion resolution. Thus, any cancellation architecture must prevent receiver saturation by sufficiently canceling the interfering signal in analog at RF before it hits the low-noise amplifier, and fully cancel the remaining residual self interference in digital by modeling all of the nonlinear distortions [60].

Considering the requirements of SI isolation and cancellation, it is quite challenging to enable the FD scheme, especially at macro BS with a large antenna number and high transmission power. Moreover, because of the wide bandwidth of 5G systems, a broadband pre-equalization in the RF domain is needed to cancel the self interference over these wide bands [163].

Other than base stations and other RAN node types, it is also challenging to handle the self interference in user devices, considering different aspects: On one hand, the UE has tighter implementation requirements in terms of complexity, cost, power consumption, and size limitation. On the other hand, due to the users' mobility and changing environment conditions, the self interference signal has a time-variant characteristic [163].

As stated previously, a self interference signal comprises a deterministic part, i.e., an amplified analog signal of digital transmit samples, and a non-deterministic part, i.e., noise and several non-linear harmonic components [163]. The latter portion requires special attention to be managed either through passive suppression or active cancellation. In practice, different techniques must be combined to cancel the self interference to the desired power level. In the following, we provide an overview of passive and active approaches to manage self interference.

The passive suppression schemes aim to attenuate the self interference using antenna directionality, cross-polarization, spatial isolation or radio frequency absorber, etc. [164]. These passive schemes are simple to realize, but they are unable to retune and mimic self interference channel variation [163]. Thus, the isolation performance is generally relatively poor, especially when the signal is constantly changing due to the dynamics of the radio channel. [61]. Therefore, in FD and in FDD with CA variants, simply relying on passive interference cancellation may not be sufficient for today's transceivers [20]. To effectively cancel SI, it is key to understand the self interference channel behavior and to employ active/adaptive cancellation approaches in addition to the passive suppression techniques [163].

The basic idea of active cancellation schemes is to produce a copy of the self interference signal using some special techniques and to subtract it from the received signal [61, 164]. According to the operation domain, active cancellation schemes can be categorized as analog, digital, or hybrid analog-digital cancellation approaches. In theory, SI can be canceled in the digital domain with low complexity, since its performance solely depends on the accuracy of baseband SI channel estimation and the modeling of non-linearity of the transceiver [165]. However, due to hardware imperfections and to avoid the receiver chain saturation, SI has to be suppressed at most in the analog domain. In the following, we further discuss the different categories of active cancellation approaches and we review some of the proposed techniques in the literature.

A) *Digital domain schemes:* implemented after the analog-to-digital converter, are one of the most straightforward of all active cancellation methods, operating exclusively in the baseband and thus benefiting from all the advantages of all-digital hardware, including simplified design and verification [20, 61]. However, compared to analog domain schemes, digital domain approaches suffer from inferior cancellation performance and reliability [20] and thus are generally used to remove residual self interference after performing analog cancellation. In the digital domain schemes, an equivalent discrete-time coupling channel is computed to reconstruct a digital SI signal which is then subtracted from the received signal in the digital domain [164, 61].

For example, in [166], an auxiliary receive chain was employed to get a digital-domain copy of the transmitted RF SI signal, including any transmitter impairments. In addition, to mitigate the effect of receiver phase noise, a common oscillator was shared between the auxiliary and ordinary receive chains, and a non-linearity estimation and suppression method was suggested to attenuate the effects of receiver non-linearity. Based on the signal configuration of an LTE full-duplex SISO system, the performance of the proposed approach was investigated numerically, and it was revealed that the SI could be attenuated to ~3 dB above the receiver noise floor, resulting in a 76% rate enhancement over conventional half-duplex systems at 20 dBm transmit power values.

Aiming to introduce an FD cellular system to 6G, a novel digital SI cancellation using a 5G-based OFDM signal was proposed in [167]. Specifically, the demodulation reference signal was reorganized to avoid UL-DMRS and DL-DMRS interference and enhance the estimation accuracy of the self interference signal and the noise power. The proposed SI digital cancellation was evaluated and proved its efficiency in terms of achieved BLER compared to the half-duplex system. However, as the signal actually transmitted is based on TDD, there are channel extrapolation components in the forward and backward slots that may compromise the channel estimation accuracy and negatively affect the block error rate performance. Therefore, the same authors further extended the study in [164], by proposing a channel extrapolation scheme and by developing a 5G PHY prototype implementing the proposed method in a software-defined radio platform. The proposed approach was validated via both computer simulations and experimental evaluation using the prototype.

For more details on digital self interference cancellation schemes, we refer the interested reader to the survey in [168], which presents an overview of all the estimation techniques used in digital cancellation such as least square estimation and digital signal processing-based self interference estimation, and summarizes the advancements to date of digital SIC approaches, including machine learning and deep learning based schemes for digital self interference cancellation.

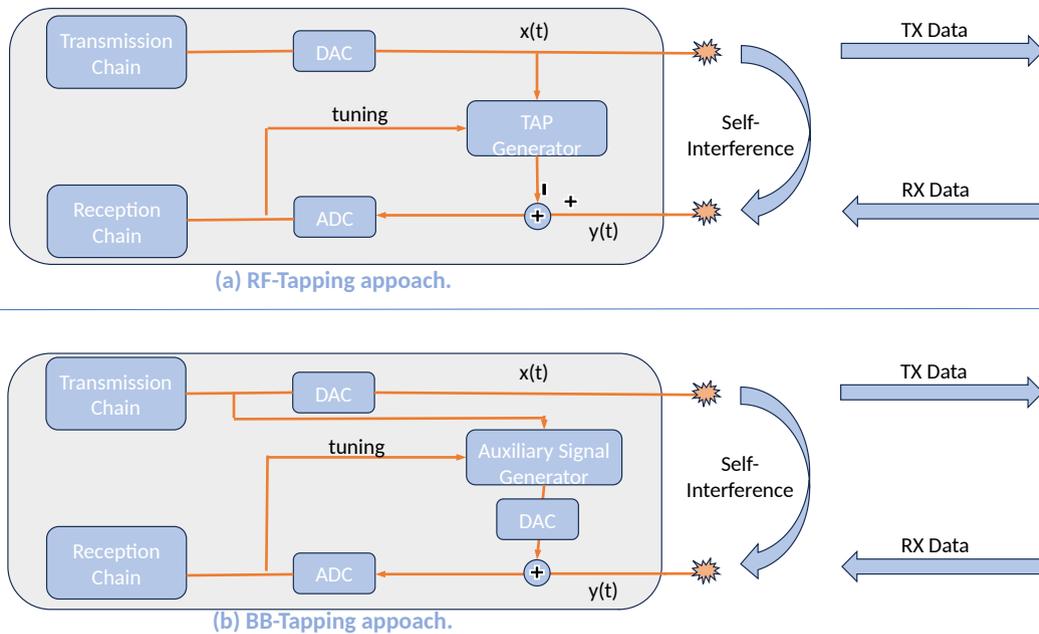

Figure 10: Analog Cancellation: RF-Tapping and Baseband-Tapping.

B) *Analog domain schemes:* based on performing SI cancellation before entering the analog-to-digital converter, are the most popular approaches for self interference cancellation. Loopback self interference signal is obtained by tapping the signal using the analog circuits and fed to the receiver as a reference signal. The CSI are then acquired at the receiver, using training-based SI cancellation approaches to estimate the loopback channel, or non-training-based SI cancellation approaches, which are more bandwidth efficient as they do not require any training phase [169]. After CSI acquisition, the reference signal is passed to a reconfigurable filter emulating the self interference channel at the required bandwidth and then fed to the RX input to cancel the self interference.

Depending on where the Tx reference signal is tapped from, the analog cancellation can be classified as RF-tapping and baseband-tapping [20, 165], as shown in Figure 10.

- The RF tapping approaches are based on electronic circuits that use several adjustable RF components to copy the TX signal and tune the delay and attenuation of the cancellation signal. These RF components can be classified into single-tap and multi-tap architectures. As shown in Figure 10 (a), the Tap Generator module adapts the attenuation, phase, and delay of each tap. While single-tap architectures are easier to implement, multi-tap architectures achieve a higher cancellation performance in multi-path environments in a wide bandwidth [20, 170].
- The BB-tapping approaches use the BB TX signal to generate a digital copy of the SI signal as shown in Figure 10 (b) via the Auxiliary Signal Generator module. After the tuning phase, which is performed in the digital domain, the digital copy is transformed into the analog domain and injected into the receiver chain to eliminate self interference [170]. This analog BB-tapping scheme is sometimes considered as a mixed signal interference suppression or a hybrid approach in the literature [20, 61].

Given the design of the necessary reconfigurable filter adjusting the canceling signal, two approaches to SI analog cancellation are possible: either time-domain cancellation (using RF or BB tapping), or frequency-domain cancellation (using RF tapping) [171]. While time-domain methods implement finite impulse response (FIR) filters with parallel delay line taps, frequency domain approaches employ parallel RF bandpass filter taps emulating the SI channel at different frequency points [172]. When the self interference channel exhibits a very large delay, frequency domain cancellers are preferred over time domain cancellers due to the elimination of delay elements in each tap, simplifying integrated circuit designs [169]. However, It was demonstrated that due to hardware losses limiting the number of possible taps, conventional electrical components are not the best choice for analog self interference cancellation [173].

Recently, optical-domain RF cancellers using optical components allowing much smaller insertion losses and more accurate delays and frequency operations, have been proposed and tested in the literature [174, 173]. Unlike conventional RF cancellers based on electrical components, the signal is processed in the optical domain, requiring the input RF signal coupled from the output of the transmitter power amplifier to be up-converted into the optical domain. Because of the wideband and tunable properties of optics, these novel approaches have shown enhanced self interference cancellation capacities over much wider bandwidths and operating frequency ranges than their RF and digital cancellation counterparts [174]. For example, in [173], to meet the requirements of 5G and beyond networks, an optical-domain active analog self interference cancellation scheme to construct sufficient effective taps within wide-band and under various hardware conditions, was investigated. The authors designed a novel tuning algorithm that reduces the usage of optical carriers saving implementation costs. With a 12-bit ADC and 20 effective taps, simulation results indicated that the self interference was efficiently mitigated in a 3GPP LTE specifications compatible system within 1GHz of operational bandwidth.

C) *Analog-Digital (Hybrid) Domain schemes:* are combining the advantages of both approaches, i.e., the high performance and reliability of analog domain schemes, and the flexibility and lower complexity of the usage of digital components [20]. Such approaches have a good cancellation performance in non-linear systems and over a wide modulated bandwidth.

While in the majority of previous studies, the focus was directed to implementing self interference cancellers in the FR1 band within limited operational, bandwidth, in [175], a hybrid SI cancellation scheme in IBFD enabled private 5G network using large-scale antenna arrays with RF beamforming, operating in FR2 band with a wider bandwidth and supporting a mixture of multi-type devices, was proposed. Particularly, the authors introduced a cost-friendly efficient multi-tap RF cancellation scheme and a frequency-domain-based digital canceller to deal with the residual SI with a reduced processing latency. They further proposed a game theoretic user allocation algorithm to mitigate the cross-link interference between UL and DL users. It was demonstrated that the proposed solution improves the spectral efficiency by 92% compared to the HD scheme, and due to the abundant resources in the FR2 band, it achieves multi-Gbps peak data rates, high reliability, and massive connectivity.

The wide bandwidth in 5G and beyond networks, requires adjusting multiple weights in a multi-tap canceller. Real-time adjustment of these weights, considered one of the remaining problems in self interference cancellation, was tackled in [176]. The authors proposed a new hybrid-analog-digital interference cancellation scheme consisting of a multi-tap and multi-stage RF and analog canceller that independently adapts to a time-varying channel in real-time via a novel on-chip hybrid-analog-digital adaptation loop. The proposed scheme supporting both FDD and IBFD operations achieves 48 and 58 dB SI cancellation across a 100 MHz BW in both modes, respectively.

**Discussion:** Table 7 provides a comparison between all previously listed papers dealing with self interference management. State-of-the-art radio frequency transceivers for mobile networks suffer from SI in FDD and IBFD operation, due to TX-RX leakage or simultaneous transmission/reception and non-idealities in the analog front-end. Different SI cancellation schemes have been proposed in the literature, from simple isolation to active analog, digital, and hybrid schemes, in frequency, time, and optical domains. Generally, there is no single approach that is capable of completely mitigating the SI and a combination of multiple methods is often required.

## 10. Intra-Cell Multi-User Interference Management

In this section, we will provide a detailed overview of intra-cell multi-user interference management. Such interference is caused by the multiple access in 5G and beyond networks. We will divide this section into three subsections dealing with the management of intra-beam interference in NOMA networks, inter-beam interference in multi-beam transmitters mMIMO networks, and inter and intra-beam interference in NOMA multi-users mMIMO networks.

### 10.1. Intra-Beam Interference Management

In contrast to orthogonal multiple access techniques, NOMA supports more UEs than the number of orthogonal resource slots using non-orthogonal resource allocation and thus improves the spectral efficiency and cell-edge users' throughput. However, serving multiple neighboring users under the same (narrow or wide) beam (in DL) and using the same non-orthogonal resources (in DL or UL), can lead to severe intra-beam (or intra-cluster) multi-user interference, and therefore advanced inter-user interference cancellation techniques are required. Generally speaking, NOMA techniques managing such interference can be categorized into power-domain and code-domain NOMA.

Several research studies have actively investigated multi-user interference management in NOMA-based networks. We briefly overview the basic concepts and we refer the reader to two comprehensive surveys in [177, 43]. Various NOMA schemes [178, 43] have been introduced and compared in terms of their operating principles, performance, and implementation challenges. In [177], a comprehensive survey of power-domain NOMA exploiting the channel gain difference between multiplexed users is presented. In power-domain NOMA, the signals from several users are superposed and the resulting signal is then sent using the same time/frequency resource. On the receiver side, multi-user detection (MUD) techniques, such as successive interference cancellation, parallel interference cancellation, and hybrid interference cancellation, are applied to detect and decode the desired signals.

It is worth noting that in UL NOMA [179], different users may independently use their maximum available power to transmit their signals using the same time-frequency resource, as long as their channel gains are sufficiently distinct [180]. The BS receives the data from all UEs simultaneously and has to apply some multi-user detection and interference cancellation algorithms to mitigate the interference. Generally, in this case, the intra-cell multi-user intra-cluster interference depends on the channel statistics of all users in the same cluster, i.e., using the same radio resource, and the transmission of weak channel users is typically more affected by strong interference arising from strong channel UE transmissions. In [181], the maximum completion time minimization for users in multi-carrier NOMA is investigated for networks and applications that have stringent latency requirements.

| REF | PROPOSED SCHEME | MAIN IDEA |
|---|---|---|
| [166] | Digital domain schemes | The use of auxiliary receive chain for SI copy generation and a common oscillator and non-linear estimation and suppression method in an LTE full-duplex SISO system |
| [167, 164] | Digital domain schemes | The reorganization of DMRS to prevent interference and enhance the SI estimation using novel digital SI cancellation scheme + a channel extrapolation and 5G PHY prototype |
| [168] | Digital domain schemes | An overview of the estimation techniques used in digital cancellation and the advancements to date of digital SIC approaches |
| [173] | Analog domain schemes | An optical-domain RF canceller using a novel tuning algorithm to construct multiple taps within wide-band and under different hardware conditions |
| [175] | Hybrid domain schemes | A cost-friendly efficient multi-tap RF cancellation scheme and a frequency-domain based digital canceller to deal with the residual SI with a reduced processing latency in 5G IBFD network operating in FR2 band |
| [176] | Hybrid domain schemes | an hybrid SI cancellation scheme consisting of a multi-tap and multi-stage RF and analog canceller adapting to a time-varying channel in real time via a novel on-chip hybrid-analog-digital adaptation loop |

Table 7: List of Papers dealing with Self Interference Management Schemes

On the other hand, in DL NOMA [182], different transmit power is allocated to intra-beam (intra-cluster) users, such as weak channel users are allocated high transmission power and thus experience low interference from strong channel users (and vice versa). To mitigate such DL interference, the receiver of every UE has to implement advanced multi-user detection and interference cancellation algorithms, which represents a cumbersome task due to its limited processing capability. [183] derived complexity reduced optimization methods for the joint power and subcarrier allocation in multi-cell multi-carrier NOMA systems. In [184], power minimization for zero-forcing based multi-cell MISO-NOMA systems is addressed with efficient joint user grouping, beamforming and power control scheme.

Generally speaking, handling multi-user interference to maximize the benefits of NOMA networks, requires :

A) *Appropriate user pairing/clustering*: which consists in dividing the users into multiple pairs or clusters using some criterion, where intra-cluster users share the same transmit source. Generally, user pairing refers to the special case of user clustering where each cluster is composed of only two users. The performance of NOMA highly depends on finding the best user pairing/clustering which is an NP-hard problem [185]. To simplify the exhaustive search over all possible matching pairs/clusters, some studies suggest some other suboptimal pairing algorithms, such as decision tree pruning, hill climbing, greedy search or matching algorithm, etc. Some of these algorithms still have high computational complexity or must be jointly designed with adaptive per-user power allocation. We overview some of the proposed approaches in the following.

In [186], two user pairing methods based on neighbor search algorithms inspired by hill climbing and simulated annealing were proposed for downlink NOMA networks. The difference between the two proposed user pairing algorithms is whether a worse solution can be tolerated to explore more possibilities in the search process [187]. Simulations showed that when combined with power allocation techniques, both could yield near-optimal sum throughput while reducing complexity relative to exhaustive search.

Most of the existing works like [186] discuss the user clustering algorithms considering two users in each cluster, i.e., user pairing, and limited analysis has been performed towards user clustering for a generalized number of users in each cluster. Furthermore, the majority of works considered user pairing/clustering under a perfect detection and interference cancellation scenario at the receiver. Therefore, to tackle these two challenges, [188] first derived bounds on power allocation and channel coefficients for user clustering under imperfect interference cancellation. Then, based on these derived bounds, the authors proposed multi-user clustering and adaptive multi-user clustering algorithms for NOMA networks. The first method considered the minimum SINR difference criterion to group the users with higher channel gain with those with lower gain values. The second algorithm is a combination of the first approach with an adaptation step to adjust the clusters by splitting, regrouping, or transmitting by orthogonal multiple access schemes. It was shown that the proposed clustering schemes improve the achievable user data rate compared to OMA and that the network performance saturates with an increase in the number of users per cluster, but does not decrease, unlike conventional user-pairing algorithms. [189] further extended the multiple possibilities of user arrangements for the proposed adaptive multi-user clustering in the previous study.

To date, the majority of clustering/pairing schemes presented in the literature do not address a fundamental problem, namely the time-varying nature of the number of users and the channel conditions. To address these characteristics of dynamic communication systems and to improve the total data rate of a NOMA DL, an adaptive and evolutionary clustering strategy inspired by a modification of the DenStream evolutionary algorithm has been proposed in [190]. The proposed method allows the online creation of new clusters and their adaptation to network temporal variations and robustness to noise. Considering a well-known power allocation scheme referred to as improved fractional strategy power allocation, the proposed clustering approach was evaluated and shown to be effective in following the network dynamics, grouping all users into clusters, and allowing for a uniform transmission rate between clusters.

B) *Designing perfect superposition coding (SC):* along with the users pairing/clustering, to be applied on the transmitter side, in a disjoint or joint manner. With SC, two or more messages are first encoded into one signal in two layers, then sent at the same time using the same resource to multiple receivers. In power-domain NOMA, UEs are grouped into several clusters and then are superposition-coded with different transmit power in every cluster. To ensure user fairness, NOMA assigns more power to UEs having lower channel gains. With this approach, multiple users are served simultaneously without degrading the quality and user throughput. We refer the interested reader to this survey [191], for more details concerning the power allocation in NOMA, along with user pairing algorithms in the literature.

C) *Error-free multi-user detection algorithms:* at the receiver side, such as successive interference cancellation are used to detect and decode the superposed information to extract the desired signals. The main idea of SIC is to order the users according to the signal strength difference and to decode the users' signals successively starting with the stronger signal which can then be removed from the combined signal to isolate the weaker signal from the residue [177]. We refer the interested reader to this comprehensive survey [78] over-viewing the various successive interference cancellation techniques with perfect and imperfect CSI for different NOMA network configurations.

Apart from conventional interference management strategies, i.e., treating interference completely as noise, as commonly used in OMA/SDMA networks, and completely decoding interference, as in NOMA, Rate-Splitting Multiple

Access (RSMA) is emerging as a novel, generic, and efficient framework for designing and optimizing non-orthogonal transmission, multiple access and interference management approaches in 5G and beyond networks [45]. Essentially, RSMA partially decodes interference and treats the residual interference as noise. Based on the rate splitting (RS) concept and linear precoding for multi-antenna multi-user transmissions, RSMA breaks the data messages into common and private portions and encodes the common portions into one or more common streams, whereas the private portions are encoded into distinct streams using either perfect or imperfect CSI. The superimposed streams are then transmitted to the receiver, which decodes the common streams, performs SIC and then decodes its respective private streams.

### 10.2. Inter-Beam Interference Management

This section will be dedicated to the in-depth study of the proposed approaches to manage the intra-cell interference caused by the use of multi-beam transmitting and receiving antennas in 5G networks. To avoid inter-beam interference, different approaches have been recently studied, including: interference-aware beam selection schemes [192, 193, 194, 195], interference-aware intra-beam resource allocation schemes [196], time-domain beam scheduling schemes [63, 197], frequency reuse scheme [198], and advanced transceivers for interference cancellation schemes [199, 200, 201].

A) *Interference-aware beam selection schemes*
Considering the potential intra-cell inter-beam interference, the study in [192] proposed an interference-aware beam selection scheme based on classifying all UEs into two sets depending on the potential inter-beam interference. First, high-power beams were chosen for UEs with low inter-beam interference. Then, for UEs suffering from high IBI levels, the authors suggested a low-complexity step-wise algorithm selecting the beams one by one and aiming to maximize the sum rate. Simulation results showed that the proposed interference-based beam selection provides a near-optimal sum rate and better energy efficiency than conventional approaches.

In [193], exploiting a certain sparsity or limited scattering environments of mm-wave channels in a TDD-based multi-user MIMO system, a low-complexity joint beam selection method for BS and user beams adopting compressed sensing without explicit channel estimation is proposed to optimize the performance managing inter-user interference. It was shown through simulations that the proposed method can provide a reasonable performance with low complexity and signaling overhead, compared to the optimal beamforming approach requiring full CSI.

The design of adaptive beamforming in multi-user MIMO systems is highly challenging as it highly depends on accurate CSI at the transmitter, which is difficult to obtain due to the rapidly changing propagation environment and feedback delay leading to a mismatch between estimated and update-to-date channels, known as channel aging. Therefore, in [194], considering multiple transmit and receive antennas case, a multi-agent deep reinforcement learning framework for joint transmit precoder and receive combiner design in DL transmission under imperfect CSI and only based on the knowledge of past channels and multi-user interference, was proposed to maximize the average information rate of all users. Specifically, the problem was cast as a sequential decision-making process, and based on stream-level, user-level, and system-level agent modeling, three DRL-based schemes, were proposed, analyzed, and evaluated. Due to the feedback delay from users, only outdated CSI information including interference information is used to formulate the observations at the agent at the beginning of each time slot. In the decision-making stage, each agent chooses one precoder and one combiner from pre-defined beamforming codebooks and receives a reward value measuring the achievable rate and quantifying the adverse impact of multi-user interference. Such design allows the agent to intelligently learn the best precoders/combiners maximizing the achievable rate while minimizing the inter-users interference. The merits of the proposed framework, especially fast convergence, robustness, and lower complexity, were shown through extensive simulations by benchmarking the different schemes against some conventional methods.

While the last study dealt with joint transmit precoding and combining under imperfect outdated CSI, in [195], the authors considered partial interfering beam feedback at the transmitter to propose two designs for the non-linear Tomlinson Harashima Precoding (THP)-based digital precoding for downlink MU-MIMO, while the best codeword from the transmitter/receiver beam codebook is selected for the analog beamformers/combiners. The first approach applied THP based on an approximate effective channel where only inter-beam interference from top-p interfering beams was considered. However, as the feedback available becomes more reduced compared to the whole set of interfering beams, the inaccuracy of the approximated channel affects the BER performance. Therefore, a second iterative method consisting of a robust THP design minimizing the transmission power under both mean square error constraints at each user and the channel error conditions was proposed. This robust THP was shown to perform better than the first non-robust THP technique, especially for poor channel conditions and in the presence of noisy feedback scenarios.

B) *Interference-aware intra-beam resource allocation schemes*
In [196], considering URLLC and eMBB users in a beam-forming heterogeneous 5G mmWave network, the authors proposed a joint QoS-aware online clustering and a deep RL-based resource allocation scheme to improve the network sum rate while taking into account the users' mobility and dynamic traffic. Specifically, to maintain efficient coverage of mobile users, Density-Based Spatial Clustering of Applications with Noise was used for grouping adjacent users in the coverage of a single beam. Such clustering was only performed when the average SINR of a beam fell below a pre-specified threshold, to avoid frequent changes in the structure and number of beams, which could result in a tricky resource allocation problem. After the clustering, an LSTM-based deep Q-learning technique was applied by each gNB to perform resource block allocation to the users within each of its beams. As intra-beam interference was avoided due to the use of OFDMA per each beam, the resource allocation was performed based on users' CQI feedback capturing inter-beam interference levels. The reward of the DRL-based algorithm was designed to account for the latency and/or the reliability depending on the different users' classes. The suggested method was compared with a baseline using K-means for clustering and priority-based proportional fairness for resource allocation. The results of the simulations showed that the proposed scheme is better than the baseline in terms of latency, reliability, and achievable rate.

C) *Time-domain beam scheduling schemes*

In [63], the authors first investigated via real-world simulations, the characteristics of inter-beam interference in a real mm-wave network with hybrid beamforming. Then, they provided insightful observations that can be summarized as follows: Single-hop adjacent beams are the most significant interfering beams, while interference from multi-hop adjacent beams can be ignored in comparison to the dominant interfering beams, so the elimination of dominant adjacent interfering beams from a set of active beams would greatly enhance network performance. Leveraging such observations, the authors suggested a practical and low-complex beam on/off scheduling employing graph theory and network utility maximization framework. Finding an optimal scheduling beam set at each time slot suffers from the curse of dimensionality, therefore, the special interference graph based on defined beam interference relations was introduced and a practical time-domain beam scheduling scheme leveraging the idea of exception of dominant interfering beams was detailed in the paper. Simulations revealed that the proposed algorithm attains a high performance with a reduced computational complexity.

In another approach, [197] proposed to manage multi-user interference using a joint multilayered user clustering and time-domain scheduling. Based on the users' location information, neighboring users are clustered together using the low computational complexity K-means algorithm. To mitigate the inter-beam interference, referred to as inter-cluster interference in this study and caused by simultaneous transmission by different adjacent clusters using the same resources, the authors proposed to perform a user-cluster layering so that clusters in each layer can be distanced from each other more than the interference-offset-distance, which is defined as the required minimal distance to maintain inter-cluster interference below a tolerable level. Then, BS performs layer-wise cluster-antenna association and layer-by-layer round-robin-type time-domain scheduling, in which transmission opportunity is sequentially assigned to each layer. It is worth noting, that the cell edge layer was referred to as the bottom layer and its scheduling timing was synchronized across neighboring cells. The performance of the proposed joint multi-layer user clustering and time-domain scheduling was validated by evaluating link capacity and inter-user fairness.

D) *Frequency reuse schemes*

In the paper [198], a greedy method maximizing the overall sum rate was used to associate users with serving beams. However, this user-beam association generates severe interference for users located in the coverage of adjacent beams. To mitigate the significant intra-cell inter-beam interference that affects users at the beam edge while maintaining the total data rate, an adaptive frequency reuse scheme was proposed. Specifically, for each beam edge user, it was proposed to allocate half of the frequency spectrum to its serving beam and the other half to the interfering beam. The simulation results confirmed that the proposed adaptive frequency reuse policy can significantly improve the beam edge users' throughput and maximize the fairness between UEs without sacrificing the total throughput.

E) *Advanced transceivers for interference cancellation schemes*

One way of eliminating inter-beam interference between multiple subarrays in multi-beam transmitters is to use an additional array to create a notch for the original beam in the desired direction. However, using the extra array just to create notches is a waste of antenna space. Therefore, a simplified two-stage beamforming approach based on cross-coupling of independent beams either in baseband, intermediate frequency (IF), or RF domain, was proposed in [199]. The main idea was to combine the signals from multiple subarrays over-the-air (OTA) by incorporating a scaled and phase-shifted version of the interfering streams to the initial beams suffering from IBI. The interference and cross-coupling signals combine destructively and mutually suppress each other in the corresponding directions. A related method may be used at the receiver. The cross-coupling coefficients can be computed from the RF beamformers and known transmission directions. While this can be simply done in a two-beam scenario, the cross-coupling coefficients may slightly change the original beams in a multi-beam cancellation. The simulation demonstrated an IBI suppression typically greater than 40 dB for randomly distributed beams and more than 4 dB improvement for the effective isotropic radiated powers in the desired directions.

With increased relative bandwidth, beam squint, i.e., frequency-dependent directivity, complicates the ability of wideband phased arrays to eliminate interference. Therefore, the same authors of the last study further analyzed the effectiveness of a component carrier-based cross-coupling-based approach for the cancellation of inter-beam interference of wideband arrays with visible squint effect in [200]. An example scenario was simulated using standard 5G NR modulated waveforms. For larger arrays with a wider relative bandwidth, simulation analysis showed that multiple cross-couplings eliminate the inter-beam interference of different users.

Large-scale arrays with wideband inter-beam interference cancellation were also investigated in [201], where a scalable

IBI cancellation scheme at intermediate frequency using an IF receiver chip supporting hybrid beamforming was introduced. Specifically, a theoretical simplified over-the-air IBI cancellation model in the presence of circuit matching and channel non-idealities was presented, along with a two-phase calibration method computing accurate and phase-optimized cancellation weights to optimize the cancellation performance. Interference rejection performance was demonstrated through the OTA measurements where 34–37-dB, 16–19-dB, and 16-dB rejections were achieved for 50–100-MHz signals, over 100-MHz bandwidth signals, and 400-MHz multicarrier 5G NR waveform respectively.

### 10.3. Intra and Inter-Beam Interference Management

The integration of both NOMA and Beamforming MU-MIMO has the potential to capture the benefits of the two key technologies, allowing two or multiple users to share a single beamforming vector and thus enhancing the spectral efficiency of the network and its performance [177]. However, despite the potential capacity gains, such integration comes with a cost: intra and inter-beam interference in the same cell. This sub-section will be dedicated to studying the mechanisms reducing such interferences, such as user clustering, adaptive beamforming, user scheduling, and power allocation.

#### A) *Interference-aware beam selection and power allocation schemes*

In [202], the sum-rate maximization problem was formulated to properly design beam selection and power allocation approaches subject to several constraints, such as a limited number of RF chains, maximum transmission power, and the users' requirements. Since the joint beam selection and power allocation problem is combinatorial with a large set of coupled and mixed integer variables, the authors in [202], proposed to decompose the original problem into two sub-problems. First, the beam selection sub-problem was solved under an equal power allocation. Three widely used intelligent searching algorithms, i.e., generic algorithm, PSO, and simulated annealing, were tested in the study. In addition, using feature selection techniques, a mutual information-based beam selection algorithm was adopted. The result of beam selection is then used as the input to the power allocation sub-problem, where two heuristic algorithms with a linear complexity to the number of users, were proposed aiming to maximize the sum rate and mitigate the intra and inter-beam interference. The first algorithm allocates power in two stages: inter-beam and intra-beam power allocation. In the second power allocation scheme, intelligent searching via PSO was used to find the optimal power allocation. It was demonstrated through numerical results that the proposed beam selection and PSO-based power allocation algorithms can effectively improve the sum rate while reducing the computational complexity compared to the two benchmark schemes.

#### B) *Interference-aware beam selection, user scheduling and power allocation schemes*

In [67], to maximize the time-averaged sum of utilities of users, the authors formulated a long-term network utility maximization constrained by a time-average transmit power, where beam selection, user scheduling, and transmit power allocation are incorporated on top of future network architecture with Multi-access Edge Computing managing a small set of BSs. They proposed a low-complex algorithm, named CRIM (CRitical user based Interference Management), to sequentially solve the aforementioned problems, where the Lyapunov optimization framework was used for the time scale decomposition, and where the interference was abstracted into a single critical user when deciding the transmit power. Extensive simulations were performed and verified that the proposed CRIM solution improves the utility performance by up to 47.4% compared to other existing algorithms.

#### C) *Interference-aware joint user-cell association and selection of number of beams*

Considering a multi-cell mm-Wave networks scenario, [203] addressed the problem of joint user-cell association and selection of the number of beams that aims to mitigate intra and inter-beam interference intra and inter-cell and maximize the aggregate network capacity. Three machine learning-based algorithms were proposed, namely: (i) transfer Q-learning (TQL) where the knowledge of an expert agent consisting of user-cell association using Q-learning is transferred to a learner agent, which builds on the expert's knowledge and carries out the difficult task of jointly associating user to appropriate cells and selecting the number of beams to cover the associated users, (ii) conventional Q-learning performing joint user-cell association and selection of number of beams, and (iii) a combination of best SINR association and Density-based Spatial Clustering of Applications with Noise (BSDC) for users clustering. The performance of the three proposed approaches was analyzed and compared under different network scenarios varying the user deployments and mobility. The simulation results confirmed the relevance of every scheme to the deployment scenarios envisaged. For mobile scenarios, TQL and Q-learning improved the rate by around 12% over BSDC for the highest traffic load, while Q-learning and BSDC outperformed TQL under stationary scenarios. The transfer-learning approach converged more rapidly than conventional Q-learning, as it utilizes samples of experience efficiently offering the advantage of offline learning of a task and transferring the knowledge to another task with online learning in the field

#### D) *Clustering, hybrid precoding and power-allocation schemes*

To improve the spectral efficiency of mMIMO-NOMA networks, the number of beams is generally fixed to the number of RF chains. Typically, a maximum of two users are associated with a beam for non-orthogonal communication, giving excellent results in terms of spectral efficiency. Nevertheless, when the number of users is significantly high, more beams and thus more RF chains are required, which is not practical because of the hardware costs involved. Therefore, it was recommended in multiple studies, to equip the BS with massive antenna arrays to meet the communication demand of all users without compromising performance and at a lower cost. The users are then grouped in clusters following an adequate strategy, and each group of multiple users is served by one beam. Adaptive hybrid beamforming and power allocation schemes are exploited to enhance the system's spectral efficiency and reduce inter and intra-beam interference.

For example, in [204], as it is impossible to implement precoding schemes for a large number of users separately, the authors first provided a cluster-based hybrid precoding approach. Specifically, they adopted the concept of chordal distance in the clustering strategy, i.e., the selection of cluster head users and the assignment of more than two users to each beam. On one hand, head users representing the principal characteristic of their clusters, have a low correlation with each other which allows the reduction of intra-beam interference. On the other hand, the users in every cluster must be highly correlated to minimize inter-beam interference, which may enhance multiplexing gains. According to the channel vectors of the cluster headset, analog precoding was designed, then a baseband precoding scheme was implemented based on the conventional ZF precoding technique. Finally, with a given total power budget, based on the Lagrange duality theorem, a dynamic power allocation approach for assigned users of each beam was proposed to improve the overall spectrum and energy efficiency.

E) *Hybrid transceiver design and interference-aware beam selection schemes*

Considering a typical single-cell downlink mm-wave network, [205] introduced the concept of code domain NOMA into beamspace MIMO to overcome the inherent limitation of existing beamspace MIMO that the number of users cannot surpass the number of RF chains. In particular, by considering intra- and inter-beam interference, the authors proposed a low-complexity beam selection scheme based on a factor graph and codebook design to achieve sum-rate maximization by exploiting the quasi-orthogonality of the beamspace channel. Furthermore, a threshold-based message-passing detection algorithm is developed to realize the hybrid transceiver design in mm-wave communications using lens antennas. While reducing the computational complexity, the proposed solution achieves high spectrum and energy efficiency and acceptable BER performance.

**Discussion:** Table 8 provides a comparison between all previously listed papers dealing with intra-cell multi-user interference management. To mitigate intra-beam interference in non-orthogonal multiple access networks, it is primordial to adopt (i) adequate user clustering strategies adapting to the dynamics of the network environment, (ii) additional advanced processing techniques for pre-interference suppression at the transmitter, including superposition coding and rate splitting, which use the prior knowledge on CSI to efficiently precode the signals, and (iii) post-interference cancellation schemes at the receiver, such as SIC, to cancel out the interference [78, 206]. The use of multi-beam antennas may induce intra-cell inter-beam interference, which can be managed through adequate beam scheduling in the time or frequency domain, interference-aware beam selection and power allocation schemes, and advanced signal processing techniques at the transceiver. Designing dynamic multi-user interference management schemes for both intra and inter-beam scenarios is of high importance to fully benefit from the performance gains of NOMA MU-m-MIMO systems. To permit the scheduling of multiple users within the same resource and handle the severe interference, such approaches should not rely on the prior knowledge of channel state information, and adapt to different configurations and dynamic changes in the environment. Robust machine learning-based signal processing schemes should be further investigated.

## 11. Inter-Numerology Interference Management

In this section, we will discuss the most recent research dealing with inter-numerology interference management in 5G NR for both non-overlapping and overlapping mixed numerologies. Many studies included theoretical models and proposed mathematical expressions of INI such as [69, 207, 208, 74, 209]. In [207], INI was studied and characterized through mathematical modeling for both discrete and continuous-time mixed-numerology systems. Reduced-form formulas for the INI were derived and explicitly revealed some influencing factors, like the pulse shape and the relative distance between subcarriers. These formulas can also be used to design mitigation techniques and interference metrics. The authors of [74, 209] derived general closed-form expressions for the fading-averaged and bandwidth-averaged INI powers in mixed-numerology 5G OFDM systems. The bounds on the bandwidth-averaged INI power demonstrate its quadratic dependence on Doppler spread and that it impacts the higher rate modulation and coding schemes of 5G NR. In the literature, various INI management techniques were proposed, such as windowing/filtering [210], flexible guard bands [71, 211, 212], interference cancellation, precoding, and equalization [69, 213, 214, 215, 40, 216, 217], resource allocation and scheduling [70, 218, 219, 220, 221, 222], and novel symbol design [73, 223].

A) *Windowing/Filtering Mechanisms*

These schemes are one of the most effective means to control the out-of-band-emission and to mitigate INI, especially in non-overlapping numerologies systems. However, designing such mechanisms can be challenging, especially to avoid SINR loss or power consumption increase. Many studies proposed windowing/filtering-based INI management schemes like [210], and many others combined the windowing/filtering mechanisms with other approaches such as [211, 69].

In [210], to mitigate the INI, the authors proposed a convolutional neural network (CNN)-based mixed numerology interference recognition approach, along with a dynamic filtering/windowing scheme that is only triggered when the inter-numerology interference appears. Such an approach permits handling the mixed numerology interference without decreasing system performance (SINR loss or power consumption increase) in normal scenarios when there is no INI.

| Ref | Intra-Beam/Inter-Beam | Proposed Scheme | Main Idea |
|---|---|---|---|
| [186] | Intra-Beam Interference | User pairing/clustering schemes | Two user pairing methods based on neighbor search algorithms: hill climbing and simulated annealing |
| [188, 189] | | | Multi-user clustering and adaptive multi-user clustering algorithms under imperfect CSI |
| [190] | | | An adaptive and evolutionary clustering strategy based on a modification of the DenStream evolutionary algorithm and considering the dynamic characteristics of communication networks |
| [191] | | Superposition coding schemes | A survey of power allocation and user pairing schemes in NOMA |
| [78] | | Multi-user detection algorithms | A survey of the various successive interference cancellation techniques with perfect and imperfect CSI |
| [45] | | Rate-Splitting and Linear Precoding | Overview of RSMA as a novel framework for multiple access and interference management |
| [192] | Inter-Beam Interference | Interference-aware beam selection schemes | Users classification into 2 groups: users with low IBI to be associated with high power beam, and users with high IBI to be associated using a low-complexity step-wise algorithm |
| [193] | | | A low-complexity joint beam selection method for BS and user beams adopting compressed sensing without explicit channel estimation |
| [194] | | | A multi-agent deep reinforcement learning framework for joint transmit precoder and receive combiner design in DL transmission under imperfect CSI |
| [195] | | | Two designs for the non-linear THP-based digital precoding for downlink MU-MIMO with partial interfering beam feedback |
| [196] | | Interference-aware intra-beam resource allocation schemes | A joint QoS-aware online clustering and a deep QL-based resource allocation scheme to improve the network sum rate while taking into account the users' mobility and dynamic traffic |
| [63] | | Time-domain beam scheduling schemes | A practical and low-complex beam on/off scheduling employing graph theory and network utility maximization framework |
| [197] | | | A joint multilayered user clustering and time-domain scheduling |
| [198] | | Frequency-reuse schemes | Adaptive frequency-reuse scheme: for each beam edge user allocate half of the frequency spectrum to the serving beam and the other half to the interfering beam |
| [199] | | Interference cancellation schemes | A simplified two-stage beamforming approach based on cross-coupling of independent beams |
| [200] | | | A component carrier-based cross-coupling based approach for wideband arrays with visible squint effect |
| [201] | | | A scalable IBI cancellation scheme at intermediate frequency using an IF receiver chip supporting hybrid beamforming |

| REF | INTRA-BEAM/ INTER-BEAM | PROPOSED SCHEME | MAIN IDEA |
|---|---|---|---|
| [202] | Inter and Intra Beam Interference | Interference-aware beam selection and power allocation schemes | Disjoint optimization using intelligent searching algorithms for beam selection sub-problem and heuristic algorithms for power allocation |
| [67] | | Interference-aware beam selection, user scheduling and power allocation schemes | a low-complex algorithm using the Lyapunov optimization for the time scale decomposition, and interference abstraction into one critical user |
| [203] | | Interference-aware joint user-cell association and selection of number of beams | Three machine learning-based algorithms: Transfer QL, conventional QL and a combination of best SINR association and BSDC for users clustering |
| [204] | | Clustering, hybrid precoding and power-allocation schemes | Chordal distance based clustering, hybrid precoding scheme and a dynamic power allocation scheme |
| [205] | | Hybrid transceiver design and interference-aware beam selection schemes | A low-complexity beam selection scheme based on factor graph and codebook design, and a threshold-based message passing detection algorithm for the hybrid transceiver design |

Table 8: List of Papers dealing with Intra-Cell Multi-User Interference Management Schemes

As mixed numerology interference operates with a power spectrum density that is similar to the in-band signal, accurately identifying inter-numerology interference when it occurs, is a key but challenging task. The deep CNN was used to recognize the hidden features in the waveform and to classify the signal samples into two groups, either with or without INI. As the generated data is a complex waveform, the authors first divided the time/frequency domain data into real/imaginary parts or amplitude/phase parts so that they can be used by standard loss functions and gradient descent algorithms in deep neural networks. The authors used Bayesian optimization and 5-fold cross-validation to determine the optimal hyper-parameters of the model. Then, they compared their built-up CNN model with three classic machine learning methods and proved it outperforms SVM, kNN, and Decision Tree algorithms in different practical scenarios, achieving over 90% accuracy. Based on the results of inter-numerology interference identification, the dynamic filtering/windowing scheme is triggered and through link level simulations, it was shown that it can increase SINR by 3~6dB depending on the scenario.

B) *Flexible Guard Bands*
Inserting a guard band (GB), as a sequence of unused contiguous subcarriers, between neighboring numerologies is one of the most common and effective ways of reducing INI. However, the common way to implement GBs is inefficient in terms of spectrum utilization. Moreover, generally, GBs are combined with other schemes such as windowing/filtering, common prefix, or adaptive scheduling.

The study in [71] assumed no overlapping numerologies in the frequency domain with equal UE powers. Based on heuristic algorithms and through simulations, it presented INI as a function of guard band allocation and multi-numerology parameters. With the performance analysis results, the GBs are shown as an effective way to minimize inter-numerology interference experienced by edge subcarriers. Few other basic inferences are made from the simulation results, such as: (i) there is more interference on the edge subcarriers, (ii) INI decreases on each subcarrier as the guard band increases, and (iii) interference decreases for higher numerologies.

While in [71], only guard bands were considered, in [211], the authors proposed an INI management scheme on the transmitter side with a cross-layer approach based on a joint optimization of adaptive time and frequency domains guards coupled with a multi-window mechanism in the physical layer. This joint optimization was realized by taking into account the SCS, the use case, and the power difference between numerologies. To decrease the need for guards, a MAC layer scheduling algorithm was also proposed to allocate the numerologies to the available spectrum, with respect to the SCS, power level, and signal-to-interference ratio. Results confirmed that the proposed guard allocation and scheduling technique can enhance the spectral efficiency of mixed numerology systems while reducing the guards.

In [212], exploiting the INI characteristics of the scalable multi-numerology structure, Memisoglu et al. introduced a new and spectrally efficient way of guard band implementation to avoid inter-numerology interference and minimize the spectral efficiency loss. More precisely, it was observed that the INI generated between adjacent non-overlapping numerologies is not random and exhibits a peculiar pattern that can be exploited to redesign the GB structure. Therefore, the authors proposed to insert data subcarriers with narrow SCS and low-order modulations in carefully chosen indexed parts of the GB such that the INI is limited. They also suggested the use of a common cyclic prefix to enhance the average bit error rate (BER) of the narrow numerology edge subcarriers within the GB region. The proposed scheme was analyzed through Monte Carlo simulations and was found to enhance the GB usage by up to 50% with the same BER performance as the traditionally implemented GB.

C) *Interference Cancellation, Precoding and Equalization Schemes*
Advanced transceivers with enhanced capacities to suppress or pre-cancel the INI are considered for non-overlapping [69, 213, 214] and overlapping mixed numerologies [215, 40, 216, 217, 42].

In [69], Zhang et.al proposed to combine the windowing mechanism with an interference cancellation scheme to mitigate INI in non-overlapping mixed numerologies Windowed-OFDM systems. In particular, a theoretical model of INI in Windowed-OFDM systems was built based on the spectral distance between subcarriers, overlapping windows, and channel frequency response. Based on derived analytical expressions, the INI was divided into a dominant deterministic part and an equivalent noise part. An IC scheme was then provided for jointly detecting the dominant victim subcarriers and interfering subcarriers to suppress the main inter-numerology interference. Numerical analysis showed the accuracy of the INI theoretical model compared to simulation results. Furthermore, the proposed interference cancellation scheme outperformed the existing receiver algorithms.

In a more recent study [213], a frequency division multiplexing-based RAN slicing framework considering the imparities in both base-band and radio frequency configurations, was proposed and analyzed from the physical layer perspective. The UL and DL system models and the relationships between the key parameters of a mixed numerology system (including SCS, symbol duration, etc.) were established for the most generic scenario. Two theorems of generalized circular convolution properties of discrete Fourier transform were presented and provided the theoretical foundation of low complexity but effective cancellation algorithms by joint detection in the UL or precoding in the DL of the proposed system model. Simulations were performed using both OFDM and Filtered-OFDM waveforms. It was shown that a 4-5 subcarrier guard band can decrease the interference to an insignificant level. Furthermore, the cancellation algorithms proved their efficiency in improving SINR performance even without additional guard bands.

Inter-numerology interference in OFDM systems was also analyzed using discrete Fourier transform equations in the frequency and time domains in [214]. INI was then modeled as weighted linear combinations of neighbor numerology's transmitted symbols and an INI weight fixed matrix was derived for all OFDM symbols. The authors proposed to suppress the inter-numerology interference by using the inverse of the computed matrix as a pre-equalizer at the transmitter side. The performance of the pre-equalization method was verified via Monte Carlo simulations, as INI was completely removed and spectral efficiency improved without needing guard bands.

While [69, 213, 214] considered non-overlapping mixed numerologies systems, the following studies investigated the use of multiple overlapping numerologies systems.

In [215], a flexible framework for multi-service systems based on sub-band filtered multicarrier transmission was established and the INI was analyzed with and without transceiver imperfections and insufficient guard interval between symbols. Based on the derived signal models, baseband processing algorithms by precoding the information symbols at the transmitter were first proposed. Assuming perfect knowledge of channel information at the transmitter side, the low-complexity cancellation algorithms using the criteria of zero-forcing and minimum mean square error can flexibly cancel the INI for arbitrary bandwidth. Furthermore, considering the synchronization errors and transceiver imperfections, enhanced channel equalization algorithms providing significant gain in terms of BER performance were also presented. Various simulation results were given to show the effectiveness of the analysis and proposed algorithms.

In another transceiver design scheme considering a mixed overlapping numerologies system, the authors in [40], first derived the interference pattern in closed-form expressions and found out that while the mean of the interference energy is maintained, its variance increases for higher SCS and badly affects decoding performance. Based on these observations, a novel transceiver design using simple cyclic shift and frequency shift operations was proposed to lower the variance of interference energy by uniformly dispersing the effect of INI across different subcarriers. Although no theoretical model was formally presented, the simulation results verified that the proposed transceiver suppresses the variance of the interference energy and improves the decoding performance.

Cheng et.al in [216, 217] first introduced an improved precoding scheme to manage the mixed numerology spectrum sharing transmissions for massive MIMO OFDM systems. While time-domain overlapping of all signals maximizes the spectrum, it introduces INI to the system. The interference was analyzed regarding many parameters such as SCS, channel selectivity, and power allocation. Closed-form theoretical expressions were derived and it was shown that INI is only generated in frequency-selective channels and only by users with large SCS to users with small SCS. Based on the derived expressions, the authors proposed an INI pre-cancellation scheme to be implemented only at the BS side so that the receiver's complexity does not increase. Through simulation results, it was proved that the proposed cancellation scheme efficiently mitigates the interference and enhances the performance of the network, at the expense of some transmit power loss and added system complexity.

D) *Resource Allocation*

A suitable resource allocation that reduces the numerology gap may reduce the inter-numerology interference. Many studies have focused on the inter-numerology interference management approach from a resource allocation perspective.

In [70], novel optimization methods for time-frequency resource allocation in a multi-user scenario employing mixed numerology and mini-slot were proposed. The authors focused on low-latency communications and thus the optimization approach aimed to increase the achievable rate of best-effort users whilst meeting the latency requirements of low-latency users. A comparative study between the mini-slot approach and the mixed numerology approach showed that both methods perform similarly in the case of flexible latency constraints and similar channel conditions. However, in the case of opposite channel conditions for the users involved, the mixed numerology approach is much better than the mini-slot approach, which is more adequate in the case of stringent latency constraints.

Another study that focused on ultra-reliable and low latency communications is provided in [218]. The authors proposed heuristic INI-aware resource allocation schemes including fractional numerology domain (FND) scheduling and power difference-based (PDB) scheduling algorithms. A machine learning-based scheduling method was also provided as a perspective. The main objective of all these scheduling schemes is to provide extra protection to uRLLC users and cell edge users while increasing fairness for the edge subcarriers. In the FND approach, each numerology sub-block is divided into one inner part for inner users and two outer parts for outer users. Inner parts are less affected by INI and therefore allocated for users who need more reliability. Outer users are sorted into two groups: non-edge and edge users which are the most impacted by inter-numerology interference. Fraction regions of each numerology sub-block are used in the scheduling mechanism. In the PDB scheme, the outer users are investigated to find the best suitable edge users of subblocks based on their power levels and bandwidth. Simulation results of the proposed algorithms with the use of a guard band demonstrated an overall reliability enhancement.

While in [218], heuristic scheduling schemes were proposed, in [219], the authors proposed a deep reinforcement learning agent using an INI-aware reward function to optimally allocate the spectrum and maximize the network throughput. The DRL agent at the Network Owner (NO) has access to the subchannel gains reported by every user to its associated Mobile Virtual Network Operator (MVNO) and the INI power measurements on each subchannel, without taking into account the channel fading variations. The idea behind this state design choice was to make the agent learn and exploit the relationship between the INI power and the channel fading variations to mitigate the INI. At each learning step, the proposed DRL agent observes the state and chooses one feasible action from its action space which corresponds to the possible sub-channel allocations per MVNO. However, as the number of resources increases, it becomes impractical to enumerate every possible allocation and approximate the optimal solution. Therefore, the same authors proposed in [220] an alternative multi-branch agent architecture to ensure agent scalability. In particular, each branch of the network is responsible for allocating the resources of only a subset of the frequency band and coordinates with the other branches via a common module that provides a shared value of the state of the environment. To augment the agent learning performance and boost its convergence to the optimal allocation policy, the selection of feasible actions was enforced via an action masking module. The analysis showed that the agent converges with success under various configurations and achieves a resource allocation scheme similar to the optimum solution in almost all cases.

Zambianco et.al further extended their idea in [221] to differentiate between eMBB and URLCC slices. The main difference with the previous paper [220] consisted of the definition of the state space and reward function. Besides the subchannel gain information of all users, the DRL-based agent observes at each time step, the buffer status of URLCC users, the minimum required number of bits of eMBB users, and the total data rate achieved in the previous time slot. This design choice permits to take into account the different requirements of different services and to take into account implicitly and without complex computation, the INI power in the data rate expression. The reward function was accordingly adapted to correspond to the total data rate minus a penalty function when the Service-Level Agreement (SLA) requirements of the users are not satisfied. The agent performance in terms of SLA satisfaction and aggregated user throughput was analyzed and compared to Round Robin and Weighted Max Rate resource allocation algorithms in two network scenarios that reflect a different network congestion level. It was proven that the proposed scheme outperforms both benchmark schemes by providing a more reliable SLA fulfillment, i.e., a higher eMBB user data rate and, a lower URLLC user delay.

Similarly to [221], [222] considered radio resource slicing problems for URLCC and eMBB in multiple numerologies 5G networks. However, Esmaeily et.al further distinguished between URLCC use cases and classified them into different URLCC traffic classes holding distinct QoS requirements. Due to stringent latency requirements of URLCC traffic, the gNB punctures previously scheduled eMBB transmissions, to immediately schedule the incoming sporadic URLCC traffic into the mini-slots. The authors investigated the puncturing technique and the mini-slots approach to meet the requirements of different URLLC classes and to increase the sum rate of the eMBB users while maintaining the minimum acceptable data rate of every eMBB user. Therefore, they defined a loss function of the eMBB user's data rate to measure the influence of puncturing by the overlapping traffic of each URLLC class, and they presented a puncturing rate threshold to limit this influence. Since the formulated resource allocation problem is mixed-integer nonlinear programming, the authors relaxed the constraints and decomposed the problem into three convex sub-problems: eMBB RBs allocation, power allocation, and URLLC traffic scheduling. The CVX toolbox was then used when solving each sub-problem. The Mixed-numerology Mini-slot-based Resource Allocation (MiMRA) proposed algorithm iteratively searches the optimal solutions for the sub-problems until convergence. The simulation results showed that the proposed algorithm MiMRA improves the total sum rate of eMBB users while ensuring the minimum data rate requirement of each eMBB user.

E) *Symbol Design*

A different approach to managing multiple numerologies using a novel cyclic prefix insertion technique can be found in [223] and [73]. While the 5G standard specified an individual CP configuration where a CP is tied to each OFDM symbol before being transmitted, the authors of the aforementioned studies introduced a new CP insertion method, called common CP, where one cyclic prefix is used to protect multiple OFDM symbols. In [73], it has been shown via simulation that the inter-numerology interference distribution can be restructured using the common CP, which leads to better interference management. In [223], the common CP configuration was mathematically analyzed and compared to conventional individual CP configuration. The analysis showed that some subcarriers of narrow and large SCS numerologies are somehow orthogonal to each other and that this orthogonality can be further increased using the common CP configuration. According to the study, the inter-numerology interference can be mitigated through alternative symbol design techniques and intelligent scheduling without using techniques such as windowing which decreases the spectral efficiency.

**Discussion:** Table 9 provides a comparison between all previously listed papers dealing with inter-numerology interference management for both non-overlapping and overlapping mixed numerologies. The multi-numerology framework is envisioned to efficiently support different service requirements and to provide flexibility to 5G and beyond communication networks. However, multiplexing different numerologies induces non-orthogonality in the system that needs to be handled. In non-overlapping mixed numerologies, multiple physical layer techniques have been proposed, such as windowing/filtering mechanisms and adding guard bands between the subcarriers. However, while these approaches have relatively reduced complexity, they decrease the spectral efficiency of the system. Furthermore, it is impossible to avoid INI in mixed overlapping numerologies systems using these techniques. Alternatively, it is recommended to apply intelligent resource allocation schemes or adaptive precoding and interference cancellation approaches.

| REF | OVERLAPPING/ NON-OVERLAPPING MIXED NUMEROLOGIES | PROPOSED SCHEME | MAIN IDEA |
|---|---|---|---|
| [210] | Non-Overlapping | Windowing/Filtering Mechanism | CNN-based mixed numerology interference recognition approach + a dynamic filtering/windowing scheme that is only triggered when the INI appears |
| [71] | | Guard Bands | Based on heuristic algorithms and through simulations, INI presented as a function of guard band allocation and multi-numerology parameters |
| [211] | | Guard Bands + Windowing | A joint optimization of adaptive time and frequency domains guards coupled with a multi-window mechanism in the physical layer and a MAC layer scheduling algorithm |
| [212] | | Guard Bands | Insertion of data subcarriers with narrow SCS and low-order modulations in carefully chosen indexed parts of the GB such that the INI is limited |
| [69] | | Windowing + Interference Cancellation | Theoretical model of INI in Windowed-OFDM systems + IC scheme for jointly detecting the dominant victim subcarriers and interfering subcarriers |
| [213] | | Interference Cancellation | low complexity interference cancellation algorithms by joint detection in the UL or precoding in the DL |
| [214] | | Equalization Scheme | Using the inverse of the computed matrix as a pre-equalizer at the transmitter side |
| [218] | | Resource Allocation | Heuristic INI-aware resource allocation schemes including FND and PDB scheduling algorithms + a machine learning-based scheduling method as a perspective |
| [73, 223] | | Symbol Design | Introduction of Common CP configuration |
| [215] | Overlapping | Precoding + Equalization schemes | Low-complexity cancellation algorithms using the criteria of ZF and MMSE |
| [40] | | Transceiver Design | A time-domain spectrum sharing transceiver design to lower the variance of interference energy by uniformly dispersing the effect of INI across different subcarriers |
| [216, 217] | | Precoding + Interference Cancellation | INI pre-cancellation scheme to be implemented only at the BS side |
| [70] | | Resource Allocation | Novel optimization methods for time-frequency resource allocation employing mixed numerology and mini-slot |
| [219, 220, 221] | | Resource Allocation | A deep reinforcement learning agent using an INI-aware reward function to optimally allocate the spectrum |
| [222] | | | Puncturing method and mini-slots approach to fulfill distinct URLLC classes' requirements and maximize eMBB users' rates |

Table 9: List of Papers dealing with Inter-Numerology Interference Management Schemes

# 12. Research Challenges and Future Research Directions

Explosive data demand growth has led to more challenging requirements such as higher data rates, greater efficiency, and more reliable, lower latency communications. To meet all of these demands and more, 5G networks incorporate multiple novel technologies, a heterogeneous and flexible architecture, and dynamic and adaptable configurations. However, despite all the advantages of this innovative design, severe interference between different network elements or within a single node remains a challenging problem to manage.

Throughout the previous sections, we have outlined the methods proposed in the literature and 3GPP specifications to address different types of interference in 4G, 5G, and beyond networks. However, there are still several challenges and open issues that have not yet been explored or need to be further studied. This section highlights these open challenges and provides an overview of future research directions in the context of interference management in 6G networks and beyond.

## 12.1. Interference Management in 5G and Beyond Networks: Open Challenges

As shown in Table 10, we begin with a summary of the main interference management challenges in 4G/5G and beyond networks and the most promising proposed schemes.

- Dense and ultra-dense multi-tier networks suffer particularly from cross-tier DL-to-DL inter-cell interference. ICIC schemes based on power-time domains with coordinated scheduling are in general the most effective. However, special attention should be paid to the cost of coordination in conventional distributed coordinated approaches, i.e., the inter-cell signaling exchange overhead. Furthermore, due to the dynamic time-varying channel conditions and users' mobility, the interference information may be incomplete and outdated. To tackle these challenges, recent emerging deep multi-agent model-free reinforcement learning algorithms can be an effective tool for real-time distributed and dynamic interference management without the need for prior knowledge of channel state information, or information exchange. Such approaches permit one to learn the dynamics of the environment and intelligently choose the optimal configuration, such as power allocation, resource scheduling, precoding, etc., maximizing a specified utility, which can be inversely proportional to the ICI levels.

- Dynamic TDD and FD networks are subject to inter-mode CLI and RI. Cross-link inter-BS interference is particularly noticeable in macro-deployments, while small-cell deployments are especially subject to cross-link UE-to-UE interference. Adequate clustering and dynamic UL/DL configuration are essential to limit the severe BS-to-BS interference in macro-deployments. However, instead of only depending on cell average data traffic, these schemes should take into account the specific users' demands. Advanced interference cancellation schemes are also beneficial for both MCs and SCs deployments. Although RI is caused in rare weather conditions, it may severely impact the network performance for a long duration, especially in coastal and plain cities near the water which are more prone to form ducts. Such interference may affect the UL data of a distant BS, or the UL RS used for DL channel quality estimation, which in turn may lead to poor DL user experience. To our knowledge, there is a lack of research studies dealing with remote interference. The majority of the works investigated the use of machine and deep learning approaches to detect the occurrence of RI using synthetic data sets, or real network side data. Few research papers studied RI management through static spatial approaches without theoretical analysis. More research work should be performed to proactively predict the duct occurrence and to automate the interference management schemes.

- The multiplexing of multiple numerologies introduces a non-orthogonality to the system resulting in inter-numerology interference. To take full advantage of the multi-numerology framework supporting different service requirements, analyzing the pattern of emitted interference to identify the different factors contributing to INI and to adequately avoid or suppress it is inevitable. Several schemes have been proposed in the literature, among them alternative symbol design, adaptive precoding, and interference cancellation, and intelligent resource allocation and scheduling are the most optimal solutions in terms of spectral efficiency, especially in overlapping mixed numerologies systems.

- Multi-beam antenna systems based on mMIMO technology are used to overcome attenuation losses of mm-wave bands. However, the space multiplexing of multiple beams causes intra and inter-cell inter-beam interference that needs to be efficiently handled. Interference-aware beam selection and beam power allocation schemes, along with advanced interference cancellation schemes are possible solutions. Coordinating the different beams in time or frequency domain reuse schemes may also avoid such interference at the cost of spectral efficiency reduction.

- Caused by the simultaneous transmission to multiple users under the same (narrow or wide) beam and using the same non-orthogonal resources, intra-beam multi-user interference has been actively investigated in NOMA-based systems. The proposed management approaches are generally grouped into pairing/clustering strategies, signal processing techniques for pre-interference suppression at the transmitter, and post-interference cancellation schemes at the receiver. Recently, a new candidate for the multiple access scheme for 6G networks, namely Rate Division Multiple Access, has emerged. By exploiting the splitting of user messages as well as the non-orthogonal transmission of differently decoded common and private messages, RSMA has a powerful interference management capability, where the amount of interference to be decoded or treated as noise is adaptively changed depending on the interference level. Further research work should be effectuated to assess its performance in different network configurations.

- Compared to other interfering signals, self interference can lead to the most severe performance degradation at the receiver and therefore should be reduced at most under the noise power level. This intra-system interference may affect both FDD and IBFD transceivers due to TX-RX leakage or simultaneous transmission/reception operation. Tremendous progress, in both industry and academia, has been made in SI cancellation schemes. Generally, a combination of analog and digital domain countermeasures is required to mitigate the self interference.

After the successful standardization of 5G networks, the focus in both academia and industry has shifted to next-generation networks. 5G kicked off flexible wireless communications through the integration of different technologies, and 6G networks are expected to extend this flexibility further [224]. In the following, and as summarized in Table 11, we present the envisioned key features in 6G networks and focus on the related interference challenges. Then, an overview of the main new approaches and paradigms for advanced interference management is provided.

## 12.2. Interference Management Challenges in 6G and beyond Networks

While 5G is rapidly becoming a commercial reality, the sixth generation of wireless systems is already attracting a great deal of interest [225]. This interest is primarily motivated by revolutionary changes in both individual and societal trends, which call for entirely automated systems and intelligent services imposing unprecedented requirements, such as providing ultra-high reliability with packet error rates as low as $10^{-8}$, Tbps data rates, and sub-ms latency simultaneously on both uplink and downlink [226, 227]. Though 5G communication systems provide significant enhancements over existing systems [228], they will likely not be able to meet the aforementioned diverse requirements and support the growing data and connection demands. Specifically, the ITU predicts that by 2030, data traffic will be reaching an astonishing 5 zettabytes per month and the number of connected devices could exceed 50 billion [229].

6G systems will rely on enhancements of existing technologies and the incorporation of new attractive features to deliver future cutting-edge services. Several papers [228, 225, 226, 230, 231, 232] provide an overview of potential 6G technological enablers such as TeraHertz, Intelligent Reflecting Surface (IRS), ultra-dense networks, cell-free massive MIMO, new multiple access schemes, unmanned aerial vehicles (UAV), etc. The integration of these disruptive technologies that will shape the future 6G networks, comes with a breed of issues and challenges that need to be effectively addressed. Interference management is still undoubtedly one of these critical challenges towards realizing 6G networks. In the following, we will highlight some of the key 6G features and their associated interference issues.

- Just as 5G systems introduced millimeter-wave frequency bands for higher data rates and to enable new services, 6G networks aim to push the limits of the frequency spectrum even further to meet greater demands [228]. Terahertz and optical frequency bands are seen as very promising technologies for the 6G era and beyond to alleviate spectrum scarcity and overcome the capacity limits of wireless networks [232]. However, despite these prospects, these bands must overcome significant interference-related problems. At higher frequencies, it is very difficult to support miniature circuits and small transceivers while guaranteeing noise and interference suppression between components [233, 231]. In addition, the ultra-wide bandwidth of available licensed and unlicensed frequencies, and the different capabilities of users, require dynamic, efficient, and self-adaptive spectrum sharing and interference management schemes between users and between different sub-bands and systems [234]. Furthermore, with the increase in the size of the antenna arrays and the wider bandwidth in 6G systems, the beam squint effect is more severe in THz which further aggravates the inter-beam interference.

- Although the multi-numerology waveform has paved the way for flexibility in networks beyond 5G, it is rather limited, and the 6G air interface will require a higher level of flexibility and reconfigurability, due to the increased heterogeneity of applications and use-case requirements in 6G. This has motivated ongoing research into a much wider range of waveforms and modulation techniques [225]. It is expected that the number of adjustable parameters and numerologies for a given waveform will increase with 6G, leading to even more pronounced inter-numerology interference than in 5G systems. Furthermore, as the design of a single waveform based on multiple numeration did not work in 5G, no single waveform will likely be able to meet the requirements of all 6G scenarios [235]. Consequently, 6G systems will probably go beyond the use of different configurations of the same parent waveform, and multiple configurable and flexible 6G waveforms will coexist within the same framework. However, the exponential increase in waveform types and settings may give rise to new forms of interference, notably inter-waveform interference [223, 236]. Controlling and mitigating such interference is likely to be a major challenge.

- Supporting the upcoming massive interconnectivity and highly diverse and revolutionary applications with far more stringent communication requirements in terms of throughput, latency, and reliability for various types of IoT devices, will call for enhanced coverage, sufficient capacity, and advanced multiple access technologies [237, 238]. Along with network ultra-densification, 6G in-X sub-networks have been identified as a promising radio concept to provide capillary wireless coverage for supporting extreme connectivity requirements [239]. In-X sub-networks are presented as highly specialized short-range low-power cells to be installed inside entities such as robots, production modules, vehicles, or human bodies [240, 241]. In-X Sub-networks can, however, lead to high device density scenarios, resulting in high interference levels that may Interference Management in 5G and Beyond Networks| INTERFERENCE TYPE | MOST SUITABLE SCHEMES | OPEN CHALLENGES & RESEARCH DIRECTIONS |
|---|---|---|
| Inter-Cell Intra-mode Interference | ICIC schemes based on power-time domains with coordinated scheduling | - Cost of coordination in distributed approaches<br>- Imperfect (incomplete/outdated) CSI<br>- Investigation of multi-agent model-free deep reinforcement learning approaches |
| Cross-Link Interference | Adequate clustering, dynamic UL/DL configuration, and advanced interference cancellation schemes | - Variable cells' data traffic patterns<br>Specific metrics based on users' demands, for clustering and dynamic configuration |
| Remote Interference | ML and DL-based approaches for RI detection | - Proactive prediction of the ducts occurrence, and automation of interference management schemes |
| Inter-Numerology Interference | Intelligent resource allocation schemes, adaptive precoding, and interference cancellation approaches | - Much more flexibility in the numerology system, and adequate interference management approaches |
| Inter-Beam Interference | Coordinated beam scheduling in time, frequency, and power domains + advanced interference cancellation schemes | - Cost of coordination, and spectral efficiency |
| Intra-Beam Multi-User Interference | Pairing/clustering, signal processing techniques, and post-interference cancellation schemes | - Investigation of novel multiple access schemes, such as RSMA |
| Self Interference | Combination of analog and digital cancellation schemes | - Complexity and cost of implementation in user devices |

Table 10: Interference Management in 5G and Beyond Networks - Summary and Open Challenges

severely limit the expected performance. Moreover, the dynamic mobility of in-X sub-networks, such as in-body and in-vehicle sub-networks, would potentially trigger highly dynamic interference behavior compared to typical cellular setups with static BSs, which poses a great challenge to radio resource management [242]. Therefore, interference management is of paramount importance in in-X sub-networks which needs to mitigate not only cellular intra and inter-sub-networks interference but also non-cellular interference types such as impulsive noise, and jamming which is one of the major potential threats to ultra-reliable applications [227, 242].

- The concept of Integrated Sensing And Communications (ISAC) is emerging as a key feature of 6G RAN which is attracting significant attention from both industry and academia [243, 244]. This is mainly motivated by the emerging applications, including autonomous cars, smart cities, and industrial Internet-of-things, requiring integration of positioning, sensing, and communication capabilities [227, 243]. By sharing the same radio resources, as well as key elements including waveform, signal processing pipelines and hardware platforms [245], ISAC enhances resource efficiency and achieves mutual benefits going beyond the two separate functionalities [243]. Nevertheless, due to the hardware and radio resource sharing, ISAC may suffer from severe mutual interference between sensing and communications, which complicates the signal processing procedure [243, 246]. In addition to mutual interference, self interference which occurs when the echo signal returns to the RX and gets interfered with self-TX transmission that is not completed yet, is a main factor impacting the sensing performance of a full-duplex ISAC system.

- To realize high-capacity high-speed worldwide and ubiquitous connectivity, 6G networks are envisioned to complement the terrestrial networks with non-terrestrial communication networks, including satellite constellation, high-altitude platforms, and unmanned aerial vehicles, etc. [247, 248, 249]. While the Space-Terrestrial Integrated Network (STIN) will play a critical role in the forthcoming 6G in extending coverage and supporting more mobile traffic in hotspots, it will likely lead to an increased level of multi-dimensional interference that may degrade the wireless links and lead to poor performance [250, 247]. First, the spectrum sharing between heterogeneous terrestrial and non-terrestrial networks will cause more congestion and bring inter-system interference. While diverse interference management techniques have been proposed for HetNets in 5G, the challenge of interference gets more complicated and challenging in vertical HetNets as cell interactions at higher altitudes are very different from those of terrestrial networks, and high-altitude LoS propagation would result in higher downlink interference affecting both terrestrial DL and UL performance [251, 252]. Hence more efficient spectrum management techniques are required [253]. Second, in a typical multi-beam scenario, it would be difficult to construct completely isolated beams while providing

Page 53 of 63

seamless coverage, which may lead to inter-beam interference [250]. As traditional interference management techniques are developed for individual application scenarios or specific types of interference, they are unable to satisfy the requirements of the STIN architecture [247].

## 12.3. Novel Interference Management Paradigms and Future Research Directions

While beyond 5G networks are envisioned to provide significantly high-speed data transmission, they will likely suffer from interference in different application scenarios because of high-level spectrum utilization, massive connectivity, and the complex dynamic wireless environment. While several novel 6G features add some challenging interference-related issues, some potential 6G technological enablers may be seen on the contrary as efficient ways to reduce or efficiently deal with such interference. In the following, we identify and present some of these novel paradigms and methods starting with the expected enhancements to the 6G new air interface including IRS, cell-free massive MIMO, and novel multiple access schemes. Then, to tackle the limitations of traditional interference management techniques, we will overview the integration of Artificial Intelligence as a 6G key paradigm and present insightful guidelines for an AI-based interference management solution.

A) *Enhanced 6G Air Interface*

Facing new challenges including various types of interference, multiple emerging technologies are expected to play their key roles in the evolution towards the next generation 6G air interface. This will be done via significant shifts from active antenna to passive reflective surfaces, from cellular to cell-free architecture, and from conventional MA schemes to new generation MA schemes. In the following, we briefly introduce these three essential paradigms in 6G RAN, their useful impact on interference generation and/or management, and the main remaining related challenges.

- Due to reflections, scattering, and fading in the propagation channel, the received signal can be uncontrollable and subject to random delays and attenuation, which may impact the channel estimation and signal processing and result in network performance degradation [234]. Furthermore, with the use of higher frequency bands, on one hand, the channel conditions are more difficult due to higher propagation loss, lower diffraction, and more blockage [235, 232]. On the other hand, as the wavelength increases, the size of each antenna element decreases, making it more difficult to design operating antenna arrays [235]. To tackle all these challenges, the technology of Intelligent Reflecting Surfaces has been recognized for dynamic configuration of the wireless propagation environment by introducing additional and controllable scattering, and for enhancing current beamforming approaches [232, 254]. It is regarded as a powerful solution for maximizing data rate and minimizing transmit power in future 6G systems, providing an efficient and cost-effective approach to cope with the co-channel interference problem [228]. With some unique and appealing characteristics, IRS changes the research of wireless communication from the conventional design of transceivers to the redesign of transmitter-environment-receiver [254]. Basically, an IRS is a thin two-dimensional surface, also known as a metasurface made up of a multitude of passive (or semi-passive) elements reflecting all incident electromagnetic signals with tunable phase shift and controllable scattering properties without the need for traditional RF chains [228, 234]. The propagation characteristics of a channel can therefore be configured and via passive beamforming, the desired signal can be added constructively to increase the received signal power, while an unwanted signal can be added destructively to remove interference and also improve the energy efficiency of the network [234]. Meanwhile, compared to conventional active MIMO arrays, the integration of IRS with back-scatter communication can balance cost, complexity, and power consumption as RF components are generally not needed, and efficiently suppress the interference induced by the active transmission. In addition, because of the simple passive structure of IRS, there is no self-generated extra circuit interference like in conventional active relays [254]. Despite all these prospects, there are still unresolved research challenges to be addressed in the areas of hardware implementation, channel modeling, evaluation, and real-time control.

- Severe inter-cell interference, frequent handovers, and the large variations in QoS between center and edge cell users, are considered as the main drawbacks of conventional cellular networks, especially in dense deployments [252, 228]. By conveniently combining the most interesting features of UDN, mMIMO, and CoMP with joint transmission/reception, the concept of user-centric cell-free massive MIMO (CF mMIMO) is viewed as a 6G key solution to turn the co-channel interference into a useful signal and achieve nearly uniform data rates and seamless handover across the coverage area [255, 235]. To achieve large-scale fading diversity in such systems, instead of traditional access points (APs) with large co-located antenna arrays, a high number of cost-effective wireless APs, each with a limited number of antennas, are distributed across a coverage zone and are connected via the front-haul connections to one or multiple central processing units to ease their cooperation [256, 235]. Furthermore, different from conventional CoMP technology, where the cellular network is divided into disjoint clusters and thus cell cluster edge UEs are subject to interference from nearby cell clusters, in user-centric CF mMIMO network, every UE is served cooperatively by a chosen set of its neighboring APs, eliminating the interference that restricts conventional ultra-dense networks and resulting in a cell-free network. While such a system can be integrated with multiple emerging technologies and has shown great potential in improving the network performance from various perspectives, the main challenge is to enable a practical deployment of massively large networks with scalable computational complexity and front-haul requirements [257].

- Originally designed for human-centric wireless networks [258], conventional multiple access schemes need a rethink in 6G, especially due to the significant gap between available radio resources and massive connections in IoT 6G networks [225]. Current systems use essentially orthogonal MA, grant-based [259], or space division MA schemes, which may not fit future 6G networks. Therefore, advanced multiple access schemes are being designed, namely Next Generation Multiple Access (NGMA), capable of serving a large number of users in a more resource and complexity-efficient way than the current ones. While the investigation of NGMA is still in the infancy stage, non-orthogonal transmission strategies, especially grant-free NOMA and RSMA, are among the most promising candidates for NGMA, enabling higher spectral and energy efficiency, user fairness, and QoS enhancements. Furthermore, RSMA, which has shown superior performance compared to spatial division MA, NOMA, and orthogonal schemes [45], provides more flexibility in managing interference by partially decoding it and partially treating it as noise and is more robust to imperfect CSI at the transmitter [225, 206]. Despite the great potential and research advances, developing advanced next-generation multiple access techniques remains an open issue. They are the fundamental enablers for efficiently accommodating massive communications via appropriate interference control and resource allocation. Besides, traffic scheduling policy is also important for future enriched user service experience, eMBB and URLLC reliability requirements [260].

B) *Artificial Intelligence*

Traditional interference management approaches will unfortunately face serious limitations in future 6G networks, for the reasons mentioned below:

- Scalability: with the explosion in the number of user devices and network nodes, it becomes impossible to obtain accurate, real-time information about the channel state between all interfering equipment. It also becomes difficult to formulate the optimization problem and find a global optimal solution without increasing the computational complexity.

- Massive heterogeneity: including different spectrum bands, different waveform settings, different network traffic loads, different DL/UL configurations, different user types and speeds, different network architectures, etc. All this configuration flexibility and rapidly changing environments require highly dynamic and adaptive interference management approaches.

Emerging machine learning and artificial intelligence approaches including deep learning [261] and deep reinforcement learning (DRL) techniques are predicted to be one of the key technology enablers to efficiently manage experienced interference and optimize resource allocation in beyond 5G networks [262]. AI algorithms can exploit the tremendous amount of generated data between heterogeneous networks and user equipment to develop dynamic and automated interference management schemes. Several works pinpointed the benefits of using AI-based schemes to manage radio resources within 5G and beyond networks such as [263]. AI is seen as a paradigm shift towards 6G, where the intelligence of the network will be implemented from the transceiver to the receiver, and from the physical layer to the application layer, to perform channel estimation and interference detection, dynamic spectrum management and resource scheduling, intelligent precoding and interference cancellation, etc [264]. In the following, we briefly present some guidelines for an AI-based interference management solution and the future scope of research.

*Guidelines for an AI-based Interference Management Solution*: To efficiently manage interference in beyond 5G networks, we envision a distributed multi-agent deep reinforcement learning-based solution that permits the different agents to interact with the dynamic environment and via a trial and error process, they would learn optimal policies to maximize certain utility while avoiding or exploiting generated interference. Modeled as a stochastic game, such a solution will have to exhibit the following characteristics:

- Multiple heterogeneous agents with varying requirements, including base stations, user devices, unmanned aerial vehicles, etc., cooperate to either avoid, suppress, or exploit the interference generated in the network.

- The solution should be hierarchical allowing the optimization of different network parameters on different multi-time scales.

- The local state information observed by each agent should take into account that the feedback can be partial and outdated due to the channel aging and the experienced delay.

- The action space of each agent should not be restricted to discrete values limiting the scope of the optimization, for instance when choosing power transmission values. Hybrid discrete-continuous spaces have to be further investigated for more reasonable modeling while controlling their complexity.

- The solution should target the optimization of multiple metrics, and therefore a multi-objective reward taking into account interference levels should be formulated.

- The solution should be scalable in terms of the number of agents or the dimension of the state and action spaces. Some possible approaches include Mean Field Reinforcement Learning for homogeneous agents, where the dynamics within different agents are represented by those between a given agent and the mean effect of the whole population or nearby agents.

- For efficient and faster learning, the agents may federate their computation and share either experiences, parameters, or policies.

- One important hurdle to be tackled is the robustness of the solution and the model's capacity to generalize to

unseen environments, requirements, and tasks. Basically, the generated policy should be transferable and to manage multiple types of interference, in different heterogeneous networks, with different users' requirements.

- The AI-based interference management solution should be explainable and its policy model understood by human experts to be considered trustworthy. Deep Neural Networks (DNN) are considered as an opaque black box which makes the interpretation of the results very difficult, especially when DNN are coupled with RL, where the MDP is integrated with hidden layer dynamics. Thus, there is a need to develop explainable DRL algorithms. One possible way is the use of Shapley values, for identifying the contribution of individual players to the outcome of a cooperative game.

- Another big challenge in integrating artificial intelligence in wireless communications, is to ensure safe decisions. An AI-based interference management solution should therefore be able to measure the risks of the output actions and exclude the risky ones leading to high interference levels and network performance degradation. This can be performed using a digital twin of the network, which is a virtual representation reflecting all the dynamics underlying the investigated system, to estimate the risk of decision-making in advance [265].

- In addition to the accurate modeling of the Markov game and its components, the choice of the DRL algorithm, its implementation, and the setting of the hyper-parameters, are crucial to an efficient solution. Ideally, the learning algorithm should have a mathematical foundation proving its convergence, optimality, and sample efficiency. There are several multi-agent DRL algorithms, including MADDPG [266], MAPPO [267], HATRPO [268], HAPPO [268], etc., each having its pros and cons.

## 13. Conclusion

In this article, we have provided a comprehensive survey of interference management in 5G and Beyond networks. After a global overview of the main 5G features and their potential impact on interference generation, we have presented a unified classification and a detailed explanation of the main encountered interference types that can degrade the performance in such systems, including remote interference and inter-numerology interference which have never been listed in previous surveys. We also provided a taxonomy categorizing existing interference management approaches and a detailed explanation of interference measurement reports and signaling in 5G NR according to the newest 3GPP specifications. For each previously identified interference, an in-depth literature review and technical discussion of appropriate management schemes were presented. To help new starters conducting research in this field, we outlined the main interference management challenges in 4G/5G and Beyond networks that have not yet been explored or need to be further studied, along with the most promising approaches. Moreover, we provided an overview of envisioned key technologies in 6G and Beyond networks with a focus on their associated interference issues. We finished the article by identifying and presenting some insights on the main novel paradigms and research directions for advanced interference management schemes in future networks. As conventional interference management approaches will likely face serious limitations, we proposed some useful guidelines for an AI-based interference management solution along with the future challenges that need to be addressed.

| Key 6G Features | 6G Interference Challenges/Opportunities | Research Directions |
|---|---|---|
| New frequency spectrum (THz, optical bands) | Interference between the transceiver miniature components<br>Inter sub-bands and systems interference<br>Severe beam squint and inter-beam interference | Dynamic, efficient, and self-adaptive spectrum sharing and interference management schemes |
| Novel waveforms and modulation techniques | More pronounced inter-numerology interference than in 5G<br>Inter-waveform interference | Advanced signal processing schemes |
| In-X sub-networks | High device density scenarios and highly dynamic interference behavior –> Cellular intra and inter-sub-networks, and non-cellular interference | AI-based proactive interference management schemes |
| Integrated Sensing And Communications | Severe mutual interference between sensing and communications<br>Self Interference impacting the sensing performance | Enhanced radio resource sharing and signal processing |
| Space-Terrestrial Integrated Network | Severe multi-dimensional interference, including inter-system and intra-system inter-beam interference | Advanced inter-beam and spectrum management techniques |
| Intelligent Reflecting Surfaces | Dynamic configuration of the wireless propagation environment –> an efficient and cost-effective approach for co-channel interference management | Open challenges related to hardware implementation, channel modeling, evaluation, and real-time control |
| Cell-free mMIMO | Turning the co-channel conventional inter-cell interference into a useful signal | Enabling a practical deployment of massively large networks with scalable computational complexity and front-haul requirements |
| Next Generation Multiple Access | Allowing massive communications and multi-user interference management | Integration with other 6G features and performance assessment |
| End-to-end Artificial Intelligence | Exploitation of the tremendous amount of generated data to develop dynamic and automated interference management schemes | Accurate modeling and implementation of an efficient, safe, scalable, explainable, robust, transferable, and multi-objective AI-based solution |

Table 11: Envisioned Key 6G Features, Challenges and Opportunities for Interference Management

Interference Management in 5G and Beyond Networks

Page 62 of 63